\documentclass[reprint,superscriptaddress,amssymb,amsmath,aip,rsi,longbibliography]{revtex4-2}

\usepackage{floatrow}
\usepackage{graphicx}
\usepackage{amssymb}
\usepackage{epstopdf}
\usepackage{upgreek}
\usepackage{dcolumn}
\usepackage{bm}
\usepackage{gensymb}
\usepackage{braket}
\usepackage{amsmath}
\usepackage{mathtools}
\usepackage{float}
\usepackage{tikz}
\usepackage[colorlinks=true,allcolors=blue]{hyperref}

\usepackage{array}
\newcolumntype{C}[1]{>{\centering\arraybackslash}m{#1}}

\usepackage{xcolor}

\usepackage[resetlabels,labeled]{multibib}

\usepackage{xfrac}

\usepackage{comment}

\expandafter\ifx\csname package@font\endcsname\relax\else
\expandafter\expandafter
\expandafter\usepackage
\expandafter\expandafter
\expandafter{\csname package@font\endcsname}%
\fi
\newcommand{\ie}{\emph{i.e.}}

\newcommand{\etc}{\emph{etc.}}

\newcommand{\micro}{$\upmu$}

\newcommand{\be}{\begin{eqnarray}}
	\newcommand{\ee}{\end{eqnarray}}
\newcommand{\bfig}{\begin{figure}}
	\newcommand{\efig}{\end{figure}}

\definecolor{myorange}{rgb}{0.97,0.59,0.27}

\DeclareFontFamily{U}{mathb}{}
\DeclareFontShape{U}{mathb}{m}{n}{
	<-5.5> mathb5
	<5.5-6.5> mathb6
	<6.5-7.5> mathb7
	<7.5-8.5> mathb8
	<8.5-9.5> mathb9
	<9.5-11.5> mathb10
	<11.5-> mathbb12
}{}

\begin{document}
	\makeatletter
	\@namedef{figure}{\killfloatstyle\def\@captype{figure}\FR@redefs
		\flrow@setlist{{figure}}%
		\columnwidth\columnwidth\edef\FBB@wd{\the\columnwidth}%
		\FRifFBOX\@@setframe\relax\@@FStrue\@float{figure}}%
	\makeatother
	
	\title{Low-noise cryogenic microwave amplifier characterization with a calibrated noise source}
	\author{M. Malnou}
	\email{maxime.malnou@nist.gov}
	\affiliation{National Institute of Standards and Technology, Boulder, Colorado 80305, USA}
	\affiliation{Department of Physics, University of Colorado, Boulder, Colorado 80309, USA}
	\author{T. F. Q. Larson}
	\affiliation{National Institute of Standards and Technology, Boulder, Colorado 80305, USA}	
	\affiliation{Department of Physics, University of Colorado, Boulder, Colorado 80309, USA}
	\author{J. D. Teufel}
	\affiliation{National Institute of Standards and Technology, Boulder, Colorado 80305, USA}	
	\author{F. Lecocq}
	\affiliation{National Institute of Standards and Technology, Boulder, Colorado 80305, USA}	
	\author{J. Aumentado}
	\affiliation{National Institute of Standards and Technology, Boulder, Colorado 80305, USA}
	
	\begin{abstract}    
		Parametric amplifiers have become a workhorse in superconducting quantum computing, however research and development of these devices has been hampered by inconsistent, and sometimes misleading noise performance characterization methodologies. The concepts behind noise characterization are deceptively simple, and there are many places where one can make mistakes, either in measurement or interpretation and analysis. In this article we cover the basics of noise performance characterization, and the special problems it presents in parametric amplifiers with limited power handling capability. We illustrate the issues with three specific examples: a high-electron mobility transistor amplifier, a Josephson traveling-wave parametric amplifier, and a Josephson parametric amplifier. We emphasize the use of a 50-$\Omega$ shot noise tunnel junction (SNTJ) as a broadband noise source, demonstrating its utility for cryogenic amplifier amplifications. These practical examples highlight the role of loss as well as the additional parametric amplifier `idler' input mode.

	\end{abstract}
	
	\maketitle
	
	\section{Introduction}
	
	
	Ultra low noise microwave amplifiers have played a critical, often transformative, role in several areas of scientific study. For example, they have allowed the widespread dissemination of superconducting quantum sensors for astronomy and astrophysics observations \cite{Galitzki2018the,McCarrick2021The,Noroozian2013high,mates2017simultaneous,Nakashima2020low,Szypryt2021design,malnou2023improved}. They have also allowed the development of circuit-based superconducting quantum information and computing research \cite{arute2019quantum,Krinner2022realizing,Acharya2023suppressing}. More recently they have boosted interest in physics beyond the standard model, via their use in axion dark matter searches \cite{brubaker2017first,du2018search,Backes2021a}.
	
	
	The majority of these ultra low noise amplifiers consist of superconducting Josephson junction-based circuits, operating as parametric amplifiers \cite{yurke1989observation,castellanos2008amplification,bergeal2010phase,macklin2015near}, that can be mounted on the mixing chamber stage of a dilution refrigerator close to the devices-under-test. These are preamplifiers, the first stage of signal amplification, and have performance that can approach the so-called quantum limit for added noise, with sufficient gain to overwhelm the remaining post-amplification chain, usually a cryogenic high electron mobility transistor (HEMT) amplifier placed at the 3--4 K stage followed by additional room temperature amplifiers to raise the signal power to suitable levels for demodulation and processing.
	
	
	The continued development of these amplifiers requires a suite of characterization tools that are capable of operating at these low temperatures with methodology that is not too cumbersome for laboratories that do not specialize in metrology. There are already methods outlined in the literature discussing how one can approach this problem for scattering parameter calibration and characterization \cite{ranzani2013two, yeh2013situ, stanley2022characterizing}, yet the problem of characterizing very low amplifier added noise is comparatively difficult. Much of the problem originates in the fact that added noise characterization is an absolute measurement, requiring a known noise power level at the input of the amplifier, while scattering parameter measurements are relative, \ie, the ratio of outgoing to incident voltage wave amplitudes.
	
	In this article, we outline the basic definitions and methodology used when characterizing the noise of nearly-quantum-limited amplifiers. We also shed light on some common pitfalls encountered in the interpretation of measurements when using a broadband calibrated noise source. We discuss critical considerations in designing an amplification chain as well as how to use a shot noise tunnel junction (SNTJ), a self-calibrating noise source, to characterize amplifier noise performance. With the SNTJ we measure the noise of commonly-used amplification chains where the first amplifier is either a HEMT, a Josephson traveling-wave parametric amplifier (JTWPA), or a Josephson parametric amplifier (JPA). Although the SNTJ is a relatively recent addition to the toolbox of noise characterization tools \cite{spietz2006shot, chang2016noise} and not as well-known as other methods, it is a  versatile noise source that is well-suited to cryogenic microwave characterization. We highlight this by comparing its operation to that of a variable temperature stage (VTS), on which the SNTJ is mounted.
	
	\section{Theory of noise measurement}
	
	\subsection{Elements in an amplification chain}
	\label{sec:ampchain}
	
	\paragraph{Units and definitions} 
	We focus on linear amplification of a white noise, which we consider as our ``signal'' \cite{beenarker2003quantum}. Randomly fluctuating, it follows a Gaussian statistics with zero mean. The noise is therefore fully characterized by its associated standard deviation, whose power spectral density $P$ (in units of W/Hz) can be measured, in practice, with a spectrum analyzer. Throughout the article we express the noise in photon-normalized units, $N=P/(h\nu)$, with $h$ the Planck constant and $\nu$ the photon frequency. Note that $N$ is \textit{not} a photon-number operator, in particular $N$ already includes vacuum fluctuations (see appendix \ref{app:op2quantapi}). In the literature the noise is also sometimes normalized to a temperature $T_N=P/k_B$ (in K), with $k_B$ the Boltzmann constant. For energies much larger than that of vacuum, $T_N$ represents the equivalent temperature of a resistor that would dissipate the same amount of noise into a matched circuit (see Eq.\,\ref{eq:Johnson}).

\begin{figure}[h!]	
		\includegraphics[scale=0.8]{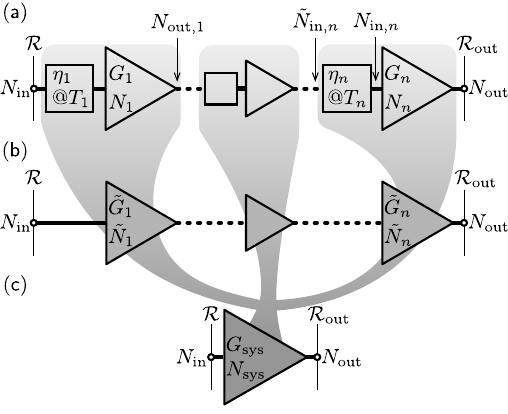}
		\caption{Schematic of an amplification chain. (a) An input signal $N_\mathrm{in}$ enters the chain at the reference plane $\mathcal{R}$ and passes through successive stages of attenuation, characterized by their transmission efficiencies $\eta_k$ and temperatures $T_k$, with $k\in\{1,...,n\}$, alternating with successive stages of amplification, characterized by their gain $G_k$ and added noise $N_k$, with $k\in\{1,...,n\}$. At the output reference plane $\mathcal{R}_\mathrm{out}$, the resulting amplified signal is $N_\mathrm{out}$. (b) Each pair of amplification and loss stage can be reduced to an effective amplifier, with gain $\Tilde{G}_k$ and added noise $\Tilde{N}_k$. (c) The whole chain itself can be reduced to an effective amplifier, with gain $G_\mathrm{sys}$ and added noise $N_\mathrm{sys}$.} 
		\label{fig:ampchain_theo}
	\end{figure}

	\paragraph{Amplification}            
	In the intuitive picture of linear amplification (corresponding to the so-called phase preserving - or phase-insensitive - case), all the components of a signal get identically amplified while traveling along an amplification chain, consisting of a succession of attenuation and amplification stages, see Fig.\,\ref{fig:ampchain_theo}. Then, the $k$-th amplifier of the chain not only amplifies an input signal by some gain $G_k$, but also adds some unavoidable noise $N_k$ on top of the signal. The input/output relation of that amplifier thus reads
	\begin{equation}
		N_{\mathrm{out},k} = G_k(N_{\mathrm{in},k}+N_k).
		\label{eq:ampeq}
	\end{equation}
	Here, $N_{\mathrm{in},k}$, $N_{\mathrm{out},k}$ and $N_k$ are all Gaussian noises, and in fact they all are in photon-normalized units. But $N_{\mathrm{in},k}$ is (or contains) a known and controlled quantity, while $N_k$ is not. That is why we will we call $N_{\mathrm{in},k}$ ($N_{\mathrm{out},k}$) the input (output) signal of the $k$-th amplifier, and we will call $N_k$ the added noise of that amplifier.
	
	
	\paragraph{Attenuation}
	A loss stage attenuates the signal, and replaces a portion of that signal by some Johnson noise $N_{T_k}$, whose value depends on the temperature $T_k$ of the stage (see Eq.\,\ref{eq:Johnson}). More precisely, the $k$-th loss stage acts as a beamsplitter and transforms an input signal $\Tilde{N}_{\mathrm{in},k}$ (equal to the output of the preceding amplifier, $N_{\mathrm{out},k-1}$) to an output signal $N_{\mathrm{in},k}$ (which corresponds to the $k$-th amplifier's input) following:
	\begin{equation}
		N_{\mathrm{in},k}=\eta_k \Tilde{N}_{\mathrm{in},k} + (1-\eta_k)N_{T_k},
		\label{eq:atteq}
	\end{equation}  
	where $0<\eta_k<1$ is called the transmission efficiency \cite{wallsmilburn2007quantum}.   
	
	\paragraph{Friis formula}
	Combining the $k$-th amplification stage (Eq.\,\ref{eq:ampeq}) with the preceding $k$-th loss stage (Eq.\,\ref{eq:atteq}), we can define an effective amplifier, with gain $\Tilde{G}_k$ and added noise $\Tilde{N}_k$, that transforms the input signal $\Tilde{N}_{\mathrm{in},k}$ to an output signal $N_{\mathrm{out},k}$ such that
	\begin{equation}
		\begin{aligned}        
			N_{\mathrm{out},k} &= \Tilde{G}_k(\Tilde{N}_{\mathrm{in},k} + \Tilde{N}_k)\\
			&=\Tilde{G}_k(N_{\mathrm{out},k-1} + \Tilde{N}_k),
			\label{eq:noutk}
		\end{aligned}
	\end{equation}
	where 
	\begin{align}
		\Tilde{G}_k&=\eta_k G_k \label{eq:tildeGk},\\
		\Tilde{N}_k&=\frac{(1-\eta_k)N_{T_k} + N_k}{\eta_k} \label{eq:tildeNk}.
	\end{align}
	Critically, it becomes apparent that the preceding loss stage degrades the performance of the amplifier, since $\Tilde{G}_k\leq G_k$ and $\Tilde{N}_k\geq N_k$.
	
	The whole chain is then defined between two reference planes: an input reference plane $\mathcal{R}$ (or simply ``the reference plane'' in the following) located where the input signal $N_\mathrm{in}$ enters the chain, and an output reference plane $\mathcal{R}_\mathrm{out}$, located where we measure the output signal $N_\mathrm{out}$ (usually inside a lab instrument). Iteratively applying Eq.\,\ref{eq:noutk}, the whole amplification chain can be reduced to a single effective amplifier, transforming $N_\mathrm{in}$ to $N_\mathrm{out}$ such that
	\begin{equation}
		N_\mathrm{out} = G_\mathrm{sys}\left(N_\mathrm{in} + N_\mathrm{sys}\right),
		\label{eq:Nout}
	\end{equation}
	where $G_\mathrm{sys}=\prod_{k=1}^n\Tilde{G}_k$ is the gain of the whole amplification chain (also called system gain) and where
	\begin{equation}        
		N_\mathrm{sys} = \Tilde{N}_1 + \frac{\Tilde{N}_2}{\Tilde{G}_1} + \frac{\Tilde{N}_3}{\Tilde{G}_1\Tilde{G}_2} + ...
		\label{eq:Nc}
	\end{equation}
	is the system-added noise. Equation \ref{eq:Nc} is the so-called Friis formula \cite{Friis1944noise}. We must ensure that $\Tilde{N}_1\gg\Tilde{N}_2/\Tilde{G}_1\gg \mathrm{etc}...$ for the amplifiers' gains to overwhelm the following noisy elements. The amplification chain is then well designed, and $N_\mathrm{sys}$ is dominated by $\Tilde{N}_1$, which accounts for both the added noise of the first amplifier and for the transmission efficiency $\eta_1$ between the reference plane and the input of that amplifier. 
	
	
	\subsection{Principle of a noise measurement}
	
	The goal of a noise measurement consists of obtaining a fair evaluation of the system-added noise $N_\mathrm{sys}$, and possibly also $N_1$, the first amplifier added noise. \textit{Both} matter, for different reasons. In fact, $N_\mathrm{sys}$ is the only quantity that can be directly measured, and it is ultimately the quantity that a user cares about. But it is chain-dependent: in particular, it depends on $\eta_1$, which likely differs from system to system. Inversely, $N_1$ is chain-independent, but heavily relies on calibration measurements. It can be calculated by subtracting out the effect of $\eta_1$ to $N_\mathrm{sys}$, i.e. by ``moving the reference plane'' of the chain. However, in many cases the question of exactly where the reference plane should be moved to is debatable: when moved up to the amplifier input, it may be discarding the loss inserted by standard microwave components mandatory for its operation. Also, any amplifier packaging inserts loss, therefore in practice any amplification stage adds some noise.
	
	Measuring a system-added noise always relies on knowing the amount of noise somewhere within the chain, usually at the input reference plane $\mathcal{R}$. Then, measuring the output noise $N_\mathrm{out}$ while varying $N_\mathrm{in}$, a least-square regression gives both $G_\mathrm{sys}$ and $N_\mathrm{sys}$ as shown by Eq.\,\ref{eq:Nout}, with a caveat when the first amplifier is parametric: in that case, the y-intercept of the regression is usually \textit{not} $N_\mathrm{sys}$, see Sec.\,\ref{sec:noiseparamp}. This measurement is called a \textit{Y-factor} \cite{Evans1977rf}, and the noise source generating $N_\mathrm{in}$ can be, for example, a hot/cold load \cite{brubaker2017first,zobrist2019wide}, a variable temperature stage (VTS) \cite{malnou2018optimal,malnou2021three,Simbierowicz2021characterizing}, a shot noise tunnel junction (SNTJ) \cite{Spietz2003primary,spietz2006shot,chang2016noise,malnou2021three}, a qubit \cite{macklin2015near}, a nanowire \cite{bergeal2010phase}, or a diode at room temperature in front of a cold attenuator\cite{cano2010ultra}. Ideally, $N_\mathrm{in}$ need only take two values, but in practice it is desirable to sweep $N_\mathrm{in}$ to calculate uncertainties on $N_\mathrm{sys}$ and $G_\mathrm{sys}$ and to get rid of experimental artifact: for example, the chain's gain may not be perfectly linear over the variation range, which cannot be diagnosed with only two points. 
	
	A variant of the Y-factor, called the `noise visibility ratio' \cite{white2015traveling,malnou2022performance} relies on knowing the system-added noise $N_\mathrm{sys}^\mathrm{off}$ when the first amplifier is turned off ($G_1=1$) while keeping $N_\mathrm{in}$ to a fixed value (for example $N_\mathrm{in}=1/2$). Then, turning the amplifier on, the output noise rises by a measurable amount $r$ such that
	\begin{equation}
		r = G_1\frac{N_\mathrm{in}+N_\mathrm{sys}^\mathrm{on}}{N_\mathrm{in}+N_\mathrm{sys}^\mathrm{off}},
	\end{equation}    
	from which we can deduce $N_\mathrm{sys}^\mathrm{on}$ (the system-added noise with the amplifier on), see appendix \ref{app:nvr}. In general, this technique is approximate, because it is often difficult to precisely know $N_\mathrm{sys}^\mathrm{off}$. In particular it can significantly depart from $N_2$, the intrinsic noise added by the second amplifier (that may be given in a datasheet) due to the loss before the second amplifier. Nonetheless, the noise rise technique is useful to obtain a quick estimate of the system-added noise.
	
	\subsection{Noise sources}
	
	Varying $N_\mathrm{in}$ necessitates the use of a noise source, and in the following, we focus on the two most versatile self-calibrated noise sources - a resistor and a SNTJ - leaving aside techniques utilizing a qubit (which can only give a narrow-band noise characterization), a nanowire (which only works in the very low temperature regime) or a noise diode (whose emitted noise is in general not tunable, and which relies on a calibrated attenuator).
	
	\subsubsection{Variable temperature resistor}    
	
	In the case of the hot/cold load, the noise source consists of two resistors at two different temperatures. A variable temperature stage (VTS) is a generalization, where the noise source is a single resistor whose temperature is made variable (it also evades issues with a possible unequal attenuation between the two resistors paths and the first amplifier). In such cases, the noise \textit{dissipated} by the resistor at temperature $T$ into a matched circuit is given by the (quantum) Johnson noise formula:
	\begin{equation}
		\begin{aligned}
			N_\mathrm{in}^\mathrm{Johnson} &= \frac{1}{e^{\sfrac{h\nu}{k_BT}}-1}+\frac{1}{2} \\ 
			&= \frac{1}{2}\coth{\left(\frac{h\nu}{2k_BT}\right)}\xrightarrow[h\nu\ll k_B T]{}\frac{k_BT}{h\nu}.
		\end{aligned}
		\label{eq:Johnson}
	\end{equation}
	When $k_B T \ll h \nu$, we have $N_\mathrm{in}^\mathrm{Johnson}=1/2$. Conversely, in the classical limit where $h \nu \ll k_B T$ (indicated the ``$\rightarrow$'') we obtain the classical Johnson noise power spectral density: $S_P^\mathrm{Johnson} = k_B T$, expressed in W/Hz (appendix \,\ref{app:spectraldensities} discusses the various expressions of noise spectral densities). Therefore in the classical limit, it is a white noise.
	
	
	\subsubsection{The shot-noise tunnel junction}
	\label{sec:SNTJtheo}
	
	A SNTJ relies on the shot noise associated to the random tunneling of electrons through a normal-insulator-normal tunnel junction, with characteristic impedance close to $50$\,\ohm{}. Like any resistor, the current-voltage relation of the SNTJ is linear; however, unlike a typical resistor, its noise increases with dc bias. A voltage bias $V$ accross the SNTJ shifts the Fermi levels on both sides of the junction, which influences the tunneling rate, and with it the shot noise level. More precisely, the noise at a frequency $\nu$ dissipated by the SNTJ into a matched circuit is equal to \cite{spietz2006shot,lecocq2017nonreciprocal}
	\begin{equation}
		\begin{aligned}
			N_\mathrm{in}^\mathrm{shot} = &\frac{eV+h\nu}{4h\nu} \coth{\left(\frac{eV+h\nu}{2k_BT_e}\right)} \\ 
			+ &\frac{eV-h\nu}{4h\nu} \coth{\left(\frac{eV-h\nu}{2k_BT_e}\right)}\\
			&\xrightarrow[\{k_BT_e , h\nu \} \ll e\lvert V\rvert]{}\frac{e\lvert V\rvert}{2h\nu},    
		\end{aligned}
		\label{eq:shot}
	\end{equation}
	where the two hyperbolic cotangent terms account for quantum corrections that arise at low temperatures, when $k_BT_e \ll h\nu$, which is the regime we focus on in this article. Here, $e$ is the electron charge and $T_e$ is the temperature of the electrons. In practice, $T_e$ will be a fit parameter. Note that Eq.\,\ref{eq:shot} presents other limit cases, depending on how $eV$, $k_BT$ and $h\nu$ compare with each other, see appendix \ref{app:sntjcases}. 
	
	At zero bias $N_\mathrm{in}^\mathrm{shot}=N_\mathrm{in}^\mathrm{Johnson}$ i.e. we are in the Johnson noise limit, and the SNTJ behaves like a $50$\,\ohm{} load. Inversely, when $h\nu \ll e\lvert V\rvert$ we are in the classical shot noise limit, and $N_\mathrm{in}^\mathrm{shot} = eV/(2h\nu)$ (see also appendix \ref{app:spectraldensities}). Converting to W/Hz, we obtain the classical shot noise power spectral density: $S_P^\mathrm{shot}=e \lvert V\rvert / 2$. Therefore in the classical limit, it is also a white noise.
	
	\subsubsection{Comparison between VTS and SNTJ}
	\label{sec:comparison}
	
\begin{figure}[h!]	
		\includegraphics[scale=1]{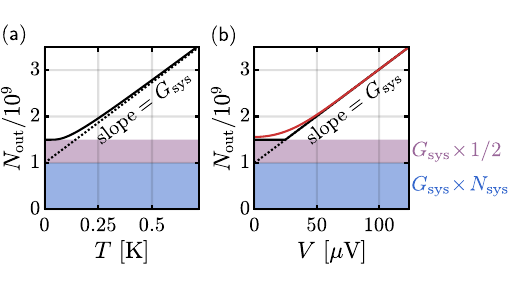}    
		\caption{Theoretical output noise $N_\mathrm{out}$ given by Eq.\,\ref{eq:Nout}, as a function of (a) the temperature $T$ of a VTS or (b) the voltage bias $V$ of a SNTJ, and calculated at $6$\,GHz. In the linear (high bias) regime, the curve is a straight line, with slope $G_\mathrm{sys}$ and y-intercept $G_\mathrm{sys}\times N_\mathrm{sys}$. We arbitrarily chose $G_\mathrm{sys}=10^9$ and $N_\mathrm{sys}=1$. When using the SNTJ, the response is plotted for $T_e=0$\,K (black), and for $T_e=0.1$\,K (red), showing that a non-zero physical temperature as a rounding effect at low biases.} 
		\label{fig:simplenoisetheo}
	\end{figure}

	Figure \ref{fig:simplenoisetheo} shows how $N_\mathrm{out}$ theoretically varies with the temperature $T$ of a VTS (a) and with the bias voltage $V$ of a SNTJ (b), for a given system gain $G_\mathrm{sys}$ and system-added noise $N_\mathrm{sys}$.  Both responses are qualitatively identical. In the simple case where $N_\mathrm{out}$ is related to $N_\mathrm{in}$ via Eq.\,\ref{eq:Nout}, the slope of the linear response (\textit{i.e.} at high bias, in the classical limit) is equal to $G_\mathrm{sys}$, and the y-intercept is equal to the output-referred system-added noise $G_\mathrm{sys}\times N_\mathrm{sys}$. At zero bias, the difference between the output noise and the y-intercept gives the unavoidable half quantum, present within the input signal.
	
	
	
	
	Although both noise sources are conceptually equivalent, using a VTS presents several advantages: (i) it is somewhat straightforward to fabricate, because it only necessitates readily available components: a calibrated thermometer, a heater, and a $50$\,\ohm{} load, all mounted on a copper support. The VTS is then weakly thermally connected to a cryostat plate, for example to the mixing chamber plate. (ii) The precision on $N_\mathrm{in}^\mathrm{Johnson}$ only depends on how well the VTS thermometer has been calibrated, and how matched the $50$\,\ohm{} load is to the readout line; two tasks that have excellent commercial solutions. In comparison, the SNTJ has not, so far, been easily accessible to the superconducting circuit community. However, it presents three major advantages compared to a VTS: (i) it allows for much faster noise measurements, because varying $N_\mathrm{in}^\mathrm{shot}$ is performed with a bias voltage $V$, while varying $N_\mathrm{in}^\mathrm{Johnson}$ with a VTS implies to change and stabilize its temperature. Thus the noise can be changed in nanoseconds instead of minutes.  (ii) It does not generate parasitic heating while, depending on the VTS thermal connection to its anchoring plate, heating up the VTS may affect the cryostat plates' temperatures, in turn modifying the effective noise inserted by the various components in the readout chain, thereby falsifying the simple linear regression of Eq.\,\ref{eq:Nout}. (iii) A SNTJ has a much higher dynamic range than a VTS, fundamentally because the noise generated by a SNTJ relies on the charge constant $e$, while a VTS relies on the Boltzmann constant $k_B$, and $k_B/e=86\,\mathrm{\micro V}/\mathrm{K}$. In other words, it is much easier for a SNTJ to generate a large noise.  For example, a SNTJ biased with 1~mV generates noise equivalent to a VTS at 11.6~K.   
	
	
	\subsection{Description of an ideal parametric amplifier}
	\label{sec:paramp}

	Parametric amplifiers play a key role in many readout applications, because their added noise can be, in theory, at the fundamental quantum limit \cite{caves1982quantum}, so when used as the first amplifier in the chain, they allow us to minimize $N_\mathrm{sys}$. During parametric amplification, photons at a `pump' frequency ($\omega_p$) are used to amplify a `signal' (frequency $\omega_s$) while also creating `idler' photons $\omega_i = n\omega_p - \omega_s$, where $n=1$ or $2$ corresponds to the so-called `3-wave' or `4-wave' mixing operations \cite{boyd2019nonlinear}. A parametric amplifier can be operated (i) as \textit{phase-insensitive} (also called ``phase-preserving''), where its gain does not depend on the signal's phase. This process unavoidably adds at least half a photon worth of noise. (ii) As \textit{phase-sensitive}, where the gain is dependent on the signal's phase. This process need not add any noise.
	
	
	
	\begin{figure}[h!]	
		\includegraphics[scale=0.9]{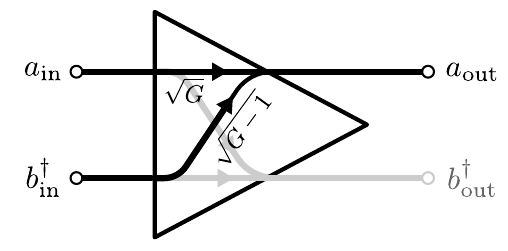} 
		\caption{Schematic of the field operators transformations, executed by a parametric amplifier. In the phase-insensitive case, the operators $a_\mathrm{in}$ and $b^\dagger_\mathrm{in}$ represent the input signal and idler modes. They transform into $a_\mathrm{out}$ and $b^\dagger_\mathrm{out}$. In the phase-sensitive case, for each Fourier component $a_\mathrm{in}(\delta)$ we have $a^\dagger_\mathrm{in}(-\delta)=b^\dagger_\mathrm{in}(-\delta)$ and $a^\dagger_\mathrm{out}(-\delta)=b^\dagger_\mathrm{out}(-\delta)$.} 
		\label{fig:amplifier_gain}
	\end{figure}    
	These modes of operation are governed by constraints on the commutation relations on field operators at the input and at the output of the amplifier \cite{yamamoto1990fundamentals}. Upon amplification, the input operator $a_\mathrm{in}$ transforms into the output operators $a_\mathrm{out}$, and the commutation relation $[a_\mathrm{in},a^\dagger_\mathrm{in}]=1$ must remain true at the output: $[a_\mathrm{out},a^\dagger_\mathrm{out}]=1$.
	
	\subsubsection{Phase-insensitive case}
	
	In the phase-insensitive case, we use an ancillary input mode $b_\mathrm{in}$ such that    
	\begin{equation}
		a_\mathrm{out} = \sqrt{G_1} a_\mathrm{in} + \sqrt{G_1-1}b^\dagger_\mathrm{in},
		\label{eq:aout}
	\end{equation}    
	where $G_1\geq1$ is the signal gain. Equivalently, $b_\mathrm{out}^\dagger$ is expressed as a function of $b_\mathrm{in}^\dagger$ and $a_\mathrm{in}$ \cite{caves1982quantum}. In parametric amplification, where pump photons (at a frequency $\omega_p$) are converted into signal ($\omega_s$) and idler ($\omega_i$) photons, this ancillary mode is at the idler frequency. Recast in terms of photon-normalized signal, Eq.\,\ref{eq:aout} yields:
	\begin{equation}
		N_{\mathrm{out},1} = G_1 N_\mathrm{in} + (G_1-1)N_\mathrm{in}^i,
		\label{eq:Nsout}
	\end{equation}
	where $N_\mathrm{in}=\braket{a^\dagger_\mathrm{in} a_\mathrm{in}} + 1/2$, and $N_\mathrm{in}^i=\braket{b^\dagger_\mathrm{in} b_\mathrm{in}} + 1/2$ are the photon-normalized inputs at the signal and idler frequencies respectively, and where $N_{\mathrm{out},1}=\braket{a^\dagger_\mathrm{out} a_\mathrm{out}} + 1/2$ is the photon-normalized amplifier output at the signal frequency, see appendix \ref{app:op2quantapi}. When $G_1\gg 1$, Eq.\,\ref{eq:Nsout} reduces to
	\begin{equation}
		N_{\mathrm{out},1} = G_1 (N_\mathrm{in} + N_\mathrm{in}^i).
		\label{eq:NsoutGgg1}
	\end{equation}
	The idler input, $N_\mathrm{in}^i$, therefore appears as some undesirable noise. Importantly, this noise does not come from an internal degree of freedom, but rather only depends on the state at the idler's input, and when the input is cold the amplifier adds the minimum amount of noise: $N_\mathrm{in}^i=1/2$. This is the quantum limit on added noise for a phase-insensitive amplifier. Note that the amplifier can be at a physically high temperature and remain quantum-limited, as long as its idler input mode is cold \cite{malnou2022performance}.
	
	As pointed out by Caves, ``quantum mechanics extracts its due twice''\cite{caves1982quantum}: looking at Eq.\,\ref{eq:NsoutGgg1}, there is also a minimum of $1/2$ quantum of fluctuations in $N_\mathrm{in}$ which, in a different context than noise measurements, can be present on top of a displacement. Added to the fluctuations in $N_\mathrm{in}^i$, it defines the \textit{standard quantum limit} (SQL) as 1 quantum worth of noise. It is the minimum amount of total noise that can be measured at the output of a phase-insensitive amplification chain.
	
	\subsubsection{Phase-sensitive case}
	Alternatively, one can exploit the correlations between signal and idler by looking at their linear combinations. This is the phase sensitive case, where the signal and idler Fourier components are centered about $\omega_0 = (\omega_s+\omega_i)/2$. Therefore, in a frame rotating at $\omega_0$ we can pair each Fourier component of the signal $a_\mathrm{in}(\delta)$ to $a^\dagger_\mathrm{in}(-\delta) = b^\dagger_\mathrm{in}(-\delta)$, where $\delta$ is the frequency detuning to $\omega_0$. It yields:
	\begin{equation}
		a_\mathrm{out}(\delta) = \sqrt{G_1(\delta)} e^{-i\theta} a_\mathrm{in}(\delta) + \sqrt{G_1(\delta)-1} e^{i\theta} a^\dagger_\mathrm{in}(-\delta),
	\end{equation}  
	where $\theta$ is the phase difference between signal and pump, and $G_1(\delta)$ remains the phase-insensitive gain (that links the input and output signal at frequency $\omega_s = \omega_0+\delta$). In the following we omit the dependence in $\delta$ to keep the equations to a simple form. The decomposition of $a_\mathrm{out}$ onto the axes of the rotating frame yields:
	\begin{align}
		X_\mathrm{out} &= (X_\mathrm{in}\cos{\theta} + Y_\mathrm{in}\sin{\theta})(\sqrt{G_1} +\sqrt{G_1-1})\\
		Y_\mathrm{out} &= (Y_\mathrm{in}\cos{\theta} - X_\mathrm{in}\sin{\theta})(\sqrt{G_1}-\sqrt{G_1-1}),
	\end{align}
	which are also called the ``quadratures'' of the signal (and the rotating frame is also called the quadrature basis). Here, $X_k = (a_k + a^\dagger_k)/\sqrt{2}$ and $Y_k = (a_k - a^\dagger_k)/(i\sqrt{2})$, with $k=\{\mathrm{in},\mathrm{out}\}$. Rotating the signal quadratures so that $\theta=0$ and assuming $G_1\gg1$, we obtain 
	\begin{align}
		X_\mathrm{out}&=X_\mathrm{in}2\sqrt{G_1} \\
		Y_\mathrm{out}&=\frac{Y_\mathrm{in}}{2\sqrt{G_1}},
	\end{align}
	respectively the amplified and squeezed quadratures. These equations show that neither the amplified nor the squeezed quadrature suffer from any added noise: the phase-sensitive amplification is noiseless. 
	
	At the output of the amplifier, the signal is measured along one quadrature $I_{\mathrm{out},1}$, that forms an angle $\alpha$ with $X_\mathrm{in}$. Therefore we have
	\begin{equation}
		I_{\mathrm{out},1} = X_\mathrm{in}2\sqrt{G_1}\cos{\alpha} + \frac{Y_\mathrm{in}}{2\sqrt{G_1}}\sin{\alpha}.
	\end{equation}
	In practice, $\alpha$ depends on the pump's phase of the parametric amplifier. In photon-normalized units it yields:
	\begin{equation}                
		N_{\mathrm{out},1} = \mathcal{G}_1(\alpha)N_\mathrm{in},
		\label{eq:PoutPS}    
	\end{equation}
	with
	\begin{equation}
		\mathcal{G}_1(\alpha) = 4G_1\cos^2{\alpha} + \frac{\sin^2{\alpha}}{4G_1},
	\end{equation}
	assuming $N_\mathrm{in}=\braket{X_\mathrm{in}^2}=\braket{Y_\mathrm{in}^2}$, and assuming uncorrelated quadrature inputs i.e. $\braket{X_\mathrm{in}Y_\mathrm{in}}=0$. Here, $\mathcal{G}_1(\alpha)$ is the effective phase-sensitive gain. When $\alpha=0$, $N_{\mathrm{out},1} = 4G N_\mathrm{in}$, and we retrieve the well-known result that $\mathcal{G}(0)$ is 4 times (or $6$\,dB) higher than $G_1$. When $\alpha=\pi/2$, $N_{\mathrm{out},1} = N_\mathrm{in}/(4G_1)$, and the input noise $N_\mathrm{in}$ has been squeezed by the amplifier.
	
	

	\subsection{Measuring the noise case of a single-input-port, non-ideal parametric amplifier}
	\label{sec:noiseparamp}
	
	\subsubsection{Phase-insensitive case}
	
\begin{figure}[h!]	
		\includegraphics[scale=1]{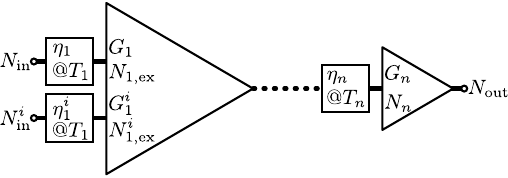}
		\caption{Schematic of an amplification chain where the first amplifier is parametric. Drawn with two input ports and one output port, we separate the input signal and idler modes. On the path from the signal input to the signal output (i.e. the signal-to-signal path), the amplifier's gain is $G_1$ and adds an excess of noise (above the quantum limit) $N_{1,\mathrm{ex}}$. On the idler-to-signal path the gain is $G_1^i$ and the excess of noise is $N_{1,\mathrm{ex}}^i$.} 
		\label{fig:paramp_chain}
	\end{figure}

	
	A major pitfall of noise measurements, especially in the context of parametric amplifiers, consists of misinterpreting Eq.\,\ref{eq:Nout} when using a wideband noise source. In fact, as we have seen in Sec.\,\ref{sec:paramp}, a phase-insensitive parametric amplifier possess two inputs, one at the signal frequency and one at the idler frequency. But, for many parametric amplifiers, these input \textit{modes} enter the amplifier via the same physical input \textit{port}. Therefore, a wideband noise source placed at the input port of the amplifier illuminates both the signal and idler modes inputs, and the amplification chain of Fig.\,\ref{fig:ampchain_theo} now has to reflect this ambivalence. In the case where the first amplifier of the chain is parametric (and when only this amplifier is parametric),  the amplification chain of Fig.\,\ref{fig:ampchain_theo} can be recast into that of Fig.\,\ref{fig:paramp_chain}, where the input of the modes now enter via two different ports. 
	
	Also, we now consider the parametric amplifier as non-ideal, therefore (i) the gain $G_1^i$ along the idler-to-signal path may not be exactly equal to $G_1-1$ (where $G_1$ is the gain along the signal-to-signal path) as it was the case in Sec.\,\ref{sec:paramp}, because of possible loss asymmetries between the signal and idler frequencies within the amplifier. (ii) We introduce two extra noise term, $N_{1,\mathrm{ex}}$ and $N_{1,\mathrm{ex}}^i$, added by the amplifier \textit{in excess above the quantum limit} along the idler-to-signal and signal-to-signal paths, respectively. They originate from internal non-idealities, \ie{} in the ideal-amplifier case $N_{1,\mathrm{ex}}=N_{1,\mathrm{ex}}^i=0$. Rewriting Eq.\,\ref{eq:Nsout} in this context, the output of the amplifier at the signal frequency is:
	\begin{equation}    
		N_{\mathrm{out},1} =\Tilde{G}_1\left(N_\mathrm{in} + \frac{\Tilde{G}_1^i}{\Tilde{G}_1}N_\mathrm{in}^i + \tilde{N}_{1,\mathrm{ex}}\right),          
		\label{eq:P1s}
	\end{equation}
	where $\Tilde{G}_1=\eta_1 G_1$ and $\Tilde{G}_1^i=\eta_1^iG_1^i$ are the effective signal-to-signal and idler-to-signal gains, respectively, and where $\tilde{N}_{1,\mathrm{ex}}$ is the effective signal-to-signal excess noise, referred to the signal input, that accounts for both $N_{1,\mathrm{ex}}$ and $N_{1,\mathrm{ex}}^i$, and for the noise added by the loss before the amplifier, see appendix \ref{app:op2quantapi}. 
	
	In principle one would have to know $\Tilde{G}_1^i/\Tilde{G}_1$ in order to fit Eq.\,\ref{eq:P1s} and extract $\tilde{N}_{1,\mathrm{ex}}$, but knowing this ratio can be experimentally difficult, because it requires to know the frequency-dependent internal loss of the amplifier. However, in most practical cases it is reasonable to assume that both gains don't vary by more than a factor $2$ of each other, an uncertainty that only has a small effect on $\tilde{N}_{1,\mathrm{ex}}$ (less than $30$\,\%), see appendix \ref{app:op2quantapi}. Therefore, a good approximation consists of assuming $\Tilde{G}_1=\Tilde{G}_1^i$. The signal at the output of the chain then takes the form:
	\begin{equation}
		N_\mathrm{out} = G_\mathrm{sys}(N_\mathrm{in} + N_\mathrm{in}^i + N_\mathrm{sys,ex}),
		\label{eq:Pcs}
	\end{equation}
	where $N_\mathrm{sys,ex}$ is the system-excess noise at the signal frequency, \textit{i.e.} noise \textit{above} the quantum limit on added noise of $N_\mathrm{sys}=1/2$ quanta. Therefore, $N_\mathrm{sys,ex}=0$ if the readout chain is quantum-limited. The output signal $N_\mathrm{out}$ depends on both the signal and idler input modes, in other words we have separated the noise intrinsic to the amplification chain, represented by $N_\mathrm{sys,ex}$, to the noise extrinsic to the amplification chain, represented by $N_\mathrm{in}^i$. When using a broadband noise source (VTS or a SNTJ) to characterize the noise of the amplification chain we illuminate $\textit{both}$ the input signal and idler ports, \textit{i.e.} we vary both $N_\mathrm{in}$ and $N_\mathrm{in}^i$ (and both given either by Eq.\,\ref{eq:Johnson} in the case of a VTS or by Eq.\,\ref{eq:shot} in the case of a SNTJ, expressed at the signal and idler frequency respectively). Therefore, the y-intercept of such a measurement is now $N_\mathrm{sys,ex}$, not $N_\mathrm{sys}$ as seen previously (see Sec.\,\ref{sec:comparison}). In turn, the total system-added noise is $N_\mathrm{sys}=N_\mathrm{in}^i+N_\mathrm{sys,ex}$, which encompasses all the amplification and loss in the chain, similarly as in Eq.\,\ref{eq:Nc}. In the ideal case where the idler input of a DUT replacing the SNTJ is cold i.e. $N_\mathrm{in}^i=1/2$ (and when $\Tilde{G}_1=\Tilde{G}_1^i$) the system-added noise would be
	\begin{equation}
		N_\mathrm{sys} = \frac{1}{2} + N_\mathrm{sys,ex}.
		\label{eq:naddsys}
	\end{equation}

	Failing to account for the idler input mode, in other words using the model represented by Eq.\,\ref{eq:Nout} to interpret the data instead of using the model represented by Eq.\,\ref{eq:Pcs} leads to under-estimating $N_\mathrm{sys}$ by more than a factor two. In fact, when $N_\mathrm{in}=N_\mathrm{in}^i$ (a good approximation for narrow-band parametric amplifiers), Eq.\,\ref{eq:Pcs} writes as 
	\begin{equation}
		N_\mathrm{out} = 2G_\mathrm{sys}\left(N_\mathrm{in} + \frac{N_\mathrm{sys,ex}}{2}\right).
		\label{eq:noutparamp}
	\end{equation}
	Therefore, the \textit{interpretation} of the slope and y-intercept of the output noise curves in Fig.\,\ref{fig:simplenoisetheo} is, when wrongly using Eq.\,\ref{eq:Nout}, $G_\mathrm{sys}$ and $G_\mathrm{sys}\times N_\mathrm{sys}$, while it should be $2G_\mathrm{sys}$ and $2G_\mathrm{sys}\times N_\mathrm{sys,ex}/2$, respectively, i.e. the y-intercept is in reality half of the output-referred system-excess noise.
	
	Note that when $\eta_1^i=0$ (achieved for example with a filter), the noise source only illuminates the signal port. In other words, the (phase-insensitive) parametric amplifier can be thought of as having only one input port at the signal frequency, while the idler input port receives some noise whose value depends on the port's temperature $T$, as given by Eq.\,\ref{eq:Johnson}. When $k_B T \ll h\nu$, this noise is half a quanta. Therefore, in this case the noise measurement can be carried out using Eqs.\,\ref{eq:Nout} and \ref{eq:Nc}, with $\Tilde{N}_1\geq1/2$.

	\subsubsection{Phase-sensitive case}
	
	In the phase-sensitive case, the ``signal'' dwells over both the signal and idler frequencies: there is no idler mode anymore, and the amplifier can once again be thought of as having only one input port receiving a signal. In this case, the input noise $N_\mathrm{in}$ is transformed according to Eq.\,\ref{eq:PoutPS}. Accounting for amplifier non-idealities via an excess of noise $N_{1,\mathrm{ex}}\geq0$, the output signal along the measurement quadrature is:
	\begin{equation}
		N_{\mathrm{out},1}(\alpha) = \Tilde{\mathcal{G}}_1(\alpha)(N_\mathrm{in} + \tilde{N}_{1,\mathrm{ex}}),
	\end{equation}
	where $\tilde{N}_{1,\mathrm{ex}} = N_{1,\mathrm{ex}}/\sqrt{\eta_1\eta_1^i}$, and $\Tilde{\mathcal{G}}_1(\alpha) = \sqrt{\eta_1\eta_1^i}\mathcal{G}_1(\alpha)$, see appendix \ref{app:op2quantaps}.
	
	When $\alpha=\pi/2$, the measurement quadrature is aligned with the amplifier's squeezed quadrature. However, only with a following phase-sensitive parametric amplifier can we measure (not infer) squeezed noise \cite{mallet2011quantum,malnou2018optimal}. If however the parametric amplifier is followed by a phase-insensitive amplifier, say a HEMT, with effective gain $\Tilde{G}_2$ and added noise $\Tilde{N}_2$, the signal at the HEMT output is:
	\begin{equation}
		N_{\mathrm{out},2}(\alpha) = \Tilde{G}_2\Tilde{\mathcal{G}}_1(\alpha) \left(N_\mathrm{in} + \tilde{N}_{1,\mathrm{ex}} + \frac{\Tilde{N}_2}{\Tilde{\mathcal{G}}_1(\alpha)}\right).
		\label{eq:pps}
	\end{equation}
	Therefore, in that situation the (non-ideal) system-added noise is
	\begin{equation}
		N_\mathrm{sys}(\alpha) = \tilde{N}_{1,\mathrm{ex}} + \frac{\Tilde{N}_2}{\Tilde{\mathcal{G}}_1(\alpha)},
		\label{eq:Naddps}
	\end{equation}
	neglecting the noise added by further amplification stages. When $\alpha=0$, the measurement quadrature is aligned with the amplified quadrature and
	\begin{equation}
		N_{\mathrm{out},2}(0) = \Tilde{G}_2\eta_1 4G_1(N_\mathrm{in} + \tilde{N}_{1,\mathrm{ex}}),
	\end{equation}
	assuming that $\Tilde{N}_2/(\eta_1 4 G_1) \ll \tilde{N}_{1,\mathrm{ex}}$. In this case, the added noise is dominated by $\tilde{N}_{1,\mathrm{ex}}$. Inversely, if $\alpha=\pi/2$:
	\begin{equation}
		N_{\mathrm{out},2}(\pi/2) = \Tilde{G}_2\frac{\eta_1}{4G_1}\left(N_\mathrm{in} + \Tilde{N}_2\frac{4G_1}{\eta_1}\right),
	\end{equation}
	and the input-referred system-added noise jumps to $\Tilde{N}_2 4G_1/\eta_1$. As $\alpha$ varies between $0$ and $\pi/2$, $N_\mathrm{sys}(\alpha)$ takes values between these two extreme cases.

	\subsection{Amplifier saturation}
	
	Not only may a real amplifier add some excess of noise above the quantum limit, but it can also saturate.  Saturation manifests as a compression of the gain: as the input signal increases, the amplifier gain diminishes. In other words, $G_1=G_1(N_\mathrm{in})$. It is usually not a desirable regime for the amplifier operate in, because its response ceases to be linear.    
	
	Physically, in parametric amplifiers, saturation can originate either from pump depletion or from the excitation of higher order nonlinearities \cite{boutin2017effect,malnou2018optimal,remm2023prapplied}. One of the metrics conventionally used to characterize saturation is the input $1$\,dB compression point $N_{1\mathrm{dB}}$, defined as the input signal (here in photon-normalized units) at which the gain has dropped by $1$\,dB (the output $1$\,dB compression point is sometimes used, but it is painfully dependent on the amplifier's gain). This point is a somewhat arbitrary criteria, not a hard limit below which the amplifier is perfectly linear, and above which it is completely nonlinear.     
	
	Amplifier saturation may heavily affect the noise curves (obtained by varying $N_\mathrm{in}$) where the system gain is proportional to the local slope. Therefore, a steadily diminishing gain translates into a steadily diminishing slope. In Sec. \ref{sec:JPApinoise} we present a way to circumvent this issue when performing a noise measurement.        
	
	\section{Broadband noise characterization of canonical amplification chains}
	
\begin{figure}[h!]	
		\includegraphics[scale=0.9]{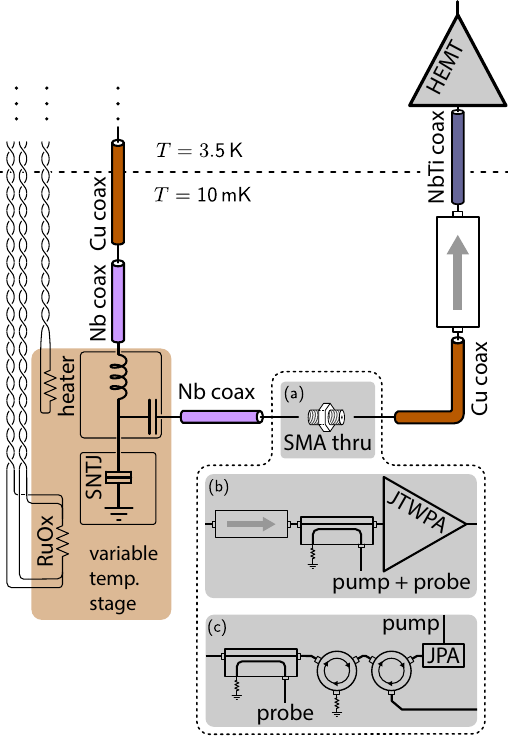}
		\caption{Schematic of the three cryogenic amplification chains, on which we performed noise measurements. At the input of the chain the noise source is double: (i) a SNTJ and (ii) a variable temperature stage (VTS), consisting of of a thermal stage whose temperature can be controlled independently to that of the mixing chamber plate on which it is anchored. We studied three configurations: (a) one where the first amplifier is the HEMT, placed at $4$\,K, (b) one where it is preceded by a JTWPA, and (c) where it is preceded by a JPA, both placed at $10$\,mK. Their presence necessitates the use of additional microwave components: circulator, isolator and directional coupler.} 
		\label{fig:ampchains}
	\end{figure}

	We illustrate how to perform a noise measurement on several amplification chains, containing microwave components typically used in superconducting circuit readout. Our noise source is double: it consists of a SNTJ, itself mounted on a VTS, see Fig.\,\ref{fig:ampchains}. This configuration offers two features: (i) it allows us to compare the system-added noise obtained when using the SNTJ to that obtained when using the VTS, because at zero bias the SNTJ acts as a $50$\,\ohm{} load (see Sec.\,\ref{sec:SNTJtheo}). (ii) It allows us to measure the loss inserted by the components of the chain that are mounted on the VTS, the SNTJ packaging and the following (commercial) bias tee, mandatory for the SNTJ's operation as a noise source. Thus, it allows us to move the reference plane of the noise measurement using the SNTJ, from the output of the SNTJ (on the chip) to the output of the bias tee. We measure the noise of three amplifications chains: (i) when a HEMT amplifier at $4$\,K is the first amplifier of the chain, (ii) when it is preceded by a Josephson Traveling Wave Parametric Amplifier (JTWPA), or (iii) by a resonant  Josephson Parametric Amplifier (JPA), placed at millikelvin temperatures.
	
	\subsection{The HEMT as first amplifier}
	\label{sec:HEMT}
	
	\begin{figure*}[htpb]	
		\includegraphics[scale=1]{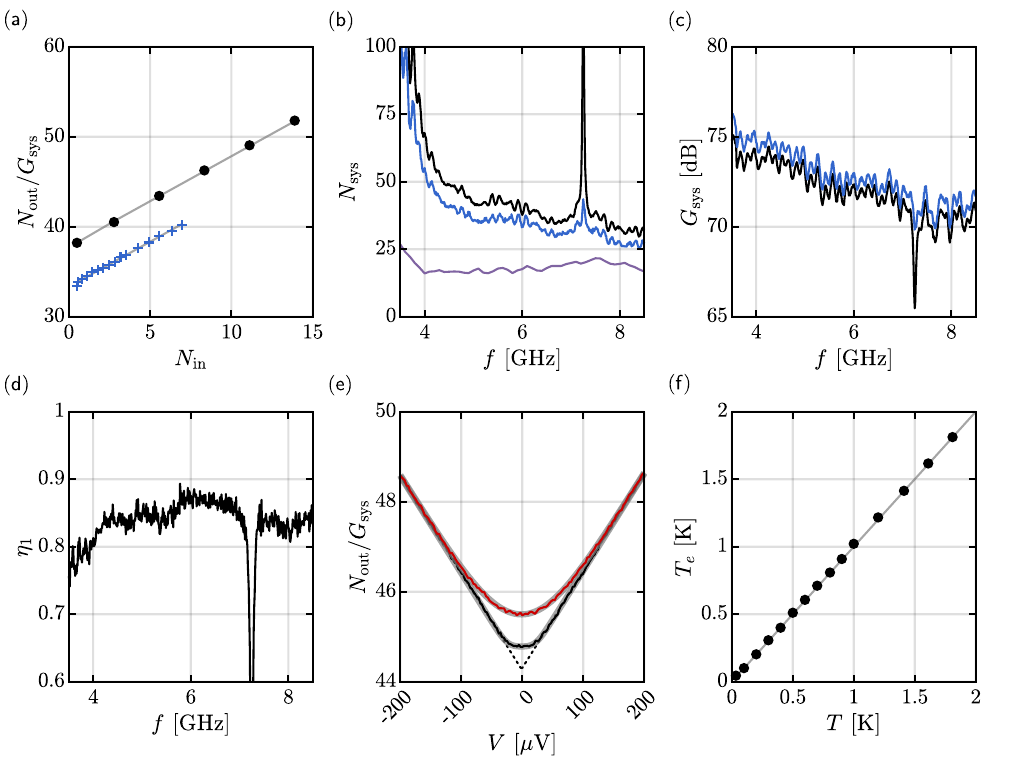}
		\caption{Noise characterization of the amplification chain where the HEMT is the first amplifier. (a) The output noise referred to the input of the chain, $N_\mathrm{out}/G_\mathrm{sys}$, is plotted as a function of the input noise $N_\mathrm{in}$. We show measurements using either the SNTJ (black markers) or the VTS (blue markers). The noise is integrated within a $8$\,MHz window, around $f=6$\,GHz. A fit of the responses (gray lines) gives $N_\mathrm{sys}$. It is plotted against frequency in (b), when using the SNTJ (black line) and when using the VTS (blue line). We also show the HEMT-added noise specified in its datasheet (purple line). The fit of the response in (a) also gives $G_\mathrm{sys}$ which we plot against frequency in (c), following the same color coding. Dividing out both system gains, we extract the transmission efficiency $\eta_1$, plotted against frequency in (d). In the regime of small SNTJ voltage biases $V$, we plot in (e) $N_\mathrm{out}/G_\mathrm{sys}$ obtained around $f=6$\,GHz, for two VTS temperatures: $T=34$\,mK (black line) and $T=300$\,mK (red line). The y-intercept of the asymptotes (dashed lines) also gives $N_\mathrm{sys}$ for this frequency. A fit of these curves (gray lines) allows us to extract the electronic temperature $T_e$, which we plot against $T$ in (f) (black markers). $T_e$ closely follows $T$ (gray line) over the whole VTS temperature range.} 
		\label{fig:hemtdata}
	\end{figure*}   
	
	When the HEMT is the first amplifier of the chain (see Fig.\,\ref{fig:ampchains}a), we can ask: is the measured system-added noise close to the intrinsic HEMT-added noise, specified in its datasheet? In Fig.\,\ref{fig:hemtdata}a we show the input-referred output noise $N_\mathrm{out}/G_\mathrm{sys}$, measured around $6$\,GHz with a spectrum analyzer (see appendix \ref{app:setup}), as a function of the input noise generated by the SNTJ (Eq.\,\ref{eq:shot}), superimposed with data obtained when using the VTS as a noise source (Eq.\,\ref{eq:Johnson}). Focusing here on the asymptotic behavior, the saturation of the HEMT is high enough so that the response remains linear over the whole excursion, up to $N_\mathrm{in}=14$ when using the SNTJ (and $N_\mathrm{in}=7$ when using the VTS), providing a reliable way to extract the system gain and added noise. A fit of the noise curves (or here, simply reading the y-intercept), see appendix \ref{app:fit}, gives $N_\mathrm{sys}$. Figure \ref{fig:hemtdata}b shows $N_\mathrm{sys}$, measured as a function of frequency, for the two situations (SNTJ and VTS). Both noise sources give an average system-added noise of $30$ to $50$ quanta between $4$ and $8$\,GHz, equivalent to a noise temperature of $10$ to $12$\,K. While it is about twice that of the HEMT-added noise specified by the datasheet, this is completely expected due to the inevitable loss between the reference plane and the HEMT itself.  
	
	The system-added noise obtained with the VTS is consistently lower to that obtained with the SNTJ, because the reference plane of the measurement with the VTS advances to the bias tee output (see Fig.\,\ref{fig:ampchains}). Conversely, the chain's gain $G_\mathrm{sys}$ is consistently lower with the SNTJ than with the VTS, see Fig.\,\ref{fig:hemtdata}c. Dividing out both gains yields the transmission efficiency of all the components on the VTS, $\eta_1=G_\mathrm{sys}^\mathrm{SNTJ}/G_\mathrm{sys}^\mathrm{VTS}$, i.e. the SNTJ packaging and bias tee's insertion loss is $\mathcal{I}_1=-10\log_{10}\eta_1$. In Fig.\,\ref{fig:hemtdata}d we show $\eta_1$ as a function of frequency. It is between $0.8$ and $0.9$, which means that the loss is between $0.5$ to $1$\,dB between $4$ and $9$\,GHz, in agreement with previous, similar measurements \cite{chang2016noise,malnou2021three}.
	
	There is however a noticeable feature around $7$\,GHz, where the system gain is significantly lower than anywhere else in frequency, see Fig.\,\ref{fig:hemtdata}c, and where the system-added noise is significantly higher, in particular when using the SNTJ, see Fig.\,\ref{fig:hemtdata}b. Consequently, the packaging insertion loss is significantly higher at this frequency, see Fig.\,\ref{fig:hemtdata}d. The origin of this feature is unclear, and its presence was unexpected; it could be due to a variation in the noise emitted by the SNTJ, or it could come from a strong impedance mismatch at a defective connection somewhere within the chain (\textit{i.e.} a broken cable). Nonetheless, it underlines the importance of wideband noise measurements to characterize readout chains: we cannot assume the response to be homogeneous and well-behaved everywhere.     
	
	Finally, focusing now on the small bias response, we measure the output noise at two VTS temperatures around $6$\,GHz, shown in Fig.\,\ref{fig:hemtdata}e (and see appendix \ref{app:setup} for the measurement method). A fit of these curves using Eq.\,\ref{eq:shot} and Eq.\,\ref{eq:Nout} (see appendix \ref{app:fit}) allows us to retrieve the electronic temperature $T_e$, which we plot against $T$, the VTS physical temperature, in Fig.\,\ref{fig:hemtdata}f. The agreement is excellent, even at the lowest temperature, so the SNTJ can also be used as a self-calibrated thermometer \cite{spietz2006shot}.
	
	
	\subsection{Using a JTWPA as a pre-amplifier}
	\label{sec:JTWPA}
	
\begin{figure}[h!]	
		\includegraphics[scale=1]{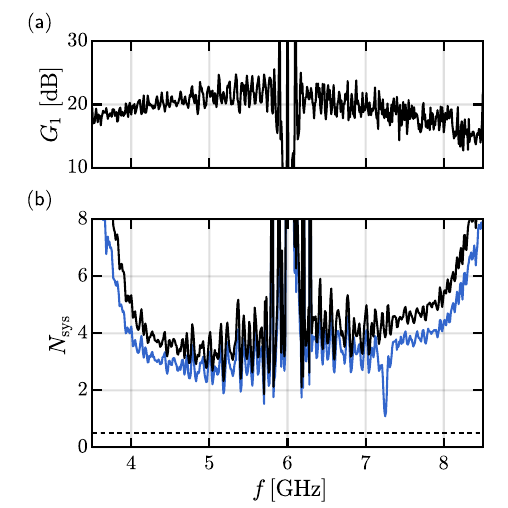}
		\caption{Noise characterization of an amplification chain where the JTWPA is the first (phase-insensitive) amplifier. (a) The JTWPA gain is measured with the VNA (ratio pump on/off). (b) The system-added noise $N_\mathrm{sys}$ (black line) is a few quanta above the quantum limit on added noise (where $N_\mathrm{sys}=0.5$, dashed line). Knowing the loss of the components mounted on the VTS, we move the reference plane from the SNTJ output to the VTS output, and show a corrected added noise (blue line).} 
		\label{fig:JTWPAdata}
	\end{figure}

	We now use the SNTJ to perform a noise measurement on the amplification chain presented in Fig.\,\ref{fig:ampchains}b, where a JTWPA has been inserted before the HEMT. The JTWPA is operated to produce $G_1=20$\,dB of wideband gain, see Fig.\,\ref{fig:JTWPAdata}a. Then, the wideband system-added noise, shown in Fig.\,\ref{fig:JTWPAdata}b (black trace), is obtained from the asymptotic behavior of the output response: the SNTJ is biased at $10$ evenly spaced values between $eV/(2k_B)=\pm0.5$\,K, and for each value we record an output spectrum. At each frequency we then fit the variation of the output noise as a function of the input noise using Eq.\,\ref{eq:Pcs} (which assumes $\Tilde{G}_1=\Tilde{G}_1^i$) and calculate $N_\mathrm{sys}$ using Eq.\,\ref{eq:naddsys}. It is the system-added noise, with reference plane at the SNTJ. From $4$ to $8$\,GHz, $N_\mathrm{sys}$ varies between $3$ and $5$ quanta, typical for such an amplification chain. The sharp increase of noise around $6$\,GHz comes from resonant features inside the JTWPA, necessary to obtain phase-matching and exponential gain via a four-wave mixing process \cite{obrien2014resonant,macklin2015near}.
	
	Knowing $\eta_1$, corresponding to the transmission efficiency of all the components mounted on the VTS (see Fig.\,\ref{fig:hemtdata}d) and assuming $\eta_1=\eta_1^i$, we can move the reference plane to the output of the VTS, and get the corrected system-excess noise (see appendix \ref{app:op2quantapi}):
	\begin{equation}
		N_\mathrm{sys,ex,corr} = \eta_1 N_\mathrm{sys,ex} - 2(1-\eta_1)N_{T_1},
		\label{eq:Ncsp}
	\end{equation}
	where the factor $2$ comes from the combination of the loss between the signal and idler frequencies, and where $N_{T_1}=0.5$, because when operating the SNTJ the VTS is kept cold. We then compute the corrected system-added noise $N_\mathrm{sys,corr}=N_\mathrm{sys,ex,corr}+1/2$. As shown in Fig.\,\ref{fig:JTWPAdata}a, $N_\mathrm{sys,corr}$ (blue trace) is lower by about $0.5$ quanta compared to $N_\mathrm{sys}$. Some components and connections remain between the JTWPA and the new reference plane, but these components are integral to the use of the JTWPA. Therefore they should be considered as part of the amplifier itself, and as such $N_\mathrm{sys,corr}$ constitutes a fair estimate of the true added noise of the amplifier.
	
	\subsection{Using a JPA as a pre-amplifier}
	
	Replacing the JTWPA with a flux-pumped JPA, we now study the amplification chain presented in Fig.\,\ref{fig:ampchains}c. In principle, the measurement and data analysis should be similar to that performed when using the JTWPA; it should even be simpler, because the JPA is a narrow-band amplifier, with a bandwidth of about $5$ to $10$\,MHz at $20$\,dB of gain. However the JPA is typically an amplifier with a very low saturation power, which complicates the noise characterization. Furthermore, the JPA is also commonly used as a phase-sensitive amplifier \cite{castellanos2008amplification,mallet2011quantum,malnou2018optimal,Backes2021a}. Below, we experimentally address both aspects.
	
	\subsubsection{Phase-insensitive noise}
	\label{sec:JPApinoise}
	
	\begin{figure*}[htpb]	
		\includegraphics[scale=1]{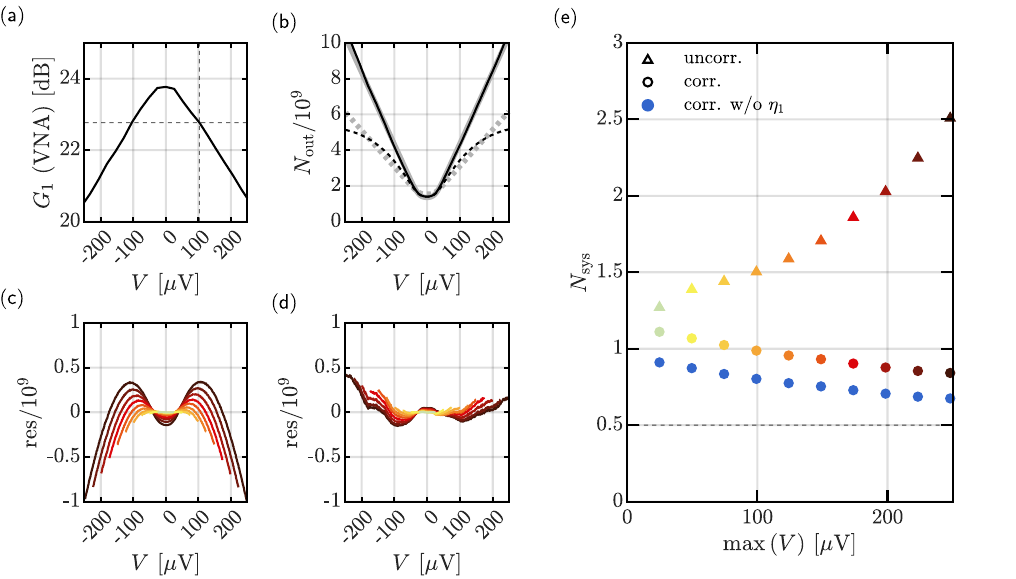}
		\caption{Noise characterization of the chain where the JPA is the first (phase-insensitive) amplifier. (a) The JPA gain is measured on the VNA (and fit to a Lorentzian) while varying the SNTJ bias (black line). The dashed lines indicate the $1$\,dB compression point. (b) The output noise as a function of the SNTJ voltage bias $V$ (black dashed line) saturates at high biases, and the fit of the curve using a linear model (gray dashed line) is not appropriate. Renormalizing the noise with the gain, the output noise regains a linear behavior (black line), and is then fitted with the linear model (gray line). (c) In the situation of uncorrected noise curve the residuals of the fit have a clear structure at high $V$. The different curves correspond to fitting various portion of the output noise. (d) In the renormalized case, the structure in the residuals dampens. (e) The system-added noise $N_\mathrm{sys}$ as a function of the fitting window's width presents two distinct behaviors: (i) when using the uncorrected data (triangles), $N_\mathrm{sys}$ diverges. (ii) When using the corrected data (circles), the variation of $N_\mathrm{sys}$ with the fitted portion of the curve dampens. When moving the reference plane to the VTS output, $N_\mathrm{sys}$ decreases by about $0.2$ quanta (blue).} 
		\label{fig:JPAPIdata}
	\end{figure*}  
	
	In Fig.\,\ref{fig:JPAPIdata}a we plot the JPA gain, measured on a vector network analyzer (VNA) with a small probe tone, as a function of the SNTJ voltage bias $V$. We measure $N_{1\mathrm{dB}}=2.1$ quanta (at $6$\,GHz, referenced to the SNTJ output) but obviously the gain steadily drops before that, even at very low signals, and continues to steadily drop beyond that. Then, in Fig.\,\ref{fig:JPAPIdata}b we show the output noise as a function of $V$ (dashed black line), which exhibits a nonlinear behavior at high bias due to amplifier saturation. Fitting this curve with the model described by Eq.\,\ref{eq:Pcs}, now wrong because it assumes a constant chain's gain, fails to produce a consistent result. As shown in Fig.\,\ref{fig:JPAPIdata}c, the residuals of the fit show a systematic deviation that grows with the portion of the curve upon which the fit is performed, and concurrently the system-added noise $N_\mathrm{sys}$ increases and diverges, see Fig.\,\ref{fig:JPAPIdata}e (since the fitted gain decreases due to compression). 
	
	To correct for this effect, we divide out the gain variation measured with the VNA from the noise curve, which then retrieves a more linear behavior, see Fig.\,\ref{fig:JPAPIdata}b. We then fit the output noise with a corrected model, close to that described by Eq.\,\ref{eq:Pcs} (see appendix \ref{app:op2quantapi}). The systematic deviation dampens. However the residuals (Fig.\,\ref{fig:JPAPIdata}d) sill have some structure when fitting a wide portion of the SNTJ noise curve. Similarly, the variation of $N_\mathrm{sys}$ as a function of the fitted portion of the curve dampens, see Fig.\,\ref{fig:JPAPIdata}e. When fitting a narrow portion we find $N_\mathrm{sys}=1.1$ quanta, but the gain is likely to be under-estimated because of the rounding of the curve around $V=0$. When fitting the whole curve we find $N_\mathrm{sys}=0.85$ quanta but the residuals show a systematic error. The true system-added noise (at low input signal) lies in between these two values.
	
	We can, once again, move the reference plane of this measurement up to the VTS output using Eq.\,\ref{eq:Ncsp}, because we know $\eta_1$ from previous measurements. Then, the corrected system-added noise, decreases by about $0.2$ quanta (see Fig.\,\ref{fig:JPAPIdata}e) , but of course remains above the quantum limit on added noise of $0.5$ quanta.

	
	\subsubsection{Phase-sensitive noise}
	
	The same cryogenic setup can be used to measure the phase-sensitive system-added noise $N_\mathrm{sys}(\alpha)$, equal to zero if the chain is ideal (see Eq.\,\ref{eq:Naddps}). At room temperature, a local oscillator (LO), whose frequency $f_{LO}=6$\,GHz is half that of the JPA's pump, now drives a mixer that overlaps the signal and idler frequency bands down to a baseband. The baseband noise is then measured on a spectrum analyzer (see apppendix \ref{app:setup}) and depends on $\alpha$, the phase difference between the phase of the LO and that of the JPA's amplified quadrature.
	
	\begin{figure}[htpb]	
		\includegraphics[scale=1]{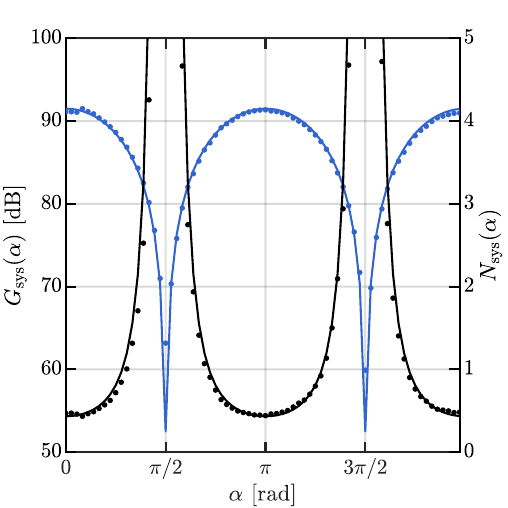}
		\caption{Phase sensitive behavior of the amplification chain, where the JPA is the first amplifier. The system gain $G_\mathrm{sys}(\alpha)$ (blue, left y-axis) and the system-added noise $N_\mathrm{sys}(\alpha)$ (black, right y-axis) are plotted as a function of $\alpha$, the angle between the measured quadrature and the JPA's amplified quadrature. The data (points) agrees with a simple predictive model (lines).} 
		\label{fig:psnoise}
	\end{figure}  
	
	The JPA saturation also affects the phase-sensitive noise. Worse, saturation is phase-dependent. In fact, in general the amplified quadrature saturates at higher power than the squeezed quadrature, as the theory and the measurements of phasor deformation have shown \cite{wustmann2013parametric,boutin2017effect,bienfait2017magnetic,malnou2018optimal,parker2022degenerate}. In comparison, saturation in phase-insensitive case can be viewed as an effective, average behavior over all the quadrature angles. To correct for this phasor deformation, one would have to measure it at various phasor amplitudes, which falls out of the scope of this paper. Instead, we restricted our fits to a fairly narrow portion of the noise curve, between $\pm1.25$ quanta, where the gain does not excessively compress, and therefore where the obtained gain and excess of noise should not be too far from their true value (at least when looking at the amplified quadrature). With that caveat in mind, Fig.\,\ref{fig:psnoise} shows the system gain $G_\mathrm{sys}(\alpha)$ (which contains $\mathcal{G}(\alpha)$) and system-added noise $N_\mathrm{sys}(\alpha)$, measured with the SNTJ. Clearly, the behavior of the data (points) as a function of $\alpha$ makes sense: we can derive a simple predictive model for $G_\mathrm{sys}(\alpha)$ and $N_\mathrm{sys}(\alpha)$ using Eq.\,\ref{eq:pps} and using the experimental HEMT-only system gain and added noise previously measured in Sec.\,\ref{sec:HEMT} (see appendix \ref{app:psnoisemodel}). This model (lines) agrees fairly well with the data. Overall, when $\alpha = 0\,\mathrm{mod}(\pi)$, the measured quadrature is the JPA's amplified quadrature, therefore the system gain is maximal while the added noise is minimal, reaching a lowest value of $0.43$\,quanta, not too far from the system-excess noise $N_\mathrm{sys,ex}=N_\mathrm{sys}-1/2$ found in the phase-insensitive case.

	\section{Conclusion}
	
	
	
	
	Calibrated noise measurements are based on simple theoretical concepts yet there are several subtle errors that can sneak into both the measurement and interpretation/analysis, especially when characterizing parametric amplifiers.  One might speculate that this is based on the idea that a `good' parametric amplifier \textit{must} be quantum limited, resulting in a literature where the majority of published noise characterization efforts report quantum-limited  behavior. While an ideal parametric amplifier can be, in principle, quantum-limited, it does not mean that (i) a real amplification chain will be and (ii) that one will rigorously measure its noise. In fact, the activation of spurious parametric processes, dielectric loss, heating, and any other non-idealities will contribute to decreasing the noise performance. Furthermore, a practical parametric amplifier requires intervening components to bias and isolate it from the quantum circuit that it's measuring. These components (isolators, circulators, directional couplers, PCBs, connectors, \etc) add insertion loss and therefore degrade the effective added noise. Even when this basic fact is acknowledged, it's not uncommon to attempt to move the reference plane to the circuit on-chip by taking manufacturer specification sheet numbers for these components-- a practice that essentially negates all other attempts at rigor. Meanwhile, the low power handling capability of many Josephson parametric amplifiers can make noise measurements hard to parse without also measuring the gain saturation properties carefully.
	
	All of these real-life complications support the need for a more flexible and reliable noise source. Compared to a traditional hot/cold load (or to its continuously variable variant, the VTS), the SNTJ enables much faster noise measurements or, equivalently, allows us to acquire many more points on the noise curve, due to its bias in voltage and not in temperature. Further developments, for example engineering an on-chip bias-tee, should make this remarkable device widely accessible and user-friendly. But the devil is in the details: when using a wideband noise source to characterize a parametric amplifier, the idler port contribution to the output noise must be considered carefully. As shown here, ignorance of this term comes at the cost of underestimating the added noise contribution, erroneously giving one the sense that their performance is closer (or even below!) the standard quantum limit.
	
	We have illustrated how to perform a noise measurement with a SNTJ on three canonical amplification chains, and shown how to move the chain's reference plane using a separate noise measurement performed with a VTS. With the HEMT as first amplifier, the system-added noise, referred to the SNTJ output, is a few tens of quanta, with a JTWPA it is a few quanta, and with a JPA it is even better yielding just a fraction of a quanta. These numbers are what an end user ultimately cares about, because they have been obtained from a realistic use of these amplifiers, including necessary lossy components. In the same vein, we strongly support explicitly stating the added noise of the whole amplification chain, uncorrected from potential noisy contributions due to loss. Discriminating between the noise coming solely from the chip and the noise coming from the extraneous, lossy components is of interest for amplifier designers, and should be carried out with a cold calibration. While it is obvious to many, it is still worthwhile stating explicitly: \textit{no practical parametric amplifier chain can ever operate at the quantum limit for added noise when referred to an input that includes required lossy components.} Perhaps, even more to the point, this is totally expected, even with a perfectly lossless on-chip amplifier circuit.
	
	\section*{Acknowledgment}
	We gratefully acknowledge Katrina Sliwa and the MIT Lincoln Laboratories, for providing us with a JTWPA. Certain commercial materials and equipment are identified in this paper to foster understanding. Such identification does not imply recommendation or endorsement by the National Institute of Standards and Technology, nor does it imply that the materials or equipment identified are necessarily the best available for the purpose.

	\appendix
	
	\section{List of variables}
	Table \ref{tab:variables} defines the noise-related variables used throughout the article.
	
	\begin{table*}[] 
		\centering
		\begin{tabular}{cl}
			\hline
			\hline
			\textbf{variable name} & \textbf{definition} \\ \hline\hline
			$N_\mathrm{in}$ & signal at the input of the amplification chain \textit{i.e.} at the reference plane $\mathcal{R}$ \\
			$N_\mathrm{out}$ & signal at the output of the amplification chain \textit{i.e.} at the reference plane $\mathcal{R}_\mathrm{out}$ \\
			$N_\mathrm{in}^i$ & idler at the input of the amplification chain \textit{i.e.} at the reference plane $\mathcal{R}$ \\
			$N_{\mathrm{in},k}$ & signal at the input of the $k$-th amplifier \\
			$\Tilde{N}_{\mathrm{in},k}$ & signal at the input of the effective $k$-th amplifier, equal to $N_{\mathrm{out},k-1}$ \\
			$N_{\mathrm{out},k}$ & signal at the output of the $k$-th amplifier \\                
			$G_k$ & signal gain of the $k$-th amplifier\\        
			$\Tilde{G}_k$ & signal gain of the $k$-th effective amplifier\\
			$G_\mathrm{sys}$ & signal gain of the whole amplification chain \textit{i.e.} system gain\\
			$G_1^i$ & idler-to-signal gain of the parametric amplifier, first amplifier in the chain\\
			$\Tilde{G}_1^i$ & idler-to-signal gain of the effective parametric amplifier, first amplifier in the chain\\
			$\mathcal{G}_1(\alpha)$ & phase-sensitive signal gain of the parametric amplifier, first amplifier in the chain\\  
			$\Tilde{\mathcal{G}}_1(\alpha)$ & phase-sensitive signal gain of the effective parametric amplifier, first amplifier in the chain\\
			
			$\eta_k$ & transmission efficiency of the $k$-th loss stage at the signal frequency\\
			$\eta_k^i$ & transmission efficiency of the $k$-th loss stage at the idler frequency\\        
			
			$N_k$ & added noise of the $k$-th amplifier, referred to its input\\          
			$\Tilde{N}_k$ & added noise of the $k$-th effective amplifier, referred to its input\\
			$N_\mathrm{sys}$ & system-added noise, referred to the input of the chain\\          
			$N_\mathrm{sys}^\mathrm{on}$ & system-added noise when the first amplifier is on, referred to the input of the chain\\
			$N_\mathrm{sys}^\mathrm{off}$ & system-added noise when the first amplifier is off, referred to the input of the chain\\
			$N_{T_k}$ & Johnson noise generated by the $k$-th loss stage at a temperature $T_k$\\                                                
			$N_{1,\mathrm{ex}}$ & signal-to-signal excess of noise of the first amplifier in the chain, referred to the signal input\\
			$N_{1,\mathrm{ex}}^i$ & idler-to-signal excess of noise of the first amplifier in the chain, referred to the idler input\\
			$\tilde{N}_{1,\mathrm{ex}}$ & effective signal-to-signal excess of noise of the first amplifier in the chain, referred to the signal input\\
			$N_\mathrm{sys,ex}$ & system-excess noise, referred to the input of the chain\\
			$N_\mathrm{in}^\mathrm{Johnson}$ & Johnson noise delivered by a resistor to a matched circuit\\
			$N_\mathrm{in}^\mathrm{shot}$ & Shot noise delivered by a SNTJ to a matched circuit\\
			\hline
		\end{tabular} 
		\caption{List of the noise-related variables used in the article. All the $N$ variables are noises, expressed in photon-normalized units (see Sec.\,\ref{sec:ampchain}). All the $G$ variables are photon-normalized gains.}
		\label{tab:variables}
	\end{table*}
	
	\section{Signal transformation with a parametric amplifier}
	\subsection{Phase-insensitive case}
	\label{app:op2quantapi}
	
	\subsubsection{Operators to noise in photon-normalized units}    
	
	Considering Eq.\,\ref{eq:aout} and its Hermitian conjugate, we calculate the average output photon number
	\begin{align}    
		\Bar{n}_\mathrm{out} &= \braket{a_\mathrm{out}^\dagger a_\mathrm{out}} \\
		&= G_1\Bar{n}_\mathrm{in} + (G_1-1)(\Bar{n}_\mathrm{in}^i + 1)\\
		&= G_1\left(\Bar{n}_\mathrm{in}+\frac{1}{2}\right) + (G_1-1)\left(\Bar{n}_\mathrm{in}^i+\frac{1}{2}\right) - \frac{1}{2} \label{eq:nsout},
	\end{align}
	where $\Bar{n}_\mathrm{in}=\braket{a_\mathrm{in}^\dagger a_\mathrm{in}}$, and $\Bar{n}_\mathrm{in}^i=\braket{b_\mathrm{in}^\dagger b_\mathrm{in}}$ are the photon-number operators. We then obtain Eq.\,\ref{eq:Nsout} from Eq.\,\ref{eq:nsout} by taking the $-1/2$ to the left-hand side:
	\begin{equation}
		N_{\mathrm{out},1} = G_1 N_\mathrm{in} + (G_1-1)N_\mathrm{in}^i,
	\end{equation}
	with $N_{\mathrm{out},1} = \Bar{n}_\mathrm{out} + 1/2$, $N_\mathrm{in} = \Bar{n}_\mathrm{in} + 1/2$ and $N_\mathrm{in}^i = \Bar{n}_\mathrm{in}^i + 1/2$.
	
	\subsubsection{Excess of noise}
	
	Let us know consider a non-ideal parametric amplifier: it has a gain $G_1$ ($G_1^i$) and adds some excess noise $N_{1,\mathrm{ex}}$ ($N_{1,\mathrm{ex}}^i$) along the signal-to-signal (idler-to-signal) path. If we also include the loss $\eta_1$ ($\eta_1^i$) along the signal (idler) path before the amplifier, we obtain at its output:
	\begin{equation}
		\begin{aligned}
			N_{\mathrm{out},1} = G_1&[\eta_1 N_\mathrm{in} + (1-\eta_1)N_{T_1} + N_{1,\mathrm{ex}}]\\
			+ G_1^i&[\eta_1^i N_\mathrm{in}^i + (1-\eta_1^i)N_{T_1} + N_{1,\mathrm{ex}}^i],
			\label{eq:Nout1app}
		\end{aligned}
	\end{equation}
	that is to say
	\begin{equation}        
		N_{\mathrm{out},1} = \Tilde{G}_1\left(N_\mathrm{in} + \frac{\Tilde{G}_1^i}{\Tilde{G}_1}N_\mathrm{in}^i + \tilde{N}_{1,\mathrm{ex}}\right),
		\label{eq:P1sapp}
	\end{equation}
	where
	\begin{equation}
		\tilde{N}_{1,\mathrm{ex}} = \frac{(1-\eta_1)N_{T_1} + N_{1,\mathrm{ex}}}{\eta_1} + \frac{\Tilde{G}_1^i}{\Tilde{G}_1} \frac{(1-\eta_1^i)N_{T_1} + N_{1,\mathrm{ex}}^i}{\eta_1^i}.  
		\label{eq:tildeNx}
	\end{equation}
	
	\subsubsection{Effect of asymmetric gains}
	
	We assume $\Tilde{G}_1^i/\Tilde{G}_1=1$. In fact, in most practical cases, we will have $1/2<\Tilde{G}_1^i/\Tilde{G}_1<2$, i.e. the effective signal-to-signal and idler-to-signal gains do not differ by more than $3$\,dB (due to loss). Therefore we can ask: how $\tilde{N}_{1,\mathrm{ex}}$ varies when $\Tilde{G}_1^i/\Tilde{G}_1$ varies between $1/2$ to $2$? For simplicity, let's consider $N_\mathrm{in}=N_\mathrm{in}^i$. Then, from Eq.\,\ref{eq:P1sapp}:
	\begin{equation}
		N_{\mathrm{out},1} = \frac{\Tilde{G}_1}{1+\frac{\Tilde{G}_1^i}{\Tilde{G}_1}}\left(N_\mathrm{in} + \frac{\tilde{N}_{1,\mathrm{ex}}}{1+\frac{\Tilde{G}_1^i}{\Tilde{G}_1}}\right),
	\end{equation}
	and the y-intercept, which corresponds to \textit{half} the excess noise, varies between $\tilde{N}_{1,\mathrm{ex}}/3$ and $2\tilde{N}_{1,\mathrm{ex}}/3$. Therefore, $\tilde{N}_{1,\mathrm{ex}}$ varies between $2\tilde{N}_{1,\mathrm{ex}}/3$ and $4\tilde{N}_{1,\mathrm{ex}}/3$.
	
	\subsubsection{Moving the reference plane}
	If we assume $\Tilde{G}_1^i/\Tilde{G}_1=1$ and $\eta_1=\eta_1^i$, Eq.\,\ref{eq:tildeNx} gives
	\begin{equation}
		\tilde{N}_{1,\mathrm{ex}} = \frac{2(1-\eta_1)+N_{1,\mathrm{ex}}+N_{1,\mathrm{ex}}^i}{\eta_1},
	\end{equation}
	therefore if we know $\eta_1$ we can remove its effect to the excess of noise added by that amplifier. More generally, if we measure a system-excess noise $N_\mathrm{sys,ex}$, we can remove the effect of $\eta_1$ and get a corrected system-excess noise:
	\begin{equation}
		N_\mathrm{sys,ex,corr} = \eta_1 N_\mathrm{sys,ex} - 2(1-\eta_1)N_{T_1}.
	\end{equation}
	
	\subsubsection{Filtering out the idler input}
	Finally, when the calibrated noise entering the idler input is filtered out, in other words when $\eta_1^i=0$ Eq.\,\ref{eq:Nout1app} reduces to
	\begin{equation}    
		\begin{aligned}
			N_{\mathrm{out},1} = &G_1[\eta_1 N_\mathrm{in} + (1-\eta_1)N_{T_1} + N_{1,\mathrm{ex}}]\\
			+ &G_1^i(N_{T_1} + N_{1,\mathrm{ex}}^i),
		\end{aligned}
	\end{equation}
	that is to say
	\begin{equation}    
		N_{\mathrm{out},1} = \tilde{G}_1\left(N_\mathrm{in} + \frac{G_1^i}{\tilde{G}_1}N_{T_1} + \tilde{N}_{1,\mathrm{ex}}\right),
	\end{equation}
	where 
	\begin{equation}
		\tilde{N}_{1,\mathrm{ex}} = \frac{(1-\eta_1)N_{T_1} + N_{1,\mathrm{ex}}}{\eta_1} + \frac{G_1^i}{\tilde{G}_1}  N_{1,\mathrm{ex}}^i.
	\end{equation}
	In other words, the parametric amplifier has an excess of noise $\tilde{N}_{1,\mathrm{ex}}$, and an added noise $G_1^i/\tilde{G}_1 N_{T_1} + \tilde{N}_{1,\mathrm{ex}}$. In the high gain regime and when the idler input is cold, an ideal amplifier is therefore still limited by the quantum limit because in that case $G_1^i/\tilde{G}_1 N_{T_1}=0.5$.
	
	\subsubsection{Added noise of a saturating parametric amplifier}
	When a parametric amplifier saturates, its effective gain $\tilde{G}_1$ becomes dependent on the input signal. When measuring the noise added by the JPA using the SNTJ, it means that $\tilde{G}_1=\tilde{G}_1(V)$, with $V$ the SNTJ voltage bias. We define $\lambda(V)$ such that:
	\begin{equation}
		\tilde{G}_1(V) = G_1\lambda(V),
	\end{equation}
	where $G_1$ is the JPA small signal gain. In other words, $\lambda(V)$ is a quantity that indicates how the JPA gain compresses (when $\lambda(V)<1$) as a function of $V$. Experimentally, $\lambda(V)$ is given by the curve in Fig.\,\ref{fig:JPAPIdata}a. The signal at the JPA output is then:
	\begin{equation}
		N_{\mathrm{out},1} = \tilde{G}_1\lambda(V)(N_\mathrm{in} + N_\mathrm{in}^i + \tilde{N}_{1,\mathrm{ex}}),
	\end{equation}
	and the signal at the HEMT output is:
	\begin{equation}
		N_{\mathrm{out},2} = G_2\tilde{G}_1\lambda(V)\left(N_\mathrm{in} + N_\mathrm{in}^i + \tilde{N}_{1,\mathrm{ex}} + \frac{\tilde{N}_2}{\tilde{G}_1\lambda(V)}\right).
	\end{equation}    
	Dividing out the signal at the output of the chain by $\lambda(V)$ we obtain
	\begin{equation}
		\frac{N_\mathrm{out}}{\lambda(V)} = G_\mathrm{sys}\left(N_\mathrm{in} + N_\mathrm{in}^i + \tilde{N}_{1,\mathrm{ex}} + \frac{\tilde{N}_2}{\tilde{G}_1\lambda(V)}\right),
		\label{eq:Noutcompr}
	\end{equation}  
	where we have approximated the system-added noise as 
	\begin{equation}
		N_\mathrm{sys}= 1/2 + \tilde{N}_{1,\mathrm{ex}}+\tilde{N}_2/(\tilde{G}_1\lambda(V)).     
		\label{eq:Nsyscompr}
	\end{equation}
	
	In general we don't know $\tilde{N}_2$, and furthermore $\tilde{N}_2/(\tilde{G}_1\lambda(V))$ cannot be neglected compared to $\tilde{N}_{1,\mathrm{ex}}$, because ideally $\tilde{N}_{1,\mathrm{ex}}=0$. But in our situation we have a good approximation of $\tilde{N}_2$ as the system-added noise of the chain where the HEMT is the first amplifier (case (a) of Fig.\,\ref{fig:ampchains}) obtained using the VTS (\ie{} blue curve in Fig.\,\ref{fig:hemtdata}b). At $6$\,GHz we find $\tilde{N}_2=33$ quanta. Then, when using the JPA, $G_1=23.8$\,dB (the small signal gain), see Fig.\,\ref{fig:JPAPIdata}a. Therefore $\tilde{N}_2/\tilde{G}_1=0.14$ quanta. We then can fit the corrected noise curves using Eq.\,\ref{eq:Noutcompr} to find $\tilde{N}_{1,\mathrm{ex}}$, from which we obtain the system-added noise (Eq.\,\ref{eq:Nsyscompr}). It is represented in Fig.\,\ref{fig:JPAPIdata} (circles) for a small signal input (\ie{} for $\lambda(V)=1$).

	\subsection{Phase-sensitive case, with pre-amplifier loss}
	\label{app:op2quantaps}
	As mentioned in the main text, in the phase sensitive case the role of the idler mode $b_\mathrm{in}$ is played by the signal mode $a_\mathrm{in}$. Then, loss is modeled as a beamsplitter interaction, transforming $a_\mathrm{in}$ such that
	\begin{align}
		a_\mathrm{in} &\rightarrow \sqrt{\eta_1} a_\mathrm{in} + \sqrt{1-\eta_1}\xi^s\\
		a_\mathrm{in}^\dagger &\rightarrow \sqrt{\eta_1^i} a_\mathrm{in}^\dagger + \sqrt{1-\eta_1^i}{\xi^i}^\dagger,
	\end{align}
	where $\xi^s$ and ${\xi^i}^\dagger$ are ladder operators for two noise modes, uncorrelated with each other and uncorrelated to to $a_\mathrm{in}$ and $a_\mathrm{in}^\dagger$ (we arbitrarily chose the creation operator for the idler noise mode, to track it easily). Also, the transmission efficiencies $\eta_1$ and $\eta_1^i$ seen by $a_\mathrm{in}$ and $a_\mathrm{in}^\dagger$ respectively are not necessarily equal, because these two modes are not at the same frequency. At the output of the amplifier we obtain
	\begin{equation}
		\begin{aligned}
			a_\mathrm{out} = &\sqrt{G_1}e^{-i\theta}\left(\sqrt{\eta_1} a_\mathrm{in} + \sqrt{1-\eta_1}\xi^s\right)\\
			+ &\sqrt{G_1-1}e^{i\theta}\left(\sqrt{\eta_1^i} a_\mathrm{in}^\dagger + \sqrt{1-\eta_1^i}{\xi^i}^\dagger\right)\\
			\label{eq:aoutps}
		\end{aligned}
	\end{equation}
	\begin{equation}
		\begin{aligned}        
			a_\mathrm{out}^\dagger = &\sqrt{G_1}e^{i\theta}\left(\sqrt{\eta_1^i} a_\mathrm{in}^\dagger + \sqrt{1-\eta_1^i}{\xi^i}^\dagger\right)\\
			+ &\sqrt{G_1-1}e^{-i\theta}\left(\sqrt{\eta_1} a_\mathrm{in} + \sqrt{1-\eta_1}\xi^s\right),
			\label{eq:aoutdaggerps}
		\end{aligned}
	\end{equation}
	from what we can calculate $X_\mathrm{out} = (a_\mathrm{out} + a_\mathrm{out}^\dagger)/\sqrt{2}$ and $Y_\mathrm{out} =  (a_\mathrm{out} - a_\mathrm{out}^\dagger)/(i\sqrt{2})$. Again, let's take $\theta=0$ to have $X_\mathrm{out}$ be the amplified quadrature and $Y_\mathrm{out}$ be the squeezed quadrature, and let's work in the high gain limit, $G_1\gg1$:
	\begin{widetext}
		\begin{align}    
			X_\mathrm{out} &= 2\sqrt{G_1}
			\left(\frac{\sqrt{\eta_1}a_\mathrm{in}+\sqrt{\eta_1^i}a_\mathrm{in}^\dagger}{\sqrt{2}} + \frac{\sqrt{1-\eta_1}\xi^s+\sqrt{1-\eta_1^i}{\xi^i}^\dagger}{\sqrt{2}}\right)\\
			Y_\mathrm{out} &= \frac{1}{2\sqrt{G_1}}
			\left(\frac{\sqrt{\eta_1}a_\mathrm{in}-\sqrt{\eta_1^i}a_\mathrm{in}^\dagger}{i\sqrt{2}} + \frac{\sqrt{1-\eta_1}\xi^s-\sqrt{1-\eta_1^i}{\xi^i}^\dagger}{i\sqrt{2}}\right).
		\end{align}    
	\end{widetext}
	All the modes being vacuum or thermal states, we then obtain:
	\begin{align}
		\braket{X_\mathrm{out}^2} &= 4 G_1 \sqrt{\eta_1 \eta_1^i} \left(\Bar{n}_\mathrm{in} + \frac{1}{2}\right) \\
		\braket{Y_\mathrm{out}^2} &= \frac{\sqrt{\eta_1 \eta_1^i}}{4 G_1}  \left(\Bar{n}_\mathrm{in} + \frac{1}{2}\right),
	\end{align}
	where $\Bar{n}_\mathrm{in}=\braket{a_\mathrm{in}^\dagger a_\mathrm{in}}$. We see that both transmission efficiencies degrade the output quadratures photon number, and in the limit case where one of them, say $\eta_1^i$ is zero, we recover the phase-insensitive case: in particular, when $\eta_1=1$ Eq.\,\ref{eq:aoutps} is is identical to Eq.\,\ref{eq:aout} (taking the now arbitrary phase $\theta$ equal to zero). At the output of the amplifier, we measure the signal in a given orientation $\alpha$:
	\begin{equation}
		\begin{aligned}
			N_{\mathrm{out},1}(\alpha) &= N_\mathrm{in}\sqrt{\eta_1 \eta_1^i}\left(4G_1\cos^2{\alpha} + \frac{\sin^2{\alpha}}{4G_1}\right)\\
			&= N_\mathrm{in}\sqrt{\eta_1 \eta_1^i}\mathcal{G}_1(\alpha),
		\end{aligned}
	\end{equation}
	where $N_\mathrm{in} = \Bar{n}_\mathrm{in}+1/2$. Finally, if we include some excess noise $N_{1,\mathrm{ex}}$, added by the amplifier, we get:
	\begin{equation}
		N_{\mathrm{out},1}(\alpha) = \sqrt{\eta_1 \eta_1^i}\mathcal{G}_1(\alpha)\left(N_\mathrm{in}+\tilde{N}_{1,\mathrm{ex}}\right),
	\end{equation}
	where $\tilde{N}_{1,\mathrm{ex}}=N_{1,\mathrm{ex}}/(\sqrt{\eta_1 \eta_1^i})$ is the input-referred excess noise.
	
	\section{Noise rise theory}
	\label{app:nvr}
	
	A noise rise - also called a noise visibility ratio - measurement allows us to estimate the first amplifier's added noise $N_1$. It consists of comparing the noise measured when the first amplifier is on, to when it is off. Such a measurement assumes some amount of noise $N_\mathrm{in}$ at the input of the chain based on its physical temperature, according to Eq.\,\ref{eq:Johnson}; at $30$\,mK and at gigahertz frequencies, we can assume $N_\mathrm{in}=1/2$. When this temperature is equal to $T_1$, \textit{i.e.} to that of the loss between the chain's input and the first amplifier, characterized by $\eta_1$ (see Fig.\ref{fig:ampchain_theo}), the chain's input reference plane $\mathcal{R}$ is effectively moved up to the input of the first amplifier, because this loss replaces noise with noise at the same power. It unveils a first caveat: a noise rise measurement does not account for the effect of the lossy elements inserted before the first amplifier (including packaging and connectors) mandatory to its operation, therefore it is an optimistic estimate of $N_1$.
	
	When the first amplifier is off we assume that it has a gain $G_1=1$, and if it is also at a temperature $T_1$, then $\mathcal{R}$ is effectively moved up to the input of the loss characterized by its transmission efficiency $\eta_2$ and its temperature $T_2>T_1$. In this case, following Eqs.\,\ref{eq:Nout}, \ref{eq:tildeGk} and \ref{eq:tildeNk} the signal at the output of the second amplifier is:
	\begin{equation}
		N_{\mathrm{out},2}^\mathrm{off} = \Tilde{G}_2(N_\mathrm{in} + \Tilde{N}_2),
	\end{equation}
	where $\Tilde{G}_2 = \eta_2G_2$ and $\Tilde{N}_2=[(1-\eta_2)N_{T_2} + N_2]/\eta_2$. One may estimate $\Tilde{G}_2$ and $\Tilde{N}_2$ with various ways, for example by heating up the whole base-temperature plate of the cryostat, effectively performing a hot/cold load measurement. But again, it reveals another caveat, because in many practical situations there is some lossy elements at temperature $T_1$ placed on the base-temperature plate, \textit{after} the first amplifier. This loss will degrade the effective gain and added noise of the second-stage amplifier, but it is not accounted for when estimating $\Tilde{G}_2$ and $\Tilde{N}_2$ in such a way.
	
	In comparison, When the first (possibly parametric) amplifier is on:
	\begin{align}
		N_{\mathrm{out},2}^\mathrm{on} &= G_1\Tilde{G}_2(N_\mathrm{in} + N_1 + \frac{\Tilde{N}_2}{G_1}) \\
		&\simeq G_1\Tilde{G}_2(N_\mathrm{in} + N_1),
	\end{align}
	where the second line assumes that $G_1$ overwhelms the following loss. Thus, we can form the ratio
	\begin{equation}
		r = \frac{N_{\mathrm{out},2}^\mathrm{on}}{N_{\mathrm{out},2}^\mathrm{off}} = G_1\frac{N_\mathrm{in} + N_1}{N_\mathrm{in} + \Tilde{N}_2},
	\end{equation}
	which, if the second amplifier's added noise overwhelms any of the following noise added by the chain, represents how much the noise rises at the output of the chain when turning on the first amplifier. The noise rise can be seen on a spectrum analyzer, for example. In practice we can have a fair estimate of $G_1$ by measuring the gain of the first amplifier on a VNA (ratio on/off traces). Knowing $\Tilde{N}_2$ we can then deduce $N_1$. In practice, all the caveats mentioned above make this technique approximate, biased toward finding a lower noise than what it truly is.
	
	\section{The various units of a noise spectral density}
	\label{app:spectraldensities}
	\subsection{Johnson noise}
	From the fluctuation-dissipation theorem, a resistor $R$ at temperature $T$ generates noise. In the classical limit, its power spectral density can be expressed in $\mathrm{V^2/Hz}$ as:
	\begin{equation}
		S_V^\mathrm{Johnson} = 4k_B T R.
	\end{equation}
	Within a given bandwidth $B$ the root mean squared voltage fluctuations $V_\mathrm{RMS}$ emitted by the resistor are:
	\begin{equation}
		\begin{aligned}            
			V_\mathrm{RMS} &= \sqrt{S_V^\mathrm{Johnson} B} \\
			&= \sqrt{4k_B T R B}.
		\end{aligned}
	\end{equation}
	Then, the power dissipated into a matched circuit is
	\begin{equation}        
		\begin{aligned}            
			P^\mathrm{Johnson} &= \frac{V_\mathrm{RMS}^2}{4R}\\
			&= k_B T B,
		\end{aligned}
	\end{equation}
	where the factor $4$ comes from a voltage divider configuration. Thus, the power spectral density, in W/Hz, dissipated into a matched circuit is:
	\begin{equation}
		S_P^\mathrm{Johnson} = k_B T.
	\end{equation}
	
	\subsection{Shot noise}
	In a SNTJ, shot noise originates from the discrete tunneling of electrons. In the classical limit, the associated power spectral density can be expressed in $\mathrm{A^2/Hz}$ as \cite{beenarker2003quantum}:
	\begin{equation}
		S_I^\mathrm{shot} = 2e \lvert I\rvert,
	\end{equation}
	with $I$ the current flowing through the SNTJ. Then, the power spectral density, in W/Hz, dissipated into a matched circuit is
	\begin{equation}
		S_P^\mathrm{shot} = \frac{e \lvert V\rvert}{2},
	\end{equation}
	where $V$ is the voltage across the SNTJ.

	\section{The various limit cases of the noise generated by a SNTJ}
	\label{app:sntjcases}
	
	The SNTJ generates a noise as given by Eq.\,\ref{eq:shot} \textit{i.e.} whose power spectral density (PSD), in W/Hz, is:
	\begin{equation}
		\begin{aligned}
			P_\mathrm{in}^\mathrm{shot} = &\frac{eV+h\nu}{4} \coth{\left(\frac{eV+h\nu}{2k_BT_e}\right)} \\ 
			+ &\frac{eV-h\nu}{4} \coth{\left(\frac{eV-h\nu}{2k_BT_e}\right)}
		\end{aligned}
		\label{eq:Pshot}
	\end{equation}
	This complicated formula may be hard to grasp, because it simultaneously depends on $\nu$, $V$, and $T_e$. But, we can consider three limit cases where the formula is simpler.
	
	When $eV=0$ (or when it is much smaller than $k_B T_e$ and $h\nu$), Eq.\,\ref{eq:Pshot} simplifies to
	\begin{equation}
		\begin{aligned}
			P_\mathrm{in}^\mathrm{shot}(V=0) = &\frac{h\nu}{2}\coth{\left(\frac{h\nu}{2k_B T_e}\right)}\\
			&\xrightarrow[k_BT_e \ll h\nu]{}\frac{h\nu}{2}\\
			&\xrightarrow[k_BT_e \gg h\nu]{}k_B T_e,
		\end{aligned}
	\end{equation}
	equal to $P_\mathrm{in}^\mathrm{Johnson}$, the PSD delivered by a resistor to a matched circuit.
	
	When $h\nu=0$ (or when it is much smaller than $e\lvert V\rvert$ and $k_B T_e$), Eq.\,\ref{eq:Pshot} simplifies to
	\begin{equation}
		\begin{aligned}
			P_\mathrm{in}^\mathrm{shot}(\nu=0) = &\frac{eV}{2}\coth{\left(\frac{eV}{2k_B T_e}\right)}\\
			&\xrightarrow[k_BT_e \ll e\lvert V\rvert]{}\frac{e\lvert V\rvert}{2}\\
			&\xrightarrow[k_BT_e \gg e\lvert V\rvert]{}k_B T_e,
		\end{aligned}
	\end{equation}
	equivalent to $P_\mathrm{in}^\mathrm{Johnson}$, with $eV$ replacing $h\nu$.
	
	Finally, when $k_B T_e=0$ (or when it is much smaller than $h\nu$ and $e\lvert V\rvert$), Eq.\,\ref{eq:Pshot} simplifies to
	\begin{equation}
		P_\mathrm{in}^\mathrm{shot}(T_e=0) =
		\begin{cases}
			\frac{h\nu}{2} &\mathrm{if\hspace{0.5cm}} e\lvert V\rvert \leq h\nu\\
			\frac{e\lvert V\rvert}{2} &\mathrm{if\hspace{0.5cm}} e\lvert V\rvert \geq h\nu,
		\end{cases}
	\end{equation}
	which is the limit case represented in Fig.\,\ref{fig:simplenoisetheo}b, in black. A non-null temperature smooths the transition between the two regimes, happening at $e\lvert V\rvert = h\nu$.
	
	\section{Experimental setup}
	\label{app:setup}
	
	\begin{figure*}[htpb!]	
		\includegraphics[scale=0.9]{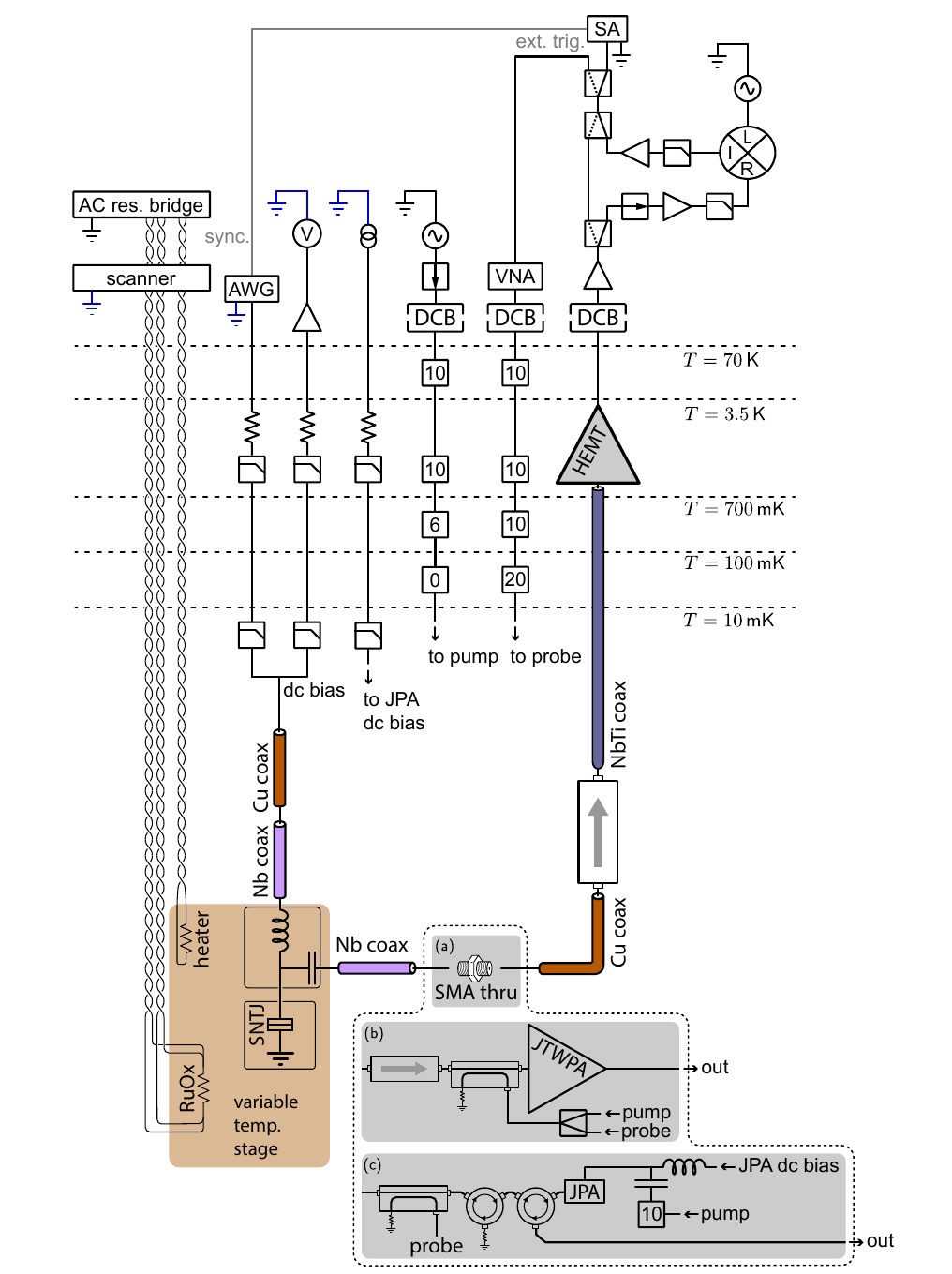}
		\caption{Schematic of the setup used to perform all the noise measurements. We distinguish the rack ground (black) from the fridge ground (blue). The VTS thermometer and heater are connected to a room temperature AC resistance bridge with twisted-pair cables. On the SNTJ and JPA dc biasing lines, there are $100$\,k\ohm{} resistors followed by $1$\,MHz low-pass filters (LPFs) at $4$\,K and $1$\,MHz LPFs at $10$\,mK. On the rf pump and VNA probe lines, microwave attenuators are placed at several thermal stages, whose values are indicated on the schematic. The VNA output tone is sent either to the JTWPA pump line, or to the JPA probe line, depending on the cryogenic configuration, (b) or (c). At the output, three switches allow us to route signals onto various equipment. In particular, the mixer is preceded by a room temperature rf mplifier and $7.2$\,GHz anti-aliasing LPF, and then followed by a baseband amplifier and $350$\,MHz LPF. See the main text for further details.} 
		\label{fig:fullsetup}
	\end{figure*}  
	
	Figure \ref{fig:fullsetup} shows a schematic of the full experimental setup, with various configurations, depending on which measurement was performed. Common to all four noise measurements presented in the main text is the VTS and SNTJ biases (on the left of the figure). An AC resistance bridge allows us to measure the VTS temperature, and to servo the VTS heater.     
	
	The fridge ground and the ground of the room temperature electronics rack only connect at a well-defined wall ground. However, the SNTJ and JPA dc biasing electronics is floated with respect to the rack ground, and only connects to the ground via the fridge ground, because the SNTJ and JPA are shorted to the fridge ground within their respective packaging. We thus avoid the formation of ground loops, that can perturb the SNTJ bias.
	
	\subsection{The VTS}
	
	The VTS consists of a copper plate, on which is mounted a calibrated resistor (ruthenium oxide), a heater (a $1$\,k\ohm{} resistor), the SNTJ and the following bias tee (Anritsu K250). We weakly couple the VTS to the mixing chamber plate via a stainless steel screw. We then use superconducting niobium microwave cables for the dc and microwave connections to the VTS, thereby preventing any thermal connection via these channels.
	
	\subsection{SNTJ impedance and biasing procedure}
	
	In principle, the SNTJ is fabricated to be $50$\,\ohm; in practice, we measure its impedance \textit{in situ} with a voltage tap. More precisely, two branches lead to the SNTJ: one delivers a current, and the other allows us to measure the voltage drop across the SNTJ. In Fig.\,\ref{fig:fullsetup}, these correspond to the branches connected to the arbitrary waveform generator (AWG) and to the voltmeter (V), respectively. The SNTJ biasing procedure is as follows: (i) we set the gain of the amplifier preceding the voltmeter to $\times100$ (verified by sending some small voltage to its input with the AWG, and reading the output with the voltmeter). (ii) Replacing the AWG with a current source, we send a known current across the SNTJ, and measure the corresponding voltage drop with the voltmeter. It gives us the SNTJ's impedance $Z_\mathrm{SNTJ}$ (neglecting the resistance of the bias tee). In our experiment, we found $Z_\mathrm{SNTJ}=49.3$\,\ohm. (iii) Using the AWG, we send a known voltage and measure the voltage drop across the SNTJ. It gives us the division ratio between the SNTJ's impedance and that of the biasing resistor placed at $4$\,K, nominally of $100$\,k\ohm. Such value permits a large division ratio ($1:2001$) which damps down possible voltage fluctuations coming from the voltage source (Keysight 33250A). Note that what we measure here is the SNTJ's dc impedance. The rf impedance might slightly deviate from the dc value, but the noise measurements are not very sensitive to such small deviations. In fact, the noise power transmitted to the transmission line varies only marginally with a reasonably small mismatch between the SNTJ and the line impedance.
	
	\subsection{Wiring of the various experiments}
	
	The number of additional input lines used to control the cryogenic setup depends on the cold configuration (a), (b) or (c). When measuring the noise of with the HEMT as first amplifier (a), no additional input line is used, and the output line is routed to the spectrum analyzer (SA), after further room temperature amplification. When using the JTWPA (b), an additional line in the cryostat is used to deliver the JTWPA pump and a VNA probe tone. At the output, a microwave switch routes signals either to the SA, or to the receiving port of the VNA. When using the JPA (c), a dc line allows us to flux-tune the JPA (operating in a 3-wave mixing mode), and another delivers the JPA pump \cite{peterson2020parametric}. At cryogenic temperatures, the dc and rf pump signals are combined before entering the pump port of the JPA. Conversely, another line is used to send a VNA probe tone, which enters via the JPA signal port (together with the noise emitted by the SNTJ). At the output, signals can be routed with microwave switches either onto the VNA, or onto the SA directly when performing a phase-insensitive noise measurement, or onto the SA after being mixed down when performing the phase-sensitive noise measurement of the amplification chain containing the JPA. In that case, the local oscillator (LO) frequency of the mixer is half that of the JPA pump.
	
	\subsection{Noise acquisition with a spectrum analyzer}
	
	When measuring the small bias response of the shot noise curves in the phase-insensitive case, the AWG outputs a low frequency ramp ($20$\,Hz), with an amplitude symmetric around $0$\,V. The SA sweep is then synchronized to the AWG ramp via its external trigger, with the SA operated in zero span mode. The SA central frequency determines the frequency around which the noise is measured, within a window defined by the SA resolution bandwidth (RBW). All our amplifiers' bandwidth is large enough, and the noise is ``white'' enough over a couple megahertz to have the RBW be $8$\,MHz. In fact, a large RBW increases the signal-to-noise ratio of the measurement. The SA sweep time is then set to $t_\mathrm{SA}=1/20$\,sec and we set the number of points per sweep at $n_\mathrm{SA}=501$. Therefore, $t_\mathrm{SA}/n_\mathrm{SA}$ is the time spend on each point. To have all the points be independent realisation of the noise, we must set the SA video bandwidth (VBW) to a value larger than this time. In practice we use $\mathrm{VBW}=3t_\mathrm{SA}/n_\mathrm{SA}$.
	
	In the phase-sensitive case (using the JPA), the SA is also configured in zero span, but this time it is centered around a frequency slightly detuned from zero ($2.6$\,MHz). In fact, in the phase-sensitive case the JPA amplification bandwidth is effectively ``folded'' onto itself, the central amplification bandwidth being referred back to the zero frequency of the SA. We also reduce the RBW to a small enough value ($1.5$\,MHz), which prevents the measured noise to be contaminated by the $1/f$ response of the SA.
	
	When measuring the wideband noise curves using the SNTJ, we set the AWG to a dc mode and send discrete biasing voltage values. Equivalently, when using the VTS we servo it to a given temperature. For each of these values (voltage or temperature), we set the SA span to cover a desired frequency range ($3.5$ to $8.5$\,GHz) while keeping $n_\mathrm{SA}=501$. We can now freely choose the VBW, as long as it is not bigger than the RBW, which we keep at $8$\,MHz (we don't expect to have features narrower than that in the spectrum). Increasing the VBW is equivalent to averaging for a longer time at each point. The equipment can also average many spectral traces before sending it to the acquisition computer. In practice, we use both types of averages (point-by-point and trace-by-trace): we choose $\mathrm{VBW}=15$\,kHz, and we choose to average over $1500$ traces, which takes about $3$ minutes.

	\section{Fit of the shot-noise curves}
	\label{app:fit}
	
	The fit of the shot noise curves is performed in two main steps: we first fit the asymptotes, where the the shot noise takes a simple form, $N_\mathrm{in}^\mathrm{shot}=e\lvert V \rvert /(2 h \nu)$ (see Eq.\,\ref{eq:shot}), linear with the bias voltage $V$ and independent to the electronic temperature $T_e$. When fitting noise coming out of a parametric amplifier, we use Eq.\,\ref{eq:Pcs}, and therefore we have two inputs, at the signal and idler frequency. We obtain three fitting parameters: an estimate of the system gain $G_\mathrm{sys}$ and added noise $N_\mathrm{sys}$ (in the case of the HEMT) or excess noise $N_\mathrm{sys,ex}$ (in the case of a parametric amplifier), and an offset voltage $V_\mathrm{offs}$, close to $0$. In fact, in principle, the positive and negative portions of the fit crosses at $V=0$, but in practice we leave it as a fitting parameter $V_\mathrm{offs}$, because the zero bias voltage delivered by the AWG may not be perfectly null. We set guesses parameters for the least square regression fit: $V_\mathrm{offs}^\mathrm{g}=0$, $N_\mathrm{sys} = 50$ (HEMT) or $N_\mathrm{sys,ex} = 2$ (JTWPA) or $N_\mathrm{sys,ex} = 0$ (JPA), and the guess for the gain is calculated from the average absolute slope between the positive and negative biases.
	
	We then fit the full curves using Eq.\,\ref{eq:shot}, where we fix $V_\mathrm{offs}$ to what has been found previously, and where we let the chain's gain and added noise vary by $\pm50$\,\% around the asymptotic values. When fitting the HEMT-only data, the guess for the electronic temperature $T_e$ is the SNTJ's physical temperature $T$. When fitting the JTWPA or the JPA data, we fix $T_e=T$, validated by the fits from the HEMT-only data.
	
	\section{Phase-sensitive noise fitting model}
	\label{app:psnoisemodel}
	
	In the phase-sensitive case, we estimate $M = \max({\mathcal{\Tilde{G}}(\alpha))}$ by subtracting $G_\mathrm{sys}^H=72$\,dB, the chain's gain at $6$\,GHz found when the HEMT is the first amplifier, see Fig.\,\ref{fig:hemtdata}c, to the maximal chain's gain presented in Fig.\,\ref{fig:psnoise}. We find $M=19.5$\,dB. We can then approximate $\mathcal{\Tilde{G}}(\alpha)\simeq M\cos^2{\alpha} + \sin^2{\alpha}/M$. It is exact along the amplified quadrature, and approximate along the squeezed quadrature, because $M$ includes $(\eta_1\eta_1^i)^{1/2}$. In Fig.\,\ref{fig:psnoise} we plot $G_\mathrm{sys}^H{\mathcal{\Tilde{G}}(\alpha)}$ (plain blue line).
	
	We estimate $N_\mathrm{sys}(\alpha)$ by taking $N_\mathrm{sys}^H=33$ the system-added noise at $6$\,GHz found when the HEMT is the first amplifier. We then extract $m = \min{(N_\mathrm{sys}}(\alpha))$ from Fig.\,\ref{fig:psnoise} and plot $m + N_\mathrm{sys}^H/\mathcal{\Tilde{G}}(\alpha) - \min{( N_\mathrm{sys}^H/\mathcal{\Tilde{G}}(\alpha))}$ (plain black line).
	
	\vspace{0.1in}	

\begin{thebibliography}{46}%
		\makeatletter
		\providecommand \@ifxundefined [1]{%
			\@ifx{#1\undefined}
		}%
		\providecommand \@ifnum [1]{%
			\ifnum #1\expandafter \@firstoftwo
			\else \expandafter \@secondoftwo
			\fi
		}%
		\providecommand \@ifx [1]{%
			\ifx #1\expandafter \@firstoftwo
			\else \expandafter \@secondoftwo
			\fi
		}%
		\providecommand \natexlab [1]{#1}%
		\providecommand \enquote  [1]{``#1''}%
		\providecommand \bibnamefont  [1]{#1}%
		\providecommand \bibfnamefont [1]{#1}%
		\providecommand \citenamefont [1]{#1}%
		\providecommand \href@noop [0]{\@secondoftwo}%
		\providecommand \href [0]{\begingroup \@sanitize@url \@href}%
		\providecommand \@href[1]{\@@startlink{#1}\@@href}%
		\providecommand \@@href[1]{\endgroup#1\@@endlink}%
		\providecommand \@sanitize@url [0]{\catcode `\\12\catcode `\$12\catcode
			`\&12\catcode `\#12\catcode `\^12\catcode `\_12\catcode `\%12\relax}%
		\providecommand \@@startlink[1]{}%
		\providecommand \@@endlink[0]{}%
		\providecommand \url  [0]{\begingroup\@sanitize@url \@url }%
		\providecommand \@url [1]{\endgroup\@href {#1}{\urlprefix }}%
		\providecommand \urlprefix  [0]{URL }%
		\providecommand \Eprint [0]{\href }%
		\providecommand \doibase [0]{https://doi.org/}%
		\providecommand \selectlanguage [0]{\@gobble}%
		\providecommand \bibinfo  [0]{\@secondoftwo}%
		\providecommand \bibfield  [0]{\@secondoftwo}%
		\providecommand \translation [1]{[#1]}%
		\providecommand \BibitemOpen [0]{}%
		\providecommand \bibitemStop [0]{}%
		\providecommand \bibitemNoStop [0]{.\EOS\space}%
		\providecommand \EOS [0]{\spacefactor3000\relax}%
		\providecommand \BibitemShut  [1]{\csname bibitem#1\endcsname}%
		\let\auto@bib@innerbib\@empty
		\bibitem [{\citenamefont {Galitzki}\ \emph {et~al.}(2018)\citenamefont
			{Galitzki}, \citenamefont {Ali}, \citenamefont {Arnold}, \citenamefont
			{Ashton}, \citenamefont {Austermann}, \citenamefont {Baccigalupi},
			\citenamefont {Baildon}, \citenamefont {Barron}, \citenamefont {Beall},
			\citenamefont {Beckman}, \citenamefont {Bruno}, \citenamefont {Bryan},
			\citenamefont {Calisse}, \citenamefont {Chesmore}, \citenamefont {Chinone},
			\citenamefont {Choi}, \citenamefont {Coppi}, \citenamefont {Crowley},
			\citenamefont {Crowley}, \citenamefont {Cukierman}, \citenamefont {Devlin},
			\citenamefont {Dicker}, \citenamefont {Dober}, \citenamefont {Duff},
			\citenamefont {Dunkley}, \citenamefont {Fabbian}, \citenamefont {Gallardo},
			\citenamefont {Gerbino}, \citenamefont {Goeckner-Wald}, \citenamefont
			{Golec}, \citenamefont {Gudmundsson}, \citenamefont {Healy}, \citenamefont
			{Henderson}, \citenamefont {Hill}, \citenamefont {Hilton}, \citenamefont
			{Ho}, \citenamefont {Howe}, \citenamefont {Hubmayr}, \citenamefont {Jeong},
			\citenamefont {Keating}, \citenamefont {Koopman}, \citenamefont {Kiuchi},
			\citenamefont {Kusaka}, \citenamefont {Lashner}, \citenamefont {Lee},
			\citenamefont {Li}, \citenamefont {Limon}, \citenamefont {Lungu},
			\citenamefont {Matsuda}, \citenamefont {Mauskopf}, \citenamefont {May},
			\citenamefont {McCallum}, \citenamefont {McMahon}, \citenamefont {Nati},
			\citenamefont {Niemack}, \citenamefont {Orlowski-Scherer}, \citenamefont
			{Parshley}, \citenamefont {Piccirillo}, \citenamefont {Rao}, \citenamefont
			{Raum}, \citenamefont {Salatino}, \citenamefont {Seibert}, \citenamefont
			{Sierra}, \citenamefont {Silva-Feaver}, \citenamefont {Simon}, \citenamefont
			{Staggs}, \citenamefont {Stevens}, \citenamefont {Suzuki}, \citenamefont
			{Teply}, \citenamefont {Thornton}, \citenamefont {Tsai}, \citenamefont
			{Ullom}, \citenamefont {Vavagiakis}, \citenamefont {Vissers}, \citenamefont
			{Westbrook}, \citenamefont {Wollack}, \citenamefont {Xu},\ and\ \citenamefont
			{Zhu}}]{Galitzki2018the}%
		\BibitemOpen
		\bibfield  {author} {\bibinfo {author} {\bibfnamefont {N.}~\bibnamefont
				{Galitzki}}, \bibinfo {author} {\bibfnamefont {A.}~\bibnamefont {Ali}},
			\bibinfo {author} {\bibfnamefont {K.~S.}\ \bibnamefont {Arnold}}, \bibinfo
			{author} {\bibfnamefont {P.~C.}\ \bibnamefont {Ashton}}, \bibinfo {author}
			{\bibfnamefont {J.~E.}\ \bibnamefont {Austermann}}, \bibinfo {author}
			{\bibfnamefont {C.}~\bibnamefont {Baccigalupi}}, \bibinfo {author}
			{\bibfnamefont {T.}~\bibnamefont {Baildon}}, \bibinfo {author} {\bibfnamefont
				{D.}~\bibnamefont {Barron}}, \bibinfo {author} {\bibfnamefont {J.~A.}\
				\bibnamefont {Beall}}, \bibinfo {author} {\bibfnamefont {S.}~\bibnamefont
				{Beckman}}, \bibinfo {author} {\bibfnamefont {S.~M.~M.}\ \bibnamefont
				{Bruno}}, \bibinfo {author} {\bibfnamefont {S.}~\bibnamefont {Bryan}},
			\bibinfo {author} {\bibfnamefont {P.~G.}\ \bibnamefont {Calisse}}, \bibinfo
			{author} {\bibfnamefont {G.~E.}\ \bibnamefont {Chesmore}}, \bibinfo {author}
			{\bibfnamefont {Y.}~\bibnamefont {Chinone}}, \bibinfo {author} {\bibfnamefont
				{S.~K.}\ \bibnamefont {Choi}}, \bibinfo {author} {\bibfnamefont
				{G.}~\bibnamefont {Coppi}}, \bibinfo {author} {\bibfnamefont {K.~D.}\
				\bibnamefont {Crowley}}, \bibinfo {author} {\bibfnamefont {K.~T.}\
				\bibnamefont {Crowley}}, \bibinfo {author} {\bibfnamefont {A.}~\bibnamefont
				{Cukierman}}, \bibinfo {author} {\bibfnamefont {M.~J.}\ \bibnamefont
				{Devlin}}, \bibinfo {author} {\bibfnamefont {S.}~\bibnamefont {Dicker}},
			\bibinfo {author} {\bibfnamefont {B.}~\bibnamefont {Dober}}, \bibinfo
			{author} {\bibfnamefont {S.~M.}\ \bibnamefont {Duff}}, \bibinfo {author}
			{\bibfnamefont {J.}~\bibnamefont {Dunkley}}, \bibinfo {author} {\bibfnamefont
				{G.}~\bibnamefont {Fabbian}}, \bibinfo {author} {\bibfnamefont {P.~A.}\
				\bibnamefont {Gallardo}}, \bibinfo {author} {\bibfnamefont {M.}~\bibnamefont
				{Gerbino}}, \bibinfo {author} {\bibfnamefont {N.}~\bibnamefont
				{Goeckner-Wald}}, \bibinfo {author} {\bibfnamefont {J.~E.}\ \bibnamefont
				{Golec}}, \bibinfo {author} {\bibfnamefont {J.~E.}\ \bibnamefont
				{Gudmundsson}}, \bibinfo {author} {\bibfnamefont {E.~E.}\ \bibnamefont
				{Healy}}, \bibinfo {author} {\bibfnamefont {S.}~\bibnamefont {Henderson}},
			\bibinfo {author} {\bibfnamefont {C.~A.}\ \bibnamefont {Hill}}, \bibinfo
			{author} {\bibfnamefont {G.~C.}\ \bibnamefont {Hilton}}, \bibinfo {author}
			{\bibfnamefont {S.-P.~P.}\ \bibnamefont {Ho}}, \bibinfo {author}
			{\bibfnamefont {L.~A.}\ \bibnamefont {Howe}}, \bibinfo {author}
			{\bibfnamefont {J.}~\bibnamefont {Hubmayr}}, \bibinfo {author} {\bibfnamefont
				{O.}~\bibnamefont {Jeong}}, \bibinfo {author} {\bibfnamefont
				{B.}~\bibnamefont {Keating}}, \bibinfo {author} {\bibfnamefont {B.~J.}\
				\bibnamefont {Koopman}}, \bibinfo {author} {\bibfnamefont {K.}~\bibnamefont
				{Kiuchi}}, \bibinfo {author} {\bibfnamefont {A.}~\bibnamefont {Kusaka}},
			\bibinfo {author} {\bibfnamefont {J.}~\bibnamefont {Lashner}}, \bibinfo
			{author} {\bibfnamefont {A.~T.}\ \bibnamefont {Lee}}, \bibinfo {author}
			{\bibfnamefont {Y.}~\bibnamefont {Li}}, \bibinfo {author} {\bibfnamefont
				{M.}~\bibnamefont {Limon}}, \bibinfo {author} {\bibfnamefont
				{M.}~\bibnamefont {Lungu}}, \bibinfo {author} {\bibfnamefont
				{F.}~\bibnamefont {Matsuda}}, \bibinfo {author} {\bibfnamefont {P.~D.}\
				\bibnamefont {Mauskopf}}, \bibinfo {author} {\bibfnamefont {A.~J.}\
				\bibnamefont {May}}, \bibinfo {author} {\bibfnamefont {N.}~\bibnamefont
				{McCallum}}, \bibinfo {author} {\bibfnamefont {J.}~\bibnamefont {McMahon}},
			\bibinfo {author} {\bibfnamefont {F.}~\bibnamefont {Nati}}, \bibinfo {author}
			{\bibfnamefont {M.~D.}\ \bibnamefont {Niemack}}, \bibinfo {author}
			{\bibfnamefont {J.~L.}\ \bibnamefont {Orlowski-Scherer}}, \bibinfo {author}
			{\bibfnamefont {S.~C.}\ \bibnamefont {Parshley}}, \bibinfo {author}
			{\bibfnamefont {L.}~\bibnamefont {Piccirillo}}, \bibinfo {author}
			{\bibfnamefont {M.~S.}\ \bibnamefont {Rao}}, \bibinfo {author} {\bibfnamefont
				{C.}~\bibnamefont {Raum}}, \bibinfo {author} {\bibfnamefont {M.}~\bibnamefont
				{Salatino}}, \bibinfo {author} {\bibfnamefont {J.~S.}\ \bibnamefont
				{Seibert}}, \bibinfo {author} {\bibfnamefont {C.}~\bibnamefont {Sierra}},
			\bibinfo {author} {\bibfnamefont {M.}~\bibnamefont {Silva-Feaver}}, \bibinfo
			{author} {\bibfnamefont {S.~M.}\ \bibnamefont {Simon}}, \bibinfo {author}
			{\bibfnamefont {S.~T.}\ \bibnamefont {Staggs}}, \bibinfo {author}
			{\bibfnamefont {J.~R.}\ \bibnamefont {Stevens}}, \bibinfo {author}
			{\bibfnamefont {A.}~\bibnamefont {Suzuki}}, \bibinfo {author} {\bibfnamefont
				{G.}~\bibnamefont {Teply}}, \bibinfo {author} {\bibfnamefont
				{R.}~\bibnamefont {Thornton}}, \bibinfo {author} {\bibfnamefont
				{C.}~\bibnamefont {Tsai}}, \bibinfo {author} {\bibfnamefont {J.~N.}\
				\bibnamefont {Ullom}}, \bibinfo {author} {\bibfnamefont {E.~M.}\ \bibnamefont
				{Vavagiakis}}, \bibinfo {author} {\bibfnamefont {M.~R.}\ \bibnamefont
				{Vissers}}, \bibinfo {author} {\bibfnamefont {B.}~\bibnamefont {Westbrook}},
			\bibinfo {author} {\bibfnamefont {E.~J.}\ \bibnamefont {Wollack}}, \bibinfo
			{author} {\bibfnamefont {Z.}~\bibnamefont {Xu}},\ and\ \bibinfo {author}
			{\bibfnamefont {N.}~\bibnamefont {Zhu}},\ }\bibfield  {title} {\enquote
			{\bibinfo {title} {{The Simons Observatory: instrument overview}},}\ }in\
		\href {https://doi.org/10.1117/12.2312985} {\emph {\bibinfo {booktitle}
				{Millimeter, Submillimeter, and Far-Infrared Detectors and Instrumentation
					for Astronomy IX}}},\ Vol.\ \bibinfo {volume} {10708},\ \bibinfo {editor}
		{edited by\ \bibinfo {editor} {\bibfnamefont {J.}~\bibnamefont {Zmuidzinas}}\
			and\ \bibinfo {editor} {\bibfnamefont {J.-R.}\ \bibnamefont {Gao}}},\
		\bibinfo {organization} {International Society for Optics and Photonics}\
		(\bibinfo  {publisher} {SPIE},\ \bibinfo {year} {2018})\ p.\ \bibinfo {pages}
		{1070804}\BibitemShut {NoStop}%
		\bibitem [{\citenamefont {McCarrick}\ \emph {et~al.}(2021)\citenamefont
			{McCarrick}, \citenamefont {Healy}, \citenamefont {Ahmed}, \citenamefont
			{Arnold}, \citenamefont {Atkins}, \citenamefont {Austermann}, \citenamefont
			{Bhandarkar}, \citenamefont {Beall}, \citenamefont {Bruno}, \citenamefont
			{Choi}, \citenamefont {Connors}, \citenamefont {Cothard}, \citenamefont
			{Crowley}, \citenamefont {Dicker}, \citenamefont {Dober}, \citenamefont
			{Duell}, \citenamefont {Duff}, \citenamefont {Dutcher}, \citenamefont
			{Frisch}, \citenamefont {Galitzki}, \citenamefont {Gralla}, \citenamefont
			{Gudmundsson}, \citenamefont {Henderson}, \citenamefont {Hilton},
			\citenamefont {Ho}, \citenamefont {Huber}, \citenamefont {Hubmayr},
			\citenamefont {Iuliano}, \citenamefont {Johnson}, \citenamefont {Kofman},
			\citenamefont {Kusaka}, \citenamefont {Lashner}, \citenamefont {Lee},
			\citenamefont {Li}, \citenamefont {Link}, \citenamefont {Lucas},
			\citenamefont {Lungu}, \citenamefont {Mates}, \citenamefont {McMahon},
			\citenamefont {Niemack}, \citenamefont {Orlowski-Scherer}, \citenamefont
			{Seibert}, \citenamefont {Silva-Feaver}, \citenamefont {Simon}, \citenamefont
			{Staggs}, \citenamefont {Suzuki}, \citenamefont {Terasaki}, \citenamefont
			{Thornton}, \citenamefont {Ullom}, \citenamefont {Vavagiakis}, \citenamefont
			{Vale}, \citenamefont {Lanen}, \citenamefont {Vissers}, \citenamefont {Wang},
			\citenamefont {Wollack}, \citenamefont {Xu}, \citenamefont {Young},
			\citenamefont {Yu}, \citenamefont {Zheng},\ and\ \citenamefont
			{Zhu}}]{McCarrick2021The}%
		\BibitemOpen
		\bibfield  {author} {\bibinfo {author} {\bibfnamefont {H.}~\bibnamefont
				{McCarrick}}, \bibinfo {author} {\bibfnamefont {E.}~\bibnamefont {Healy}},
			\bibinfo {author} {\bibfnamefont {Z.}~\bibnamefont {Ahmed}}, \bibinfo
			{author} {\bibfnamefont {K.}~\bibnamefont {Arnold}}, \bibinfo {author}
			{\bibfnamefont {Z.}~\bibnamefont {Atkins}}, \bibinfo {author} {\bibfnamefont
				{J.~E.}\ \bibnamefont {Austermann}}, \bibinfo {author} {\bibfnamefont
				{T.}~\bibnamefont {Bhandarkar}}, \bibinfo {author} {\bibfnamefont {J.~A.}\
				\bibnamefont {Beall}}, \bibinfo {author} {\bibfnamefont {S.~M.}\ \bibnamefont
				{Bruno}}, \bibinfo {author} {\bibfnamefont {S.~K.}\ \bibnamefont {Choi}},
			\bibinfo {author} {\bibfnamefont {J.}~\bibnamefont {Connors}}, \bibinfo
			{author} {\bibfnamefont {N.~F.}\ \bibnamefont {Cothard}}, \bibinfo {author}
			{\bibfnamefont {K.~D.}\ \bibnamefont {Crowley}}, \bibinfo {author}
			{\bibfnamefont {S.}~\bibnamefont {Dicker}}, \bibinfo {author} {\bibfnamefont
				{B.}~\bibnamefont {Dober}}, \bibinfo {author} {\bibfnamefont {C.~J.}\
				\bibnamefont {Duell}}, \bibinfo {author} {\bibfnamefont {S.~M.}\ \bibnamefont
				{Duff}}, \bibinfo {author} {\bibfnamefont {D.}~\bibnamefont {Dutcher}},
			\bibinfo {author} {\bibfnamefont {J.~C.}\ \bibnamefont {Frisch}}, \bibinfo
			{author} {\bibfnamefont {N.}~\bibnamefont {Galitzki}}, \bibinfo {author}
			{\bibfnamefont {M.~B.}\ \bibnamefont {Gralla}}, \bibinfo {author}
			{\bibfnamefont {J.~E.}\ \bibnamefont {Gudmundsson}}, \bibinfo {author}
			{\bibfnamefont {S.~W.}\ \bibnamefont {Henderson}}, \bibinfo {author}
			{\bibfnamefont {G.~C.}\ \bibnamefont {Hilton}}, \bibinfo {author}
			{\bibfnamefont {S.-P.~P.}\ \bibnamefont {Ho}}, \bibinfo {author}
			{\bibfnamefont {Z.~B.}\ \bibnamefont {Huber}}, \bibinfo {author}
			{\bibfnamefont {J.}~\bibnamefont {Hubmayr}}, \bibinfo {author} {\bibfnamefont
				{J.}~\bibnamefont {Iuliano}}, \bibinfo {author} {\bibfnamefont {B.~R.}\
				\bibnamefont {Johnson}}, \bibinfo {author} {\bibfnamefont {A.~M.}\
				\bibnamefont {Kofman}}, \bibinfo {author} {\bibfnamefont {A.}~\bibnamefont
				{Kusaka}}, \bibinfo {author} {\bibfnamefont {J.}~\bibnamefont {Lashner}},
			\bibinfo {author} {\bibfnamefont {A.~T.}\ \bibnamefont {Lee}}, \bibinfo
			{author} {\bibfnamefont {Y.}~\bibnamefont {Li}}, \bibinfo {author}
			{\bibfnamefont {M.~J.}\ \bibnamefont {Link}}, \bibinfo {author}
			{\bibfnamefont {T.~J.}\ \bibnamefont {Lucas}}, \bibinfo {author}
			{\bibfnamefont {M.}~\bibnamefont {Lungu}}, \bibinfo {author} {\bibfnamefont
				{J.~A.~B.}\ \bibnamefont {Mates}}, \bibinfo {author} {\bibfnamefont {J.~J.}\
				\bibnamefont {McMahon}}, \bibinfo {author} {\bibfnamefont {M.~D.}\
				\bibnamefont {Niemack}}, \bibinfo {author} {\bibfnamefont {J.}~\bibnamefont
				{Orlowski-Scherer}}, \bibinfo {author} {\bibfnamefont {J.}~\bibnamefont
				{Seibert}}, \bibinfo {author} {\bibfnamefont {M.}~\bibnamefont
				{Silva-Feaver}}, \bibinfo {author} {\bibfnamefont {S.~M.}\ \bibnamefont
				{Simon}}, \bibinfo {author} {\bibfnamefont {S.}~\bibnamefont {Staggs}},
			\bibinfo {author} {\bibfnamefont {A.}~\bibnamefont {Suzuki}}, \bibinfo
			{author} {\bibfnamefont {T.}~\bibnamefont {Terasaki}}, \bibinfo {author}
			{\bibfnamefont {R.}~\bibnamefont {Thornton}}, \bibinfo {author}
			{\bibfnamefont {J.~N.}\ \bibnamefont {Ullom}}, \bibinfo {author}
			{\bibfnamefont {E.~M.}\ \bibnamefont {Vavagiakis}}, \bibinfo {author}
			{\bibfnamefont {L.~R.}\ \bibnamefont {Vale}}, \bibinfo {author}
			{\bibfnamefont {J.~V.}\ \bibnamefont {Lanen}}, \bibinfo {author}
			{\bibfnamefont {M.~R.}\ \bibnamefont {Vissers}}, \bibinfo {author}
			{\bibfnamefont {Y.}~\bibnamefont {Wang}}, \bibinfo {author} {\bibfnamefont
				{E.~J.}\ \bibnamefont {Wollack}}, \bibinfo {author} {\bibfnamefont
				{Z.}~\bibnamefont {Xu}}, \bibinfo {author} {\bibfnamefont {E.}~\bibnamefont
				{Young}}, \bibinfo {author} {\bibfnamefont {C.}~\bibnamefont {Yu}}, \bibinfo
			{author} {\bibfnamefont {K.}~\bibnamefont {Zheng}},\ and\ \bibinfo {author}
			{\bibfnamefont {N.}~\bibnamefont {Zhu}},\ }\bibfield  {title} {\enquote
			{\bibinfo {title} {The simons observatory microwave squid multiplexing
					detector module design},}\ }\href {https://doi.org/10.3847/1538-4357/ac2232}
		{\bibfield  {journal} {\bibinfo  {journal} {The Astrophysical Journal}\
			}\textbf {\bibinfo {volume} {922}},\ \bibinfo {pages} {38} (\bibinfo {year}
			{2021})}\BibitemShut {NoStop}%
		\bibitem [{\citenamefont {Noroozian}\ \emph {et~al.}(2013)\citenamefont
			{Noroozian}, \citenamefont {Mates}, \citenamefont {Bennett}, \citenamefont
			{Brevik}, \citenamefont {Fowler}, \citenamefont {Gao}, \citenamefont
			{Hilton}, \citenamefont {Horansky}, \citenamefont {Irwin}, \citenamefont
			{Kang}, \citenamefont {Schmidt}, \citenamefont {Vale},\ and\ \citenamefont
			{Ullom}}]{Noroozian2013high}%
		\BibitemOpen
		\bibfield  {author} {\bibinfo {author} {\bibfnamefont {O.}~\bibnamefont
				{Noroozian}}, \bibinfo {author} {\bibfnamefont {J.~A.~B.}\ \bibnamefont
				{Mates}}, \bibinfo {author} {\bibfnamefont {D.~A.}\ \bibnamefont {Bennett}},
			\bibinfo {author} {\bibfnamefont {J.~A.}\ \bibnamefont {Brevik}}, \bibinfo
			{author} {\bibfnamefont {J.~W.}\ \bibnamefont {Fowler}}, \bibinfo {author}
			{\bibfnamefont {J.}~\bibnamefont {Gao}}, \bibinfo {author} {\bibfnamefont
				{G.~C.}\ \bibnamefont {Hilton}}, \bibinfo {author} {\bibfnamefont {R.~D.}\
				\bibnamefont {Horansky}}, \bibinfo {author} {\bibfnamefont {K.~D.}\
				\bibnamefont {Irwin}}, \bibinfo {author} {\bibfnamefont {Z.}~\bibnamefont
				{Kang}}, \bibinfo {author} {\bibfnamefont {D.~R.}\ \bibnamefont {Schmidt}},
			\bibinfo {author} {\bibfnamefont {L.~R.}\ \bibnamefont {Vale}},\ and\
			\bibinfo {author} {\bibfnamefont {J.~N.}\ \bibnamefont {Ullom}},\ }\bibfield
		{title} {\enquote {\bibinfo {title} {High-resolution gamma-ray spectroscopy
					with a microwave-multiplexed transition-edge sensor array},}\ }\href
		{https://doi.org/10.1063/1.4829156} {\bibfield  {journal} {\bibinfo
				{journal} {Applied Physics Letters}\ }\textbf {\bibinfo {volume} {103}},\
			\bibinfo {pages} {202602} (\bibinfo {year} {2013})}\BibitemShut {NoStop}%
		\bibitem [{\citenamefont {Mates}\ \emph {et~al.}(2017)\citenamefont {Mates},
			\citenamefont {Becker}, \citenamefont {Bennett}, \citenamefont {Dober},
			\citenamefont {Gard}, \citenamefont {Hays-Wehle}, \citenamefont {Fowler},
			\citenamefont {Hilton}, \citenamefont {Reintsema}, \citenamefont {Schmidt},
			\citenamefont {Swetz}, \citenamefont {Vale},\ and\ \citenamefont
			{Ullom}}]{mates2017simultaneous}%
		\BibitemOpen
		\bibfield  {author} {\bibinfo {author} {\bibfnamefont {J.~A.~B.}\
				\bibnamefont {Mates}}, \bibinfo {author} {\bibfnamefont {D.~T.}\ \bibnamefont
				{Becker}}, \bibinfo {author} {\bibfnamefont {D.~A.}\ \bibnamefont {Bennett}},
			\bibinfo {author} {\bibfnamefont {B.~J.}\ \bibnamefont {Dober}}, \bibinfo
			{author} {\bibfnamefont {J.~D.}\ \bibnamefont {Gard}}, \bibinfo {author}
			{\bibfnamefont {J.~P.}\ \bibnamefont {Hays-Wehle}}, \bibinfo {author}
			{\bibfnamefont {J.~W.}\ \bibnamefont {Fowler}}, \bibinfo {author}
			{\bibfnamefont {G.~C.}\ \bibnamefont {Hilton}}, \bibinfo {author}
			{\bibfnamefont {C.~D.}\ \bibnamefont {Reintsema}}, \bibinfo {author}
			{\bibfnamefont {D.~R.}\ \bibnamefont {Schmidt}}, \bibinfo {author}
			{\bibfnamefont {D.~S.}\ \bibnamefont {Swetz}}, \bibinfo {author}
			{\bibfnamefont {L.~R.}\ \bibnamefont {Vale}},\ and\ \bibinfo {author}
			{\bibfnamefont {J.~N.}\ \bibnamefont {Ullom}},\ }\bibfield  {title} {\enquote
			{\bibinfo {title} {Simultaneous readout of 128 x-ray and gamma-ray
					transition-edge microcalorimeters using microwave squid multiplexing},}\
		}\href {https://doi.org/10.1063/1.4986222} {\bibfield  {journal} {\bibinfo
				{journal} {Applied Physics Letters}\ }\textbf {\bibinfo {volume} {111}},\
			\bibinfo {pages} {062601} (\bibinfo {year} {2017})},\ \Eprint
		{https://arxiv.org/abs/https://doi.org/10.1063/1.4986222}
		{https://doi.org/10.1063/1.4986222} \BibitemShut {NoStop}%
		\bibitem [{\citenamefont {Nakashima}\ \emph {et~al.}(2020)\citenamefont
			{Nakashima}, \citenamefont {Hirayama}, \citenamefont {Kohjiro}, \citenamefont
			{Yamamori}, \citenamefont {Nagasawa}, \citenamefont {Sato}, \citenamefont
			{Yamada}, \citenamefont {Hayakawa}, \citenamefont {Yamasaki}, \citenamefont
			{Mitsuda}, \citenamefont {Nagayoshi}, \citenamefont {Akamatsu}, \citenamefont
			{Gottardi}, \citenamefont {Taralli}, \citenamefont {Bruijn}, \citenamefont
			{Ridder}, \citenamefont {Gao},\ and\ \citenamefont {den
				Herder}}]{Nakashima2020low}%
		\BibitemOpen
		\bibfield  {author} {\bibinfo {author} {\bibfnamefont {Y.}~\bibnamefont
				{Nakashima}}, \bibinfo {author} {\bibfnamefont {F.}~\bibnamefont {Hirayama}},
			\bibinfo {author} {\bibfnamefont {S.}~\bibnamefont {Kohjiro}}, \bibinfo
			{author} {\bibfnamefont {H.}~\bibnamefont {Yamamori}}, \bibinfo {author}
			{\bibfnamefont {S.}~\bibnamefont {Nagasawa}}, \bibinfo {author}
			{\bibfnamefont {A.}~\bibnamefont {Sato}}, \bibinfo {author} {\bibfnamefont
				{S.}~\bibnamefont {Yamada}}, \bibinfo {author} {\bibfnamefont
				{R.}~\bibnamefont {Hayakawa}}, \bibinfo {author} {\bibfnamefont {N.~Y.}\
				\bibnamefont {Yamasaki}}, \bibinfo {author} {\bibfnamefont {K.}~\bibnamefont
				{Mitsuda}}, \bibinfo {author} {\bibfnamefont {K.}~\bibnamefont {Nagayoshi}},
			\bibinfo {author} {\bibfnamefont {H.}~\bibnamefont {Akamatsu}}, \bibinfo
			{author} {\bibfnamefont {L.}~\bibnamefont {Gottardi}}, \bibinfo {author}
			{\bibfnamefont {E.}~\bibnamefont {Taralli}}, \bibinfo {author} {\bibfnamefont
				{M.~P.}\ \bibnamefont {Bruijn}}, \bibinfo {author} {\bibfnamefont {M.~L.}\
				\bibnamefont {Ridder}}, \bibinfo {author} {\bibfnamefont {J.~R.}\
				\bibnamefont {Gao}},\ and\ \bibinfo {author} {\bibfnamefont {J.~W.~A.}\
				\bibnamefont {den Herder}},\ }\bibfield  {title} {\enquote {\bibinfo {title}
				{Low-noise microwave squid multiplexed readout of 38 x-ray transition-edge
					sensor microcalorimeters},}\ }\href {https://doi.org/10.1063/5.0016333}
		{\bibfield  {journal} {\bibinfo  {journal} {Applied Physics Letters}\
			}\textbf {\bibinfo {volume} {117}},\ \bibinfo {pages} {122601} (\bibinfo
			{year} {2020})}\BibitemShut {NoStop}%
		\bibitem [{\citenamefont {Szypryt}\ \emph {et~al.}(2021)\citenamefont
			{Szypryt}, \citenamefont {Bennett}, \citenamefont {Boone}, \citenamefont
			{Dagel}, \citenamefont {Dalton}, \citenamefont {Doriese}, \citenamefont
			{Durkin}, \citenamefont {Fowler}, \citenamefont {Garboczi}, \citenamefont
			{Gard}, \citenamefont {Hilton}, \citenamefont {Imrek}, \citenamefont
			{Jimenez}, \citenamefont {Kotsubo}, \citenamefont {Larson}, \citenamefont
			{Levine}, \citenamefont {Mates}, \citenamefont {McArthur}, \citenamefont
			{Morgan}, \citenamefont {Nakamura}, \citenamefont {O’Neil}, \citenamefont
			{Ortiz}, \citenamefont {Pappas}, \citenamefont {Reintsema}, \citenamefont
			{Schmidt}, \citenamefont {Swetz}, \citenamefont {Thompson}, \citenamefont
			{Ullom}, \citenamefont {Walker}, \citenamefont {Weber}, \citenamefont
			{Wessels},\ and\ \citenamefont {Wheeler}}]{Szypryt2021design}%
		\BibitemOpen
		\bibfield  {author} {\bibinfo {author} {\bibfnamefont {P.}~\bibnamefont
				{Szypryt}}, \bibinfo {author} {\bibfnamefont {D.~A.}\ \bibnamefont
				{Bennett}}, \bibinfo {author} {\bibfnamefont {W.~J.}\ \bibnamefont {Boone}},
			\bibinfo {author} {\bibfnamefont {A.~L.}\ \bibnamefont {Dagel}}, \bibinfo
			{author} {\bibfnamefont {G.}~\bibnamefont {Dalton}}, \bibinfo {author}
			{\bibfnamefont {W.~B.}\ \bibnamefont {Doriese}}, \bibinfo {author}
			{\bibfnamefont {M.}~\bibnamefont {Durkin}}, \bibinfo {author} {\bibfnamefont
				{J.~W.}\ \bibnamefont {Fowler}}, \bibinfo {author} {\bibfnamefont {E.~J.}\
				\bibnamefont {Garboczi}}, \bibinfo {author} {\bibfnamefont {J.~D.}\
				\bibnamefont {Gard}}, \bibinfo {author} {\bibfnamefont {G.~C.}\ \bibnamefont
				{Hilton}}, \bibinfo {author} {\bibfnamefont {J.}~\bibnamefont {Imrek}},
			\bibinfo {author} {\bibfnamefont {E.~S.}\ \bibnamefont {Jimenez}}, \bibinfo
			{author} {\bibfnamefont {V.~Y.}\ \bibnamefont {Kotsubo}}, \bibinfo {author}
			{\bibfnamefont {K.}~\bibnamefont {Larson}}, \bibinfo {author} {\bibfnamefont
				{Z.~H.}\ \bibnamefont {Levine}}, \bibinfo {author} {\bibfnamefont {J.~A.~B.}\
				\bibnamefont {Mates}}, \bibinfo {author} {\bibfnamefont {D.}~\bibnamefont
				{McArthur}}, \bibinfo {author} {\bibfnamefont {K.~M.}\ \bibnamefont
				{Morgan}}, \bibinfo {author} {\bibfnamefont {N.}~\bibnamefont {Nakamura}},
			\bibinfo {author} {\bibfnamefont {G.~C.}\ \bibnamefont {O’Neil}}, \bibinfo
			{author} {\bibfnamefont {N.~J.}\ \bibnamefont {Ortiz}}, \bibinfo {author}
			{\bibfnamefont {C.~G.}\ \bibnamefont {Pappas}}, \bibinfo {author}
			{\bibfnamefont {C.~D.}\ \bibnamefont {Reintsema}}, \bibinfo {author}
			{\bibfnamefont {D.~R.}\ \bibnamefont {Schmidt}}, \bibinfo {author}
			{\bibfnamefont {D.~S.}\ \bibnamefont {Swetz}}, \bibinfo {author}
			{\bibfnamefont {K.~R.}\ \bibnamefont {Thompson}}, \bibinfo {author}
			{\bibfnamefont {J.~N.}\ \bibnamefont {Ullom}}, \bibinfo {author}
			{\bibfnamefont {C.}~\bibnamefont {Walker}}, \bibinfo {author} {\bibfnamefont
				{J.~C.}\ \bibnamefont {Weber}}, \bibinfo {author} {\bibfnamefont {A.~L.}\
				\bibnamefont {Wessels}},\ and\ \bibinfo {author} {\bibfnamefont {J.~W.}\
				\bibnamefont {Wheeler}},\ }\bibfield  {title} {\enquote {\bibinfo {title}
				{Design of a 3000-pixel transition-edge sensor x-ray spectrometer for
					microcircuit tomography},}\ }\href
		{https://doi.org/10.1109/TASC.2021.3052723} {\bibfield  {journal} {\bibinfo
				{journal} {IEEE Transactions on Applied Superconductivity}\ }\textbf
			{\bibinfo {volume} {31}},\ \bibinfo {pages} {1--5} (\bibinfo {year}
			{2021})}\BibitemShut {NoStop}%
		\bibitem [{\citenamefont {Malnou}\ \emph {et~al.}(2023)\citenamefont {Malnou},
			\citenamefont {Mates}, \citenamefont {Vissers}, \citenamefont {Vale},
			\citenamefont {Schmidt}, \citenamefont {Bennett}, \citenamefont {Gao},\ and\
			\citenamefont {Ullom}}]{malnou2023improved}%
		\BibitemOpen
		\bibfield  {author} {\bibinfo {author} {\bibfnamefont {M.}~\bibnamefont
				{Malnou}}, \bibinfo {author} {\bibfnamefont {J.~A.~B.}\ \bibnamefont
				{Mates}}, \bibinfo {author} {\bibfnamefont {M.~R.}\ \bibnamefont {Vissers}},
			\bibinfo {author} {\bibfnamefont {L.~R.}\ \bibnamefont {Vale}}, \bibinfo
			{author} {\bibfnamefont {D.~R.}\ \bibnamefont {Schmidt}}, \bibinfo {author}
			{\bibfnamefont {D.~A.}\ \bibnamefont {Bennett}}, \bibinfo {author}
			{\bibfnamefont {J.}~\bibnamefont {Gao}},\ and\ \bibinfo {author}
			{\bibfnamefont {J.~N.}\ \bibnamefont {Ullom}},\ }\bibfield  {title} {\enquote
			{\bibinfo {title} {{Improved microwave SQUID multiplexer readout using a
						kinetic-inductance traveling-wave parametric amplifier}},}\ }\href
		{https://doi.org/10.1063/5.0149646} {\bibfield  {journal} {\bibinfo
				{journal} {Applied Physics Letters}\ }\textbf {\bibinfo {volume} {122}},\
			\bibinfo {pages} {214001} (\bibinfo {year} {2023})}\BibitemShut {NoStop}%
		\bibitem [{\citenamefont {Arute}\ \emph {et~al.}(2019)\citenamefont {Arute},
			\citenamefont {Arya}, \citenamefont {Babbush}, \citenamefont {Bacon},
			\citenamefont {Bardin}, \citenamefont {Barends}, \citenamefont {Biswas},
			\citenamefont {Boixo}, \citenamefont {Brandao}, \citenamefont {Buell},
			\citenamefont {Burkett}, \citenamefont {Chen}, \citenamefont {Chen},
			\citenamefont {Chiaro}, \citenamefont {Collins}, \citenamefont {Courtney},
			\citenamefont {Dunsworth}, \citenamefont {Farhi}, \citenamefont {Foxen},
			\citenamefont {Fowler}, \citenamefont {Gidney}, \citenamefont {Giustina},
			\citenamefont {Graff}, \citenamefont {Guerin}, \citenamefont {Habegger},
			\citenamefont {Harrigan}, \citenamefont {Hartmann}, \citenamefont {Ho},
			\citenamefont {Hoffmann}, \citenamefont {Huang}, \citenamefont {Humble},
			\citenamefont {Isakov}, \citenamefont {Jeffrey}, \citenamefont {Jiang},
			\citenamefont {Kafri}, \citenamefont {Kechedzhi}, \citenamefont {Kelly},
			\citenamefont {Klimov}, \citenamefont {Knysh}, \citenamefont {Korotkov},
			\citenamefont {Kostritsa}, \citenamefont {Landhuis}, \citenamefont
			{Lindmark}, \citenamefont {Lucero}, \citenamefont {Lyakh}, \citenamefont
			{Mandr{\`a}}, \citenamefont {McClean}, \citenamefont {McEwen}, \citenamefont
			{Megrant}, \citenamefont {Mi}, \citenamefont {Michielsen}, \citenamefont
			{Mohseni}, \citenamefont {Mutus}, \citenamefont {Naaman}, \citenamefont
			{Neeley}, \citenamefont {Neill}, \citenamefont {Niu}, \citenamefont {Ostby},
			\citenamefont {Petukhov}, \citenamefont {Platt}, \citenamefont {Quintana},
			\citenamefont {Rieffel}, \citenamefont {Roushan}, \citenamefont {Rubin},
			\citenamefont {Sank}, \citenamefont {Satzinger}, \citenamefont {Smelyanskiy},
			\citenamefont {Sung}, \citenamefont {Trevithick}, \citenamefont
			{Vainsencher}, \citenamefont {Villalonga}, \citenamefont {White},
			\citenamefont {Yao}, \citenamefont {Yeh}, \citenamefont {Zalcman},
			\citenamefont {Neven},\ and\ \citenamefont {Martinis}}]{arute2019quantum}%
		\BibitemOpen
		\bibfield  {author} {\bibinfo {author} {\bibfnamefont {F.}~\bibnamefont
				{Arute}}, \bibinfo {author} {\bibfnamefont {K.}~\bibnamefont {Arya}},
			\bibinfo {author} {\bibfnamefont {R.}~\bibnamefont {Babbush}}, \bibinfo
			{author} {\bibfnamefont {D.}~\bibnamefont {Bacon}}, \bibinfo {author}
			{\bibfnamefont {J.~C.}\ \bibnamefont {Bardin}}, \bibinfo {author}
			{\bibfnamefont {R.}~\bibnamefont {Barends}}, \bibinfo {author} {\bibfnamefont
				{R.}~\bibnamefont {Biswas}}, \bibinfo {author} {\bibfnamefont
				{S.}~\bibnamefont {Boixo}}, \bibinfo {author} {\bibfnamefont {F.~G. S.~L.}\
				\bibnamefont {Brandao}}, \bibinfo {author} {\bibfnamefont {D.~A.}\
				\bibnamefont {Buell}}, \bibinfo {author} {\bibfnamefont {B.}~\bibnamefont
				{Burkett}}, \bibinfo {author} {\bibfnamefont {Y.}~\bibnamefont {Chen}},
			\bibinfo {author} {\bibfnamefont {Z.}~\bibnamefont {Chen}}, \bibinfo {author}
			{\bibfnamefont {B.}~\bibnamefont {Chiaro}}, \bibinfo {author} {\bibfnamefont
				{R.}~\bibnamefont {Collins}}, \bibinfo {author} {\bibfnamefont
				{W.}~\bibnamefont {Courtney}}, \bibinfo {author} {\bibfnamefont
				{A.}~\bibnamefont {Dunsworth}}, \bibinfo {author} {\bibfnamefont
				{E.}~\bibnamefont {Farhi}}, \bibinfo {author} {\bibfnamefont
				{B.}~\bibnamefont {Foxen}}, \bibinfo {author} {\bibfnamefont
				{A.}~\bibnamefont {Fowler}}, \bibinfo {author} {\bibfnamefont
				{C.}~\bibnamefont {Gidney}}, \bibinfo {author} {\bibfnamefont
				{M.}~\bibnamefont {Giustina}}, \bibinfo {author} {\bibfnamefont
				{R.}~\bibnamefont {Graff}}, \bibinfo {author} {\bibfnamefont
				{K.}~\bibnamefont {Guerin}}, \bibinfo {author} {\bibfnamefont
				{S.}~\bibnamefont {Habegger}}, \bibinfo {author} {\bibfnamefont {M.~P.}\
				\bibnamefont {Harrigan}}, \bibinfo {author} {\bibfnamefont {M.~J.}\
				\bibnamefont {Hartmann}}, \bibinfo {author} {\bibfnamefont {A.}~\bibnamefont
				{Ho}}, \bibinfo {author} {\bibfnamefont {M.}~\bibnamefont {Hoffmann}},
			\bibinfo {author} {\bibfnamefont {T.}~\bibnamefont {Huang}}, \bibinfo
			{author} {\bibfnamefont {T.~S.}\ \bibnamefont {Humble}}, \bibinfo {author}
			{\bibfnamefont {S.~V.}\ \bibnamefont {Isakov}}, \bibinfo {author}
			{\bibfnamefont {E.}~\bibnamefont {Jeffrey}}, \bibinfo {author} {\bibfnamefont
				{Z.}~\bibnamefont {Jiang}}, \bibinfo {author} {\bibfnamefont
				{D.}~\bibnamefont {Kafri}}, \bibinfo {author} {\bibfnamefont
				{K.}~\bibnamefont {Kechedzhi}}, \bibinfo {author} {\bibfnamefont
				{J.}~\bibnamefont {Kelly}}, \bibinfo {author} {\bibfnamefont {P.~V.}\
				\bibnamefont {Klimov}}, \bibinfo {author} {\bibfnamefont {S.}~\bibnamefont
				{Knysh}}, \bibinfo {author} {\bibfnamefont {A.}~\bibnamefont {Korotkov}},
			\bibinfo {author} {\bibfnamefont {F.}~\bibnamefont {Kostritsa}}, \bibinfo
			{author} {\bibfnamefont {D.}~\bibnamefont {Landhuis}}, \bibinfo {author}
			{\bibfnamefont {M.}~\bibnamefont {Lindmark}}, \bibinfo {author}
			{\bibfnamefont {E.}~\bibnamefont {Lucero}}, \bibinfo {author} {\bibfnamefont
				{D.}~\bibnamefont {Lyakh}}, \bibinfo {author} {\bibfnamefont
				{S.}~\bibnamefont {Mandr{\`a}}}, \bibinfo {author} {\bibfnamefont {J.~R.}\
				\bibnamefont {McClean}}, \bibinfo {author} {\bibfnamefont {M.}~\bibnamefont
				{McEwen}}, \bibinfo {author} {\bibfnamefont {A.}~\bibnamefont {Megrant}},
			\bibinfo {author} {\bibfnamefont {X.}~\bibnamefont {Mi}}, \bibinfo {author}
			{\bibfnamefont {K.}~\bibnamefont {Michielsen}}, \bibinfo {author}
			{\bibfnamefont {M.}~\bibnamefont {Mohseni}}, \bibinfo {author} {\bibfnamefont
				{J.}~\bibnamefont {Mutus}}, \bibinfo {author} {\bibfnamefont
				{O.}~\bibnamefont {Naaman}}, \bibinfo {author} {\bibfnamefont
				{M.}~\bibnamefont {Neeley}}, \bibinfo {author} {\bibfnamefont
				{C.}~\bibnamefont {Neill}}, \bibinfo {author} {\bibfnamefont {M.~Y.}\
				\bibnamefont {Niu}}, \bibinfo {author} {\bibfnamefont {E.}~\bibnamefont
				{Ostby}}, \bibinfo {author} {\bibfnamefont {A.}~\bibnamefont {Petukhov}},
			\bibinfo {author} {\bibfnamefont {J.~C.}\ \bibnamefont {Platt}}, \bibinfo
			{author} {\bibfnamefont {C.}~\bibnamefont {Quintana}}, \bibinfo {author}
			{\bibfnamefont {E.~G.}\ \bibnamefont {Rieffel}}, \bibinfo {author}
			{\bibfnamefont {P.}~\bibnamefont {Roushan}}, \bibinfo {author} {\bibfnamefont
				{N.~C.}\ \bibnamefont {Rubin}}, \bibinfo {author} {\bibfnamefont
				{D.}~\bibnamefont {Sank}}, \bibinfo {author} {\bibfnamefont {K.~J.}\
				\bibnamefont {Satzinger}}, \bibinfo {author} {\bibfnamefont {V.}~\bibnamefont
				{Smelyanskiy}}, \bibinfo {author} {\bibfnamefont {K.~J.}\ \bibnamefont
				{Sung}}, \bibinfo {author} {\bibfnamefont {M.~D.}\ \bibnamefont
				{Trevithick}}, \bibinfo {author} {\bibfnamefont {A.}~\bibnamefont
				{Vainsencher}}, \bibinfo {author} {\bibfnamefont {B.}~\bibnamefont
				{Villalonga}}, \bibinfo {author} {\bibfnamefont {T.}~\bibnamefont {White}},
			\bibinfo {author} {\bibfnamefont {Z.~J.}\ \bibnamefont {Yao}}, \bibinfo
			{author} {\bibfnamefont {P.}~\bibnamefont {Yeh}}, \bibinfo {author}
			{\bibfnamefont {A.}~\bibnamefont {Zalcman}}, \bibinfo {author} {\bibfnamefont
				{H.}~\bibnamefont {Neven}},\ and\ \bibinfo {author} {\bibfnamefont {J.~M.}\
				\bibnamefont {Martinis}},\ }\bibfield  {title} {\enquote {\bibinfo {title}
				{Quantum supremacy using a programmable superconducting processor},}\ }\href
		{https://doi.org/10.1038/s41586-019-1666-5} {\bibfield  {journal} {\bibinfo
				{journal} {Nature}\ }\textbf {\bibinfo {volume} {574}},\ \bibinfo {pages}
			{505--510} (\bibinfo {year} {2019})}\BibitemShut {NoStop}%
		\bibitem [{\citenamefont {Krinner}\ \emph {et~al.}(2022)\citenamefont
			{Krinner}, \citenamefont {Lacroix}, \citenamefont {Remm}, \citenamefont
			{Di~Paolo}, \citenamefont {Genois}, \citenamefont {Leroux}, \citenamefont
			{Hellings}, \citenamefont {Lazar}, \citenamefont {Swiadek}, \citenamefont
			{Herrmann}, \citenamefont {Norris}, \citenamefont {Andersen}, \citenamefont
			{M{\"u}ller}, \citenamefont {Blais}, \citenamefont {Eichler},\ and\
			\citenamefont {Wallraff}}]{Krinner2022realizing}%
		\BibitemOpen
		\bibfield  {author} {\bibinfo {author} {\bibfnamefont {S.}~\bibnamefont
				{Krinner}}, \bibinfo {author} {\bibfnamefont {N.}~\bibnamefont {Lacroix}},
			\bibinfo {author} {\bibfnamefont {A.}~\bibnamefont {Remm}}, \bibinfo {author}
			{\bibfnamefont {A.}~\bibnamefont {Di~Paolo}}, \bibinfo {author}
			{\bibfnamefont {E.}~\bibnamefont {Genois}}, \bibinfo {author} {\bibfnamefont
				{C.}~\bibnamefont {Leroux}}, \bibinfo {author} {\bibfnamefont
				{C.}~\bibnamefont {Hellings}}, \bibinfo {author} {\bibfnamefont
				{S.}~\bibnamefont {Lazar}}, \bibinfo {author} {\bibfnamefont
				{F.}~\bibnamefont {Swiadek}}, \bibinfo {author} {\bibfnamefont
				{J.}~\bibnamefont {Herrmann}}, \bibinfo {author} {\bibfnamefont {G.~J.}\
				\bibnamefont {Norris}}, \bibinfo {author} {\bibfnamefont {C.~K.}\
				\bibnamefont {Andersen}}, \bibinfo {author} {\bibfnamefont {M.}~\bibnamefont
				{M{\"u}ller}}, \bibinfo {author} {\bibfnamefont {A.}~\bibnamefont {Blais}},
			\bibinfo {author} {\bibfnamefont {C.}~\bibnamefont {Eichler}},\ and\ \bibinfo
			{author} {\bibfnamefont {A.}~\bibnamefont {Wallraff}},\ }\bibfield  {title}
		{\enquote {\bibinfo {title} {Realizing repeated quantum error correction in a
					distance-three surface code},}\ }\href
		{https://doi.org/10.1038/s41586-022-04566-8} {\bibfield  {journal} {\bibinfo
				{journal} {Nature}\ }\textbf {\bibinfo {volume} {605}},\ \bibinfo {pages}
			{669--674} (\bibinfo {year} {2022})}\BibitemShut {NoStop}%
		\bibitem [{\citenamefont {Acharya}\ \emph {et~al.}(2023)\citenamefont {Acharya}
			\emph {et~al.}}]{Acharya2023suppressing}%
		\BibitemOpen
		\bibfield  {author} {\bibinfo {author} {\bibfnamefont {R.}~\bibnamefont
				{Acharya}} \emph {et~al.},\ }\bibfield  {title} {\enquote {\bibinfo {title}
				{Suppressing quantum errors by scaling a surface code logical qubit},}\
		}\href {https://doi.org/10.1038/s41586-022-05434-1} {\bibfield  {journal}
			{\bibinfo  {journal} {Nature}\ }\textbf {\bibinfo {volume} {614}},\ \bibinfo
			{pages} {676--681} (\bibinfo {year} {2023})}\BibitemShut {NoStop}%
		\bibitem [{\citenamefont {Brubaker}\ \emph {et~al.}(2017)\citenamefont
			{Brubaker}, \citenamefont {Zhong}, \citenamefont {Gurevich}, \citenamefont
			{Cahn}, \citenamefont {Lamoreaux}, \citenamefont {Simanovskaia},
			\citenamefont {Root}, \citenamefont {Lewis}, \citenamefont {Al~Kenany},
			\citenamefont {Backes}, \citenamefont {Urdinaran}, \citenamefont {Rapidis},
			\citenamefont {Shokair}, \citenamefont {van Bibber}, \citenamefont {Palken},
			\citenamefont {Malnou}, \citenamefont {Kindel}, \citenamefont {Anil},
			\citenamefont {Lehnert},\ and\ \citenamefont {Carosi}}]{brubaker2017first}%
		\BibitemOpen
		\bibfield  {author} {\bibinfo {author} {\bibfnamefont {B.~M.}\ \bibnamefont
				{Brubaker}}, \bibinfo {author} {\bibfnamefont {L.}~\bibnamefont {Zhong}},
			\bibinfo {author} {\bibfnamefont {Y.~V.}\ \bibnamefont {Gurevich}}, \bibinfo
			{author} {\bibfnamefont {S.~B.}\ \bibnamefont {Cahn}}, \bibinfo {author}
			{\bibfnamefont {S.~K.}\ \bibnamefont {Lamoreaux}}, \bibinfo {author}
			{\bibfnamefont {M.}~\bibnamefont {Simanovskaia}}, \bibinfo {author}
			{\bibfnamefont {J.~R.}\ \bibnamefont {Root}}, \bibinfo {author}
			{\bibfnamefont {S.~M.}\ \bibnamefont {Lewis}}, \bibinfo {author}
			{\bibfnamefont {S.}~\bibnamefont {Al~Kenany}}, \bibinfo {author}
			{\bibfnamefont {K.~M.}\ \bibnamefont {Backes}}, \bibinfo {author}
			{\bibfnamefont {I.}~\bibnamefont {Urdinaran}}, \bibinfo {author}
			{\bibfnamefont {N.~M.}\ \bibnamefont {Rapidis}}, \bibinfo {author}
			{\bibfnamefont {T.~M.}\ \bibnamefont {Shokair}}, \bibinfo {author}
			{\bibfnamefont {K.~A.}\ \bibnamefont {van Bibber}}, \bibinfo {author}
			{\bibfnamefont {D.~A.}\ \bibnamefont {Palken}}, \bibinfo {author}
			{\bibfnamefont {M.}~\bibnamefont {Malnou}}, \bibinfo {author} {\bibfnamefont
				{W.~F.}\ \bibnamefont {Kindel}}, \bibinfo {author} {\bibfnamefont {M.~A.}\
				\bibnamefont {Anil}}, \bibinfo {author} {\bibfnamefont {K.~W.}\ \bibnamefont
				{Lehnert}},\ and\ \bibinfo {author} {\bibfnamefont {G.}~\bibnamefont
				{Carosi}},\ }\bibfield  {title} {\enquote {\bibinfo {title} {First results
					from a microwave cavity axion search at $24\text{ }\text{
					}\ensuremath{\mu}\mathrm{eV}$},}\ }\href
		{https://doi.org/10.1103/PhysRevLett.118.061302} {\bibfield  {journal}
			{\bibinfo  {journal} {Phys. Rev. Lett.}\ }\textbf {\bibinfo {volume} {118}},\
			\bibinfo {pages} {061302} (\bibinfo {year} {2017})}\BibitemShut {NoStop}%
		\bibitem [{\citenamefont {Du}\ \emph {et~al.}(2018)\citenamefont {Du},
			\citenamefont {Force}, \citenamefont {Khatiwada}, \citenamefont {Lentz},
			\citenamefont {Ottens}, \citenamefont {Rosenberg}, \citenamefont {Rybka},
			\citenamefont {Carosi}, \citenamefont {Woollett}, \citenamefont {Bowring},
			\citenamefont {Chou}, \citenamefont {Sonnenschein}, \citenamefont {Wester},
			\citenamefont {Boutan}, \citenamefont {Oblath}, \citenamefont {Bradley},
			\citenamefont {Daw}, \citenamefont {Dixit}, \citenamefont {Clarke},
			\citenamefont {O'Kelley}, \citenamefont {Crisosto}, \citenamefont {Gleason},
			\citenamefont {Jois}, \citenamefont {Sikivie}, \citenamefont {Stern},
			\citenamefont {Sullivan}, \citenamefont {Tanner},\ and\ \citenamefont
			{Hilton}}]{du2018search}%
		\BibitemOpen
		\bibfield  {author} {\bibinfo {author} {\bibfnamefont {N.}~\bibnamefont
				{Du}}, \bibinfo {author} {\bibfnamefont {N.}~\bibnamefont {Force}}, \bibinfo
			{author} {\bibfnamefont {R.}~\bibnamefont {Khatiwada}}, \bibinfo {author}
			{\bibfnamefont {E.}~\bibnamefont {Lentz}}, \bibinfo {author} {\bibfnamefont
				{R.}~\bibnamefont {Ottens}}, \bibinfo {author} {\bibfnamefont {L.~J.}\
				\bibnamefont {Rosenberg}}, \bibinfo {author} {\bibfnamefont {G.}~\bibnamefont
				{Rybka}}, \bibinfo {author} {\bibfnamefont {G.}~\bibnamefont {Carosi}},
			\bibinfo {author} {\bibfnamefont {N.}~\bibnamefont {Woollett}}, \bibinfo
			{author} {\bibfnamefont {D.}~\bibnamefont {Bowring}}, \bibinfo {author}
			{\bibfnamefont {A.~S.}\ \bibnamefont {Chou}}, \bibinfo {author}
			{\bibfnamefont {A.}~\bibnamefont {Sonnenschein}}, \bibinfo {author}
			{\bibfnamefont {W.}~\bibnamefont {Wester}}, \bibinfo {author} {\bibfnamefont
				{C.}~\bibnamefont {Boutan}}, \bibinfo {author} {\bibfnamefont {N.~S.}\
				\bibnamefont {Oblath}}, \bibinfo {author} {\bibfnamefont {R.}~\bibnamefont
				{Bradley}}, \bibinfo {author} {\bibfnamefont {E.~J.}\ \bibnamefont {Daw}},
			\bibinfo {author} {\bibfnamefont {A.~V.}\ \bibnamefont {Dixit}}, \bibinfo
			{author} {\bibfnamefont {J.}~\bibnamefont {Clarke}}, \bibinfo {author}
			{\bibfnamefont {S.~R.}\ \bibnamefont {O'Kelley}}, \bibinfo {author}
			{\bibfnamefont {N.}~\bibnamefont {Crisosto}}, \bibinfo {author}
			{\bibfnamefont {J.~R.}\ \bibnamefont {Gleason}}, \bibinfo {author}
			{\bibfnamefont {S.}~\bibnamefont {Jois}}, \bibinfo {author} {\bibfnamefont
				{P.}~\bibnamefont {Sikivie}}, \bibinfo {author} {\bibfnamefont
				{I.}~\bibnamefont {Stern}}, \bibinfo {author} {\bibfnamefont {N.~S.}\
				\bibnamefont {Sullivan}}, \bibinfo {author} {\bibfnamefont {D.~B.}\
				\bibnamefont {Tanner}},\ and\ \bibinfo {author} {\bibfnamefont {G.~C.}\
				\bibnamefont {Hilton}} (\bibinfo {collaboration} {ADMX Collaboration}),\
		}\bibfield  {title} {\enquote {\bibinfo {title} {Search for invisible axion
					dark matter with the axion dark matter experiment},}\ }\href
		{https://doi.org/10.1103/PhysRevLett.120.151301} {\bibfield  {journal}
			{\bibinfo  {journal} {Phys. Rev. Lett.}\ }\textbf {\bibinfo {volume} {120}},\
			\bibinfo {pages} {151301} (\bibinfo {year} {2018})}\BibitemShut {NoStop}%
		\bibitem [{\citenamefont {Backes}\ \emph {et~al.}(2021)\citenamefont {Backes},
			\citenamefont {Palken}, \citenamefont {Kenany}, \citenamefont {Brubaker},
			\citenamefont {Cahn}, \citenamefont {Droster}, \citenamefont {Hilton},
			\citenamefont {Ghosh}, \citenamefont {Jackson}, \citenamefont {Lamoreaux},
			\citenamefont {Leder}, \citenamefont {Lehnert}, \citenamefont {Lewis},
			\citenamefont {Malnou}, \citenamefont {Maruyama}, \citenamefont {Rapidis},
			\citenamefont {Simanovskaia}, \citenamefont {Singh}, \citenamefont {Speller},
			\citenamefont {Urdinaran}, \citenamefont {Vale}, \citenamefont {van
				Assendelft}, \citenamefont {van Bibber},\ and\ \citenamefont
			{Wang}}]{Backes2021a}%
		\BibitemOpen
		\bibfield  {author} {\bibinfo {author} {\bibfnamefont {K.~M.}\ \bibnamefont
				{Backes}}, \bibinfo {author} {\bibfnamefont {D.~A.}\ \bibnamefont {Palken}},
			\bibinfo {author} {\bibfnamefont {S.~A.}\ \bibnamefont {Kenany}}, \bibinfo
			{author} {\bibfnamefont {B.~M.}\ \bibnamefont {Brubaker}}, \bibinfo {author}
			{\bibfnamefont {S.~B.}\ \bibnamefont {Cahn}}, \bibinfo {author}
			{\bibfnamefont {A.}~\bibnamefont {Droster}}, \bibinfo {author} {\bibfnamefont
				{G.~C.}\ \bibnamefont {Hilton}}, \bibinfo {author} {\bibfnamefont
				{S.}~\bibnamefont {Ghosh}}, \bibinfo {author} {\bibfnamefont
				{H.}~\bibnamefont {Jackson}}, \bibinfo {author} {\bibfnamefont {S.~K.}\
				\bibnamefont {Lamoreaux}}, \bibinfo {author} {\bibfnamefont {A.~F.}\
				\bibnamefont {Leder}}, \bibinfo {author} {\bibfnamefont {K.~W.}\ \bibnamefont
				{Lehnert}}, \bibinfo {author} {\bibfnamefont {S.~M.}\ \bibnamefont {Lewis}},
			\bibinfo {author} {\bibfnamefont {M.}~\bibnamefont {Malnou}}, \bibinfo
			{author} {\bibfnamefont {R.~H.}\ \bibnamefont {Maruyama}}, \bibinfo {author}
			{\bibfnamefont {N.~M.}\ \bibnamefont {Rapidis}}, \bibinfo {author}
			{\bibfnamefont {M.}~\bibnamefont {Simanovskaia}}, \bibinfo {author}
			{\bibfnamefont {S.}~\bibnamefont {Singh}}, \bibinfo {author} {\bibfnamefont
				{D.~H.}\ \bibnamefont {Speller}}, \bibinfo {author} {\bibfnamefont
				{I.}~\bibnamefont {Urdinaran}}, \bibinfo {author} {\bibfnamefont {L.~R.}\
				\bibnamefont {Vale}}, \bibinfo {author} {\bibfnamefont {E.~C.}\ \bibnamefont
				{van Assendelft}}, \bibinfo {author} {\bibfnamefont {K.}~\bibnamefont {van
					Bibber}},\ and\ \bibinfo {author} {\bibfnamefont {H.}~\bibnamefont {Wang}},\
		}\bibfield  {title} {\enquote {\bibinfo {title} {A quantum enhanced search
					for dark matter axions},}\ }\href
		{https://doi.org/10.1038/s41586-021-03226-7} {\bibfield  {journal} {\bibinfo
				{journal} {Nature}\ }\textbf {\bibinfo {volume} {590}},\ \bibinfo {pages}
			{238--242} (\bibinfo {year} {2021})}\BibitemShut {NoStop}%
		\bibitem [{\citenamefont {Yurke}\ \emph {et~al.}(1989)\citenamefont {Yurke},
			\citenamefont {Corruccini}, \citenamefont {Kaminsky}, \citenamefont {Rupp},
			\citenamefont {Smith}, \citenamefont {Silver}, \citenamefont {Simon},\ and\
			\citenamefont {Whittaker}}]{yurke1989observation}%
		\BibitemOpen
		\bibfield  {author} {\bibinfo {author} {\bibfnamefont {B.}~\bibnamefont
				{Yurke}}, \bibinfo {author} {\bibfnamefont {L.~R.}\ \bibnamefont
				{Corruccini}}, \bibinfo {author} {\bibfnamefont {P.~G.}\ \bibnamefont
				{Kaminsky}}, \bibinfo {author} {\bibfnamefont {L.~W.}\ \bibnamefont {Rupp}},
			\bibinfo {author} {\bibfnamefont {A.~D.}\ \bibnamefont {Smith}}, \bibinfo
			{author} {\bibfnamefont {A.~H.}\ \bibnamefont {Silver}}, \bibinfo {author}
			{\bibfnamefont {R.~W.}\ \bibnamefont {Simon}},\ and\ \bibinfo {author}
			{\bibfnamefont {E.~A.}\ \bibnamefont {Whittaker}},\ }\bibfield  {title}
		{\enquote {\bibinfo {title} {Observation of parametric amplification and
					deamplification in a {J}osephson parametric amplifier},}\ }\href
		{https://doi.org/10.1103/PhysRevA.39.2519} {\bibfield  {journal} {\bibinfo
				{journal} {Phys. Rev. A}\ }\textbf {\bibinfo {volume} {39}},\ \bibinfo
			{pages} {2519--2533} (\bibinfo {year} {1989})}\BibitemShut {NoStop}%
		\bibitem [{\citenamefont {Castellanos-Beltran}\ \emph
			{et~al.}(2008)\citenamefont {Castellanos-Beltran}, \citenamefont {Irwin},
			\citenamefont {Hilton}, \citenamefont {Vale},\ and\ \citenamefont
			{Lehnert}}]{castellanos2008amplification}%
		\BibitemOpen
		\bibfield  {author} {\bibinfo {author} {\bibfnamefont {M.~A.}\ \bibnamefont
				{Castellanos-Beltran}}, \bibinfo {author} {\bibfnamefont {K.~D.}\
				\bibnamefont {Irwin}}, \bibinfo {author} {\bibfnamefont {G.~C.}\ \bibnamefont
				{Hilton}}, \bibinfo {author} {\bibfnamefont {L.~R.}\ \bibnamefont {Vale}},\
			and\ \bibinfo {author} {\bibfnamefont {K.~W.}\ \bibnamefont {Lehnert}},\
		}\bibfield  {title} {\enquote {\bibinfo {title} {Amplification and squeezing
					of quantum noise with a tunable {J}osephson metamaterial},}\ }\href
		{http://dx.doi.org/10.1038/nphys1090} {\bibfield  {journal} {\bibinfo
				{journal} {Nature Physics}\ }\textbf {\bibinfo {volume} {4}},\ \bibinfo
			{pages} {929--931} (\bibinfo {year} {2008})}\BibitemShut {NoStop}%
		\bibitem [{\citenamefont {Bergeal}\ \emph {et~al.}(2010)\citenamefont
			{Bergeal}, \citenamefont {Schackert}, \citenamefont {Metcalfe}, \citenamefont
			{Vijay}, \citenamefont {Manucharyan}, \citenamefont {Frunzio}, \citenamefont
			{Prober}, \citenamefont {Schoelkopf}, \citenamefont {Girvin},\ and\
			\citenamefont {Devoret}}]{bergeal2010phase}%
		\BibitemOpen
		\bibfield  {author} {\bibinfo {author} {\bibfnamefont {N.}~\bibnamefont
				{Bergeal}}, \bibinfo {author} {\bibfnamefont {F.}~\bibnamefont {Schackert}},
			\bibinfo {author} {\bibfnamefont {M.}~\bibnamefont {Metcalfe}}, \bibinfo
			{author} {\bibfnamefont {R.}~\bibnamefont {Vijay}}, \bibinfo {author}
			{\bibfnamefont {V.~E.}\ \bibnamefont {Manucharyan}}, \bibinfo {author}
			{\bibfnamefont {L.}~\bibnamefont {Frunzio}}, \bibinfo {author} {\bibfnamefont
				{D.~E.}\ \bibnamefont {Prober}}, \bibinfo {author} {\bibfnamefont {R.~J.}\
				\bibnamefont {Schoelkopf}}, \bibinfo {author} {\bibfnamefont {S.~M.}\
				\bibnamefont {Girvin}},\ and\ \bibinfo {author} {\bibfnamefont {M.~H.}\
				\bibnamefont {Devoret}},\ }\bibfield  {title} {\enquote {\bibinfo {title}
				{Phase-preserving amplification near the quantum limit with a josephson ring
					modulator},}\ }\href {https://doi.org/https://doi.org/10.1038/nature09035}
		{\bibfield  {journal} {\bibinfo  {journal} {Nature}\ }\textbf {\bibinfo
				{volume} {465}},\ \bibinfo {pages} {64--68} (\bibinfo {year}
			{2010})}\BibitemShut {NoStop}%
		\bibitem [{\citenamefont {Macklin}\ \emph {et~al.}(2015)\citenamefont
			{Macklin}, \citenamefont {O{\textquoteright}Brien}, \citenamefont {Hover},
			\citenamefont {Schwartz}, \citenamefont {Bolkhovsky}, \citenamefont {Zhang},
			\citenamefont {Oliver},\ and\ \citenamefont {Siddiqi}}]{macklin2015near}%
		\BibitemOpen
		\bibfield  {author} {\bibinfo {author} {\bibfnamefont {C.}~\bibnamefont
				{Macklin}}, \bibinfo {author} {\bibfnamefont {K.}~\bibnamefont
				{O{\textquoteright}Brien}}, \bibinfo {author} {\bibfnamefont
				{D.}~\bibnamefont {Hover}}, \bibinfo {author} {\bibfnamefont {M.~E.}\
				\bibnamefont {Schwartz}}, \bibinfo {author} {\bibfnamefont {V.}~\bibnamefont
				{Bolkhovsky}}, \bibinfo {author} {\bibfnamefont {X.}~\bibnamefont {Zhang}},
			\bibinfo {author} {\bibfnamefont {W.~D.}\ \bibnamefont {Oliver}},\ and\
			\bibinfo {author} {\bibfnamefont {I.}~\bibnamefont {Siddiqi}},\ }\bibfield
		{title} {\enquote {\bibinfo {title} {A near{\textendash}quantum-limited
					josephson traveling-wave parametric amplifier},}\ }\href
		{https://doi.org/10.1126/science.aaa8525} {\bibfield  {journal} {\bibinfo
				{journal} {Science}\ }\textbf {\bibinfo {volume} {350}},\ \bibinfo {pages}
			{307--310} (\bibinfo {year} {2015})}\BibitemShut {NoStop}%
		\bibitem [{\citenamefont {Ranzani}\ \emph {et~al.}(2013)\citenamefont
			{Ranzani}, \citenamefont {Spietz}, \citenamefont {Popovic},\ and\
			\citenamefont {Aumentado}}]{ranzani2013two}%
		\BibitemOpen
		\bibfield  {author} {\bibinfo {author} {\bibfnamefont {L.}~\bibnamefont
				{Ranzani}}, \bibinfo {author} {\bibfnamefont {L.}~\bibnamefont {Spietz}},
			\bibinfo {author} {\bibfnamefont {Z.}~\bibnamefont {Popovic}},\ and\ \bibinfo
			{author} {\bibfnamefont {J.}~\bibnamefont {Aumentado}},\ }\bibfield  {title}
		{\enquote {\bibinfo {title} {Two-port microwave calibration at millikelvin
					temperatures},}\ }\href@noop {} {\bibfield  {journal} {\bibinfo  {journal}
				{Review of scientific instruments}\ }\textbf {\bibinfo {volume} {84}}
			(\bibinfo {year} {2013})}\BibitemShut {NoStop}%
		\bibitem [{\citenamefont {Yeh}\ and\ \citenamefont
			{Anlage}(2013)}]{yeh2013situ}%
		\BibitemOpen
		\bibfield  {author} {\bibinfo {author} {\bibfnamefont {J.-H.}\ \bibnamefont
				{Yeh}}\ and\ \bibinfo {author} {\bibfnamefont {S.~M.}\ \bibnamefont
				{Anlage}},\ }\bibfield  {title} {\enquote {\bibinfo {title} {In situ
					broadband cryogenic calibration for two-port superconducting microwave
					resonators},}\ }\href@noop {} {\bibfield  {journal} {\bibinfo  {journal}
				{Review of Scientific Instruments}\ }\textbf {\bibinfo {volume} {84}}
			(\bibinfo {year} {2013})}\BibitemShut {NoStop}%
		\bibitem [{\citenamefont {Stanley}\ \emph {et~al.}(2022)\citenamefont
			{Stanley}, \citenamefont {De~Graaf}, \citenamefont {H{\"o}nigl-Decrinis},
			\citenamefont {Lindstr{\"o}m},\ and\ \citenamefont
			{Ridler}}]{stanley2022characterizing}%
		\BibitemOpen
		\bibfield  {author} {\bibinfo {author} {\bibfnamefont {M.}~\bibnamefont
				{Stanley}}, \bibinfo {author} {\bibfnamefont {S.}~\bibnamefont {De~Graaf}},
			\bibinfo {author} {\bibfnamefont {T.}~\bibnamefont {H{\"o}nigl-Decrinis}},
			\bibinfo {author} {\bibfnamefont {T.}~\bibnamefont {Lindstr{\"o}m}},\ and\
			\bibinfo {author} {\bibfnamefont {N.~M.}\ \bibnamefont {Ridler}},\ }\bibfield
		{title} {\enquote {\bibinfo {title} {Characterizing scattering parameters of
					superconducting quantum integrated circuits at milli-kelvin temperatures},}\
		}\href@noop {} {\bibfield  {journal} {\bibinfo  {journal} {IEEE Access}\
			}\textbf {\bibinfo {volume} {10}},\ \bibinfo {pages} {43376--43386} (\bibinfo
			{year} {2022})}\BibitemShut {NoStop}%
		\bibitem [{\citenamefont {Spietz}, \citenamefont {Schoelkopf},\ and\
			\citenamefont {Pari}(2006)}]{spietz2006shot}%
		\BibitemOpen
		\bibfield  {author} {\bibinfo {author} {\bibfnamefont {L.}~\bibnamefont
				{Spietz}}, \bibinfo {author} {\bibfnamefont {R.~J.}\ \bibnamefont
				{Schoelkopf}},\ and\ \bibinfo {author} {\bibfnamefont {P.}~\bibnamefont
				{Pari}},\ }\bibfield  {title} {\enquote {\bibinfo {title} {Shot noise
					thermometry down to 10mk},}\ }\href {https://doi.org/10.1063/1.2382736}
		{\bibfield  {journal} {\bibinfo  {journal} {Applied Physics Letters}\
			}\textbf {\bibinfo {volume} {89}},\ \bibinfo {pages} {183123} (\bibinfo
			{year} {2006})}\BibitemShut {NoStop}%
		\bibitem [{\citenamefont {Chang}\ \emph {et~al.}(2016)\citenamefont {Chang},
			\citenamefont {Aumentado}, \citenamefont {Wong},\ and\ \citenamefont
			{Bardin}}]{chang2016noise}%
		\BibitemOpen
		\bibfield  {author} {\bibinfo {author} {\bibfnamefont {S.-W.}\ \bibnamefont
				{Chang}}, \bibinfo {author} {\bibfnamefont {J.}~\bibnamefont {Aumentado}},
			\bibinfo {author} {\bibfnamefont {W.-T.}\ \bibnamefont {Wong}},\ and\
			\bibinfo {author} {\bibfnamefont {J.}~\bibnamefont {Bardin}},\ }\bibfield
		{title} {\enquote {\bibinfo {title} {Noise measurement of cryogenic low noise
					amplifiers using a tunnel-junction shot-noise source},}\ }in\ \href
		{https://doi.org/10.1109/MWSYM.2016.7538226} {\emph {\bibinfo {booktitle}
				{2016 IEEE MTT-S International Microwave Symposium (IMS)}}}\ (\bibinfo
		{organization} {IEEE},\ \bibinfo {year} {2016})\ pp.\ \bibinfo {pages}
		{1--4}\BibitemShut {NoStop}%
		\bibitem [{\citenamefont {Beenakker}\ and\ \citenamefont
			{Schönenberger}(2003)}]{beenarker2003quantum}%
		\BibitemOpen
		\bibfield  {author} {\bibinfo {author} {\bibfnamefont {C.}~\bibnamefont
				{Beenakker}}\ and\ \bibinfo {author} {\bibfnamefont {C.}~\bibnamefont
				{Schönenberger}},\ }\bibfield  {title} {\enquote {\bibinfo {title} {{Quantum
						Shot Noise}},}\ }\href {https://doi.org/10.1063/1.1583532} {\bibfield
			{journal} {\bibinfo  {journal} {Physics Today}\ }\textbf {\bibinfo {volume}
				{56}},\ \bibinfo {pages} {37--42} (\bibinfo {year} {2003})}\BibitemShut
		{NoStop}%
		\bibitem [{\citenamefont {Walls}\ and\ \citenamefont
			{Milburn}(2011)}]{wallsmilburn2007quantum}%
		\BibitemOpen
		\bibfield  {author} {\bibinfo {author} {\bibfnamefont {D.~F.}\ \bibnamefont
				{Walls}}\ and\ \bibinfo {author} {\bibfnamefont {G.~J.}\ \bibnamefont
				{Milburn}},\ }\href@noop {} {\emph {\bibinfo {title} {Quantum Optics, 2nd
					Edition}}}\ (\bibinfo  {publisher} {Springer Berlin, Heidelberg},\ \bibinfo
		{year} {2011})\BibitemShut {NoStop}%
		\bibitem [{\citenamefont {Friis}(1944)}]{Friis1944noise}%
		\BibitemOpen
		\bibfield  {author} {\bibinfo {author} {\bibfnamefont {H.}~\bibnamefont
				{Friis}},\ }\bibfield  {title} {\enquote {\bibinfo {title} {Noise figures of
					radio receivers},}\ }\href {https://doi.org/10.1109/JRPROC.1944.232049}
		{\bibfield  {journal} {\bibinfo  {journal} {Proceedings of the IRE}\ }\textbf
			{\bibinfo {volume} {32}},\ \bibinfo {pages} {419--422} (\bibinfo {year}
			{1944})}\BibitemShut {NoStop}%
		\bibitem [{\citenamefont {{Evans}}\ and\ \citenamefont
			{{McLeish}}(1977)}]{Evans1977rf}%
		\BibitemOpen
		\bibfield  {author} {\bibinfo {author} {\bibfnamefont {G.}~\bibnamefont
				{{Evans}}}\ and\ \bibinfo {author} {\bibfnamefont {C.~W.}\ \bibnamefont
				{{McLeish}}},\ }\href {https://ui.adsabs.harvard.edu/abs/1977ah...book.....E}
		{\emph {\bibinfo {title} {{RF Radiometer Handbook}}}}\ (\bibinfo {year}
		{1977})\ \bibinfo {note} {provided by the SAO/NASA Astrophysics Data
			System}\BibitemShut {NoStop}%
		\bibitem [{\citenamefont {Zobrist}\ \emph {et~al.}(2019)\citenamefont
			{Zobrist}, \citenamefont {Eom}, \citenamefont {Day}, \citenamefont {Mazin},
			\citenamefont {Meeker}, \citenamefont {Bumble}, \citenamefont {LeDuc},
			\citenamefont {Coiffard}, \citenamefont {Szypryt}, \citenamefont {Fruitwala},
			\citenamefont {Lipartito},\ and\ \citenamefont
			{Bockstiegel}}]{zobrist2019wide}%
		\BibitemOpen
		\bibfield  {author} {\bibinfo {author} {\bibfnamefont {N.}~\bibnamefont
				{Zobrist}}, \bibinfo {author} {\bibfnamefont {B.~H.}\ \bibnamefont {Eom}},
			\bibinfo {author} {\bibfnamefont {P.}~\bibnamefont {Day}}, \bibinfo {author}
			{\bibfnamefont {B.~A.}\ \bibnamefont {Mazin}}, \bibinfo {author}
			{\bibfnamefont {S.~R.}\ \bibnamefont {Meeker}}, \bibinfo {author}
			{\bibfnamefont {B.}~\bibnamefont {Bumble}}, \bibinfo {author} {\bibfnamefont
				{H.~G.}\ \bibnamefont {LeDuc}}, \bibinfo {author} {\bibfnamefont
				{G.}~\bibnamefont {Coiffard}}, \bibinfo {author} {\bibfnamefont
				{P.}~\bibnamefont {Szypryt}}, \bibinfo {author} {\bibfnamefont
				{N.}~\bibnamefont {Fruitwala}}, \bibinfo {author} {\bibfnamefont
				{I.}~\bibnamefont {Lipartito}},\ and\ \bibinfo {author} {\bibfnamefont
				{C.}~\bibnamefont {Bockstiegel}},\ }\bibfield  {title} {\enquote {\bibinfo
				{title} {{Wide-band parametric amplifier readout and resolution of optical
						microwave kinetic inductance detectors}},}\ }\href
		{https://doi.org/10.1063/1.5098469} {\bibfield  {journal} {\bibinfo
				{journal} {Applied Physics Letters}\ }\textbf {\bibinfo {volume} {115}}
			(\bibinfo {year} {2019}),\ 10.1063/1.5098469},\ \bibinfo {note}
		{042601}\BibitemShut {NoStop}%
		\bibitem [{\citenamefont {Malnou}\ \emph {et~al.}(2018)\citenamefont {Malnou},
			\citenamefont {Palken}, \citenamefont {Vale}, \citenamefont {Hilton},\ and\
			\citenamefont {Lehnert}}]{malnou2018optimal}%
		\BibitemOpen
		\bibfield  {author} {\bibinfo {author} {\bibfnamefont {M.}~\bibnamefont
				{Malnou}}, \bibinfo {author} {\bibfnamefont {D.~A.}\ \bibnamefont {Palken}},
			\bibinfo {author} {\bibfnamefont {L.~R.}\ \bibnamefont {Vale}}, \bibinfo
			{author} {\bibfnamefont {G.~C.}\ \bibnamefont {Hilton}},\ and\ \bibinfo
			{author} {\bibfnamefont {K.~W.}\ \bibnamefont {Lehnert}},\ }\bibfield
		{title} {\enquote {\bibinfo {title} {Optimal operation of a josephson
					parametric amplifier for vacuum squeezing},}\ }\href
		{https://doi.org/10.1103/PhysRevApplied.9.044023} {\bibfield  {journal}
			{\bibinfo  {journal} {Phys. Rev. Appl.}\ }\textbf {\bibinfo {volume} {9}},\
			\bibinfo {pages} {044023} (\bibinfo {year} {2018})}\BibitemShut {NoStop}%
		\bibitem [{\citenamefont {Malnou}\ \emph {et~al.}(2021)\citenamefont {Malnou},
			\citenamefont {Vissers}, \citenamefont {Wheeler}, \citenamefont {Aumentado},
			\citenamefont {Hubmayr}, \citenamefont {Ullom},\ and\ \citenamefont
			{Gao}}]{malnou2021three}%
		\BibitemOpen
		\bibfield  {author} {\bibinfo {author} {\bibfnamefont {M.}~\bibnamefont
				{Malnou}}, \bibinfo {author} {\bibfnamefont {M.}~\bibnamefont {Vissers}},
			\bibinfo {author} {\bibfnamefont {J.}~\bibnamefont {Wheeler}}, \bibinfo
			{author} {\bibfnamefont {J.}~\bibnamefont {Aumentado}}, \bibinfo {author}
			{\bibfnamefont {J.}~\bibnamefont {Hubmayr}}, \bibinfo {author} {\bibfnamefont
				{J.}~\bibnamefont {Ullom}},\ and\ \bibinfo {author} {\bibfnamefont
				{J.}~\bibnamefont {Gao}},\ }\bibfield  {title} {\enquote {\bibinfo {title}
				{Three-wave mixing kinetic inductance traveling-wave amplifier with
					near-quantum-limited noise performance},}\ }\href
		{https://doi.org/10.1103/PRXQuantum.2.010302} {\bibfield  {journal} {\bibinfo
				{journal} {PRX Quantum}\ }\textbf {\bibinfo {volume} {2}},\ \bibinfo {pages}
			{010302} (\bibinfo {year} {2021})}\BibitemShut {NoStop}%
		\bibitem [{\citenamefont {Simbierowicz}\ \emph {et~al.}(2021)\citenamefont
			{Simbierowicz}, \citenamefont {Vesterinen}, \citenamefont {Milem},
			\citenamefont {Lintunen}, \citenamefont {Oksanen}, \citenamefont {Roschier},
			\citenamefont {Grönberg}, \citenamefont {Hassel}, \citenamefont
			{Gunnarsson},\ and\ \citenamefont {Lake}}]{Simbierowicz2021characterizing}%
		\BibitemOpen
		\bibfield  {author} {\bibinfo {author} {\bibfnamefont {S.}~\bibnamefont
				{Simbierowicz}}, \bibinfo {author} {\bibfnamefont {V.}~\bibnamefont
				{Vesterinen}}, \bibinfo {author} {\bibfnamefont {J.}~\bibnamefont {Milem}},
			\bibinfo {author} {\bibfnamefont {A.}~\bibnamefont {Lintunen}}, \bibinfo
			{author} {\bibfnamefont {M.}~\bibnamefont {Oksanen}}, \bibinfo {author}
			{\bibfnamefont {L.}~\bibnamefont {Roschier}}, \bibinfo {author}
			{\bibfnamefont {L.}~\bibnamefont {Grönberg}}, \bibinfo {author}
			{\bibfnamefont {J.}~\bibnamefont {Hassel}}, \bibinfo {author} {\bibfnamefont
				{D.}~\bibnamefont {Gunnarsson}},\ and\ \bibinfo {author} {\bibfnamefont
				{R.~E.}\ \bibnamefont {Lake}},\ }\bibfield  {title} {\enquote {\bibinfo
				{title} {{Characterizing cryogenic amplifiers with a matched
						temperature-variable noise source}},}\ }\href
		{https://doi.org/10.1063/5.0028951} {\bibfield  {journal} {\bibinfo
				{journal} {Review of Scientific Instruments}\ }\textbf {\bibinfo {volume}
				{92}} (\bibinfo {year} {2021}),\ 10.1063/5.0028951},\ \bibinfo {note}
		{034708},\ \Eprint
		{https://arxiv.org/abs/https://pubs.aip.org/aip/rsi/article-pdf/doi/10.1063/5.0028951/13865090/034708\_1\_online.pdf}
		{https://pubs.aip.org/aip/rsi/article-pdf/doi/10.1063/5.0028951/13865090/034708\_1\_online.pdf}
		\BibitemShut {NoStop}%
		\bibitem [{\citenamefont {Spietz}\ \emph {et~al.}(2003)\citenamefont {Spietz},
			\citenamefont {Lehnert}, \citenamefont {Siddiqi},\ and\ \citenamefont
			{Schoelkopf}}]{Spietz2003primary}%
		\BibitemOpen
		\bibfield  {author} {\bibinfo {author} {\bibfnamefont {L.}~\bibnamefont
				{Spietz}}, \bibinfo {author} {\bibfnamefont {K.~W.}\ \bibnamefont {Lehnert}},
			\bibinfo {author} {\bibfnamefont {I.}~\bibnamefont {Siddiqi}},\ and\ \bibinfo
			{author} {\bibfnamefont {R.~J.}\ \bibnamefont {Schoelkopf}},\ }\bibfield
		{title} {\enquote {\bibinfo {title} {Primary electronic thermometry using the
					shot noise of a tunnel junction},}\ }\href
		{https://doi.org/10.1126/science.1084647} {\bibfield  {journal} {\bibinfo
				{journal} {Science}\ }\textbf {\bibinfo {volume} {300}},\ \bibinfo {pages}
			{1929--1932} (\bibinfo {year} {2003})}\BibitemShut {NoStop}%
		\bibitem [{\citenamefont {Cano}, \citenamefont {Wadefalk},\ and\ \citenamefont
			{Gallego-Puyol}(2010)}]{cano2010ultra}%
		\BibitemOpen
		\bibfield  {author} {\bibinfo {author} {\bibfnamefont {J.~L.}\ \bibnamefont
				{Cano}}, \bibinfo {author} {\bibfnamefont {N.}~\bibnamefont {Wadefalk}},\
			and\ \bibinfo {author} {\bibfnamefont {J.~D.}\ \bibnamefont
				{Gallego-Puyol}},\ }\bibfield  {title} {\enquote {\bibinfo {title}
				{Ultra-wideband chip attenuator for precise noise measurements at cryogenic
					temperatures},}\ }\href {https://doi.org/10.1109/TMTT.2010.2058276}
		{\bibfield  {journal} {\bibinfo  {journal} {IEEE Transactions on Microwave
					Theory and Techniques}\ }\textbf {\bibinfo {volume} {58}},\ \bibinfo {pages}
			{2504--2510} (\bibinfo {year} {2010})}\BibitemShut {NoStop}%
		\bibitem [{\citenamefont {White}\ \emph {et~al.}(2015)\citenamefont {White},
			\citenamefont {Mutus}, \citenamefont {Hoi}, \citenamefont {Barends},
			\citenamefont {Campbell}, \citenamefont {Chen}, \citenamefont {Chen},
			\citenamefont {Chiaro}, \citenamefont {Dunsworth}, \citenamefont {Jeffrey},
			\citenamefont {Kelly}, \citenamefont {Megrant}, \citenamefont {Neill},
			\citenamefont {O'Malley}, \citenamefont {Roushan}, \citenamefont {Sank},
			\citenamefont {Vainsencher}, \citenamefont {Wenner}, \citenamefont
			{Chaudhuri}, \citenamefont {Gao},\ and\ \citenamefont
			{Martinis}}]{white2015traveling}%
		\BibitemOpen
		\bibfield  {author} {\bibinfo {author} {\bibfnamefont {T.~C.}\ \bibnamefont
				{White}}, \bibinfo {author} {\bibfnamefont {J.~Y.}\ \bibnamefont {Mutus}},
			\bibinfo {author} {\bibfnamefont {I.-C.}\ \bibnamefont {Hoi}}, \bibinfo
			{author} {\bibfnamefont {R.}~\bibnamefont {Barends}}, \bibinfo {author}
			{\bibfnamefont {B.}~\bibnamefont {Campbell}}, \bibinfo {author}
			{\bibfnamefont {Y.}~\bibnamefont {Chen}}, \bibinfo {author} {\bibfnamefont
				{Z.}~\bibnamefont {Chen}}, \bibinfo {author} {\bibfnamefont {B.}~\bibnamefont
				{Chiaro}}, \bibinfo {author} {\bibfnamefont {A.}~\bibnamefont {Dunsworth}},
			\bibinfo {author} {\bibfnamefont {E.}~\bibnamefont {Jeffrey}}, \bibinfo
			{author} {\bibfnamefont {J.}~\bibnamefont {Kelly}}, \bibinfo {author}
			{\bibfnamefont {A.}~\bibnamefont {Megrant}}, \bibinfo {author} {\bibfnamefont
				{C.}~\bibnamefont {Neill}}, \bibinfo {author} {\bibfnamefont {P.~J.~J.}\
				\bibnamefont {O'Malley}}, \bibinfo {author} {\bibfnamefont {P.}~\bibnamefont
				{Roushan}}, \bibinfo {author} {\bibfnamefont {D.}~\bibnamefont {Sank}},
			\bibinfo {author} {\bibfnamefont {A.}~\bibnamefont {Vainsencher}}, \bibinfo
			{author} {\bibfnamefont {J.}~\bibnamefont {Wenner}}, \bibinfo {author}
			{\bibfnamefont {S.}~\bibnamefont {Chaudhuri}}, \bibinfo {author}
			{\bibfnamefont {J.}~\bibnamefont {Gao}},\ and\ \bibinfo {author}
			{\bibfnamefont {J.~M.}\ \bibnamefont {Martinis}},\ }\bibfield  {title}
		{\enquote {\bibinfo {title} {Traveling wave parametric amplifier with
					josephson junctions using minimal resonator phase matching},}\ }\href
		{https://doi.org/10.1063/1.4922348} {\bibfield  {journal} {\bibinfo
				{journal} {Applied Physics Letters}\ }\textbf {\bibinfo {volume} {106}},\
			\bibinfo {pages} {242601} (\bibinfo {year} {2015})}\BibitemShut {NoStop}%
		\bibitem [{\citenamefont {Malnou}\ \emph {et~al.}(2022)\citenamefont {Malnou},
			\citenamefont {Aumentado}, \citenamefont {Vissers}, \citenamefont {Wheeler},
			\citenamefont {Hubmayr}, \citenamefont {Ullom},\ and\ \citenamefont
			{Gao}}]{malnou2022performance}%
		\BibitemOpen
		\bibfield  {author} {\bibinfo {author} {\bibfnamefont {M.}~\bibnamefont
				{Malnou}}, \bibinfo {author} {\bibfnamefont {J.}~\bibnamefont {Aumentado}},
			\bibinfo {author} {\bibfnamefont {M.}~\bibnamefont {Vissers}}, \bibinfo
			{author} {\bibfnamefont {J.}~\bibnamefont {Wheeler}}, \bibinfo {author}
			{\bibfnamefont {J.}~\bibnamefont {Hubmayr}}, \bibinfo {author} {\bibfnamefont
				{J.}~\bibnamefont {Ullom}},\ and\ \bibinfo {author} {\bibfnamefont
				{J.}~\bibnamefont {Gao}},\ }\bibfield  {title} {\enquote {\bibinfo {title}
				{Performance of a kinetic inductance traveling-wave parametric amplifier at 4
					kelvin: Toward an alternative to semiconductor amplifiers},}\ }\href
		{https://doi.org/10.1103/PhysRevApplied.17.044009} {\bibfield  {journal}
			{\bibinfo  {journal} {Phys. Rev. Appl.}\ }\textbf {\bibinfo {volume} {17}},\
			\bibinfo {pages} {044009} (\bibinfo {year} {2022})}\BibitemShut {NoStop}%
		\bibitem [{\citenamefont {Lecocq}\ \emph {et~al.}(2017)\citenamefont {Lecocq},
			\citenamefont {Ranzani}, \citenamefont {Peterson}, \citenamefont {Cicak},
			\citenamefont {Simmonds}, \citenamefont {Teufel},\ and\ \citenamefont
			{Aumentado}}]{lecocq2017nonreciprocal}%
		\BibitemOpen
		\bibfield  {author} {\bibinfo {author} {\bibfnamefont {F.}~\bibnamefont
				{Lecocq}}, \bibinfo {author} {\bibfnamefont {L.}~\bibnamefont {Ranzani}},
			\bibinfo {author} {\bibfnamefont {G.~A.}\ \bibnamefont {Peterson}}, \bibinfo
			{author} {\bibfnamefont {K.}~\bibnamefont {Cicak}}, \bibinfo {author}
			{\bibfnamefont {R.~W.}\ \bibnamefont {Simmonds}}, \bibinfo {author}
			{\bibfnamefont {J.~D.}\ \bibnamefont {Teufel}},\ and\ \bibinfo {author}
			{\bibfnamefont {J.}~\bibnamefont {Aumentado}},\ }\bibfield  {title} {\enquote
			{\bibinfo {title} {Nonreciprocal microwave signal processing with a
					field-programmable josephson amplifier},}\ }\href
		{https://doi.org/10.1103/PhysRevApplied.7.024028} {\bibfield  {journal}
			{\bibinfo  {journal} {Phys. Rev. Appl.}\ }\textbf {\bibinfo {volume} {7}},\
			\bibinfo {pages} {024028} (\bibinfo {year} {2017})}\BibitemShut {NoStop}%
		\bibitem [{\citenamefont {Caves}(1982)}]{caves1982quantum}%
		\BibitemOpen
		\bibfield  {author} {\bibinfo {author} {\bibfnamefont {C.~M.}\ \bibnamefont
				{Caves}},\ }\bibfield  {title} {\enquote {\bibinfo {title} {Quantum limits on
					noise in linear amplifiers},}\ }\href
		{https://doi.org/10.1103/PhysRevD.26.1817} {\bibfield  {journal} {\bibinfo
				{journal} {Phys. Rev. D}\ }\textbf {\bibinfo {volume} {26}},\ \bibinfo
			{pages} {1817--1839} (\bibinfo {year} {1982})}\BibitemShut {NoStop}%
		\bibitem [{\citenamefont {Boyd}(2019)}]{boyd2019nonlinear}%
		\BibitemOpen
		\bibfield  {author} {\bibinfo {author} {\bibfnamefont {R.~W.}\ \bibnamefont
				{Boyd}},\ }\href@noop {} {\emph {\bibinfo {title} {Nonlinear Optics}}}\
		(\bibinfo  {publisher} {Academic press},\ \bibinfo {year} {2019})\BibitemShut
		{NoStop}%
		\bibitem [{\citenamefont {Yamamoto}\ and\ \citenamefont
			{Mukai}(1990)}]{yamamoto1990fundamentals}%
		\BibitemOpen
		\bibfield  {author} {\bibinfo {author} {\bibfnamefont {Y.}~\bibnamefont
				{Yamamoto}}\ and\ \bibinfo {author} {\bibfnamefont {T.}~\bibnamefont
				{Mukai}},\ }\bibfield  {title} {\enquote {\bibinfo {title} {Fundamentals of
					optical amplifiers},}\ }\href@noop {} {\bibfield  {journal} {\bibinfo
				{journal} {Optical and quantum electronics}\ }\textbf {\bibinfo {volume}
				{21}},\ \bibinfo {pages} {S1--S14} (\bibinfo {year} {1990})}\BibitemShut
		{NoStop}%
		\bibitem [{\citenamefont {Mallet}\ \emph {et~al.}(2011)\citenamefont {Mallet},
			\citenamefont {Castellanos-Beltran}, \citenamefont {Ku}, \citenamefont
			{Glancy}, \citenamefont {Knill}, \citenamefont {Irwin}, \citenamefont
			{Hilton}, \citenamefont {Vale},\ and\ \citenamefont
			{Lehnert}}]{mallet2011quantum}%
		\BibitemOpen
		\bibfield  {author} {\bibinfo {author} {\bibfnamefont {F.}~\bibnamefont
				{Mallet}}, \bibinfo {author} {\bibfnamefont {M.~A.}\ \bibnamefont
				{Castellanos-Beltran}}, \bibinfo {author} {\bibfnamefont {H.~S.}\
				\bibnamefont {Ku}}, \bibinfo {author} {\bibfnamefont {S.}~\bibnamefont
				{Glancy}}, \bibinfo {author} {\bibfnamefont {E.}~\bibnamefont {Knill}},
			\bibinfo {author} {\bibfnamefont {K.~D.}\ \bibnamefont {Irwin}}, \bibinfo
			{author} {\bibfnamefont {G.~C.}\ \bibnamefont {Hilton}}, \bibinfo {author}
			{\bibfnamefont {L.~R.}\ \bibnamefont {Vale}},\ and\ \bibinfo {author}
			{\bibfnamefont {K.~W.}\ \bibnamefont {Lehnert}},\ }\bibfield  {title}
		{\enquote {\bibinfo {title} {Quantum state tomography of an itinerant
					squeezed microwave field},}\ }\href
		{https://doi.org/10.1103/PhysRevLett.106.220502} {\bibfield  {journal}
			{\bibinfo  {journal} {Phys. Rev. Lett.}\ }\textbf {\bibinfo {volume} {106}},\
			\bibinfo {pages} {220502} (\bibinfo {year} {2011})}\BibitemShut {NoStop}%
		\bibitem [{\citenamefont {Boutin}\ \emph {et~al.}(2017)\citenamefont {Boutin},
			\citenamefont {Toyli}, \citenamefont {Venkatramani}, \citenamefont {Eddins},
			\citenamefont {Siddiqi},\ and\ \citenamefont {Blais}}]{boutin2017effect}%
		\BibitemOpen
		\bibfield  {author} {\bibinfo {author} {\bibfnamefont {S.}~\bibnamefont
				{Boutin}}, \bibinfo {author} {\bibfnamefont {D.~M.}\ \bibnamefont {Toyli}},
			\bibinfo {author} {\bibfnamefont {A.~V.}\ \bibnamefont {Venkatramani}},
			\bibinfo {author} {\bibfnamefont {A.~W.}\ \bibnamefont {Eddins}}, \bibinfo
			{author} {\bibfnamefont {I.}~\bibnamefont {Siddiqi}},\ and\ \bibinfo {author}
			{\bibfnamefont {A.}~\bibnamefont {Blais}},\ }\bibfield  {title} {\enquote
			{\bibinfo {title} {Effect of higher-order nonlinearities on amplification and
					squeezing in josephson parametric amplifiers},}\ }\href
		{https://doi.org/10.1103/PhysRevApplied.8.054030} {\bibfield  {journal}
			{\bibinfo  {journal} {Phys. Rev. Appl.}\ }\textbf {\bibinfo {volume} {8}},\
			\bibinfo {pages} {054030} (\bibinfo {year} {2017})}\BibitemShut {NoStop}%
		\bibitem [{\citenamefont {Remm}\ \emph {et~al.}(2023)\citenamefont {Remm},
			\citenamefont {Krinner}, \citenamefont {Lacroix}, \citenamefont {Hellings},
			\citenamefont {Swiadek}, \citenamefont {Norris}, \citenamefont {Eichler},\
			and\ \citenamefont {Wallraff}}]{remm2023prapplied}%
		\BibitemOpen
		\bibfield  {author} {\bibinfo {author} {\bibfnamefont {A.}~\bibnamefont
				{Remm}}, \bibinfo {author} {\bibfnamefont {S.}~\bibnamefont {Krinner}},
			\bibinfo {author} {\bibfnamefont {N.}~\bibnamefont {Lacroix}}, \bibinfo
			{author} {\bibfnamefont {C.}~\bibnamefont {Hellings}}, \bibinfo {author}
			{\bibfnamefont {F.~m.~c.}\ \bibnamefont {Swiadek}}, \bibinfo {author}
			{\bibfnamefont {G.~J.}\ \bibnamefont {Norris}}, \bibinfo {author}
			{\bibfnamefont {C.}~\bibnamefont {Eichler}},\ and\ \bibinfo {author}
			{\bibfnamefont {A.}~\bibnamefont {Wallraff}},\ }\bibfield  {title} {\enquote
			{\bibinfo {title} {Intermodulation distortion in a josephson traveling-wave
					parametric amplifier},}\ }\href
		{https://doi.org/10.1103/PhysRevApplied.20.034027} {\bibfield  {journal}
			{\bibinfo  {journal} {Phys. Rev. Appl.}\ }\textbf {\bibinfo {volume} {20}},\
			\bibinfo {pages} {034027} (\bibinfo {year} {2023})}\BibitemShut {NoStop}%
		\bibitem [{\citenamefont {O'Brien}\ \emph {et~al.}(2014)\citenamefont
			{O'Brien}, \citenamefont {Macklin}, \citenamefont {Siddiqi},\ and\
			\citenamefont {Zhang}}]{obrien2014resonant}%
		\BibitemOpen
		\bibfield  {author} {\bibinfo {author} {\bibfnamefont {K.}~\bibnamefont
				{O'Brien}}, \bibinfo {author} {\bibfnamefont {C.}~\bibnamefont {Macklin}},
			\bibinfo {author} {\bibfnamefont {I.}~\bibnamefont {Siddiqi}},\ and\ \bibinfo
			{author} {\bibfnamefont {X.}~\bibnamefont {Zhang}},\ }\bibfield  {title}
		{\enquote {\bibinfo {title} {Resonant phase matching of josephson junction
					traveling wave parametric amplifiers},}\ }\href
		{https://doi.org/10.1103/PhysRevLett.113.157001} {\bibfield  {journal}
			{\bibinfo  {journal} {Phys. Rev. Lett.}\ }\textbf {\bibinfo {volume} {113}},\
			\bibinfo {pages} {157001} (\bibinfo {year} {2014})}\BibitemShut {NoStop}%
		\bibitem [{\citenamefont {Wustmann}\ and\ \citenamefont
			{Shumeiko}(2013)}]{wustmann2013parametric}%
		\BibitemOpen
		\bibfield  {author} {\bibinfo {author} {\bibfnamefont {W.}~\bibnamefont
				{Wustmann}}\ and\ \bibinfo {author} {\bibfnamefont {V.}~\bibnamefont
				{Shumeiko}},\ }\bibfield  {title} {\enquote {\bibinfo {title} {Parametric
					resonance in tunable superconducting cavities},}\ }\href
		{https://doi.org/10.1103/PhysRevB.87.184501} {\bibfield  {journal} {\bibinfo
				{journal} {Phys. Rev. B}\ }\textbf {\bibinfo {volume} {87}},\ \bibinfo
			{pages} {184501} (\bibinfo {year} {2013})}\BibitemShut {NoStop}%
		\bibitem [{\citenamefont {Bienfait}\ \emph {et~al.}(2017)\citenamefont
			{Bienfait}, \citenamefont {Campagne-Ibarcq}, \citenamefont {Kiilerich},
			\citenamefont {Zhou}, \citenamefont {Probst}, \citenamefont {Pla},
			\citenamefont {Schenkel}, \citenamefont {Vion}, \citenamefont {Esteve},
			\citenamefont {Morton}, \citenamefont {Moelmer},\ and\ \citenamefont
			{Bertet}}]{bienfait2017magnetic}%
		\BibitemOpen
		\bibfield  {author} {\bibinfo {author} {\bibfnamefont {A.}~\bibnamefont
				{Bienfait}}, \bibinfo {author} {\bibfnamefont {P.}~\bibnamefont
				{Campagne-Ibarcq}}, \bibinfo {author} {\bibfnamefont {A.~H.}\ \bibnamefont
				{Kiilerich}}, \bibinfo {author} {\bibfnamefont {X.}~\bibnamefont {Zhou}},
			\bibinfo {author} {\bibfnamefont {S.}~\bibnamefont {Probst}}, \bibinfo
			{author} {\bibfnamefont {J.~J.}\ \bibnamefont {Pla}}, \bibinfo {author}
			{\bibfnamefont {T.}~\bibnamefont {Schenkel}}, \bibinfo {author}
			{\bibfnamefont {D.}~\bibnamefont {Vion}}, \bibinfo {author} {\bibfnamefont
				{D.}~\bibnamefont {Esteve}}, \bibinfo {author} {\bibfnamefont {J.~J.~L.}\
				\bibnamefont {Morton}}, \bibinfo {author} {\bibfnamefont {K.}~\bibnamefont
				{Moelmer}},\ and\ \bibinfo {author} {\bibfnamefont {P.}~\bibnamefont
				{Bertet}},\ }\bibfield  {title} {\enquote {\bibinfo {title} {Magnetic
					resonance with squeezed microwaves},}\ }\href
		{https://doi.org/10.1103/PhysRevX.7.041011} {\bibfield  {journal} {\bibinfo
				{journal} {Phys. Rev. X}\ }\textbf {\bibinfo {volume} {7}},\ \bibinfo {pages}
			{041011} (\bibinfo {year} {2017})}\BibitemShut {NoStop}%
		\bibitem [{\citenamefont {Parker}\ \emph {et~al.}(2022)\citenamefont {Parker},
			\citenamefont {Savytskyi}, \citenamefont {Vine}, \citenamefont {Laucht},
			\citenamefont {Duty}, \citenamefont {Morello}, \citenamefont {Grimsmo},\ and\
			\citenamefont {Pla}}]{parker2022degenerate}%
		\BibitemOpen
		\bibfield  {author} {\bibinfo {author} {\bibfnamefont {D.~J.}\ \bibnamefont
				{Parker}}, \bibinfo {author} {\bibfnamefont {M.}~\bibnamefont {Savytskyi}},
			\bibinfo {author} {\bibfnamefont {W.}~\bibnamefont {Vine}}, \bibinfo {author}
			{\bibfnamefont {A.}~\bibnamefont {Laucht}}, \bibinfo {author} {\bibfnamefont
				{T.}~\bibnamefont {Duty}}, \bibinfo {author} {\bibfnamefont {A.}~\bibnamefont
				{Morello}}, \bibinfo {author} {\bibfnamefont {A.~L.}\ \bibnamefont
				{Grimsmo}},\ and\ \bibinfo {author} {\bibfnamefont {J.~J.}\ \bibnamefont
				{Pla}},\ }\bibfield  {title} {\enquote {\bibinfo {title} {Degenerate
					parametric amplification via three-wave mixing using kinetic inductance},}\
		}\href {https://doi.org/10.1103/PhysRevApplied.17.034064} {\bibfield
			{journal} {\bibinfo  {journal} {Phys. Rev. Appl.}\ }\textbf {\bibinfo
				{volume} {17}},\ \bibinfo {pages} {034064} (\bibinfo {year}
			{2022})}\BibitemShut {NoStop}%
		\bibitem [{\citenamefont {Peterson}(2020)}]{peterson2020parametric}%
		\BibitemOpen
		\bibfield  {author} {\bibinfo {author} {\bibfnamefont {G.~A.}\ \bibnamefont
				{Peterson}},\ }\emph {\bibinfo {title} {Parametric Coupling between
				Microwaves and Motion in Quantum Circuits: Fundamental Limits and
				Applications}},\ \href@noop {} {Ph.D. thesis},\ \bibinfo  {school}
		{University of Colorado at Boulder} (\bibinfo {year} {2020})\BibitemShut
		{NoStop}%
	\end{thebibliography}
	%

\end{document}